\documentclass[twocolumn]{aastex63}
\usepackage{lineno}
\usepackage{xcolor}
\usepackage{xspace}
\usepackage{comment}
\usepackage{graphicx}
\usepackage[utf8]{inputenc}
\graphicspath{ {figures/} }

\definecolor{tiffany}{RGB}{79, 166, 158}

\newcommand{\eventname}{GW190814\xspace}

\newcommand{\aprilevent}{GW190412\xspace}

\newcommand{\lho}{{LIGO Hanford}\xspace}
\newcommand{\llo}{{LIGO Livingston}\xspace}
\newcommand{\LHO}{{LIGO Hanford}\xspace}
\newcommand{\LLO}{{LIGO Livingston}\xspace}
\newcommand{\virgo}{{Virgo}\xspace}

\newcommand{\lalinference}{{\sc LALInference\xspace}}

\newcommand{\bilby}{{\sc bilby\xspace}}
\newcommand{\pbilby}{{\sc pbilby\xspace}}
\newcommand{\gstlal}{{\sc GstLAL\xspace}}
\newcommand{\pycbc}{{\sc PyCBC\xspace}}

\newcommand{\cwb}{{\sc cWB\xspace}}
\newcommand{\bayeswave}{{\sc BayesWave\xspace}}
\newcommand{\mbta}{{\sc MBTA\xspace}}
\newcommand{\spiir}{{\sc SPIIR\xspace}}
\newcommand{\bayestar}{{\sc BAYESTAR\xspace}}

\begin{document}
\def\Mtotalsource{\,25.8_{-0.9}^{+1.0}}
\def\Mtotalsourcemedian{\,25.8}
\def\Mtotalsourceupper{\,26.8}
\def\Mtotalsourcelower{\,24.9}
\def\Mchirpsource{\,6.09_{-0.06}^{+0.06}}
\def\Mchirpsourcemedian{\,6.09}
\def\Mchirpsourceupper{\,6.15}
\def\Mchirpsourcelower{\,6.03}
\def\Monesource{\ensuremath{23.2_{-1.0}^{+1.1}}}
\def\Monesourcemedian{\,23.0}
\def\Monesourceupper{\,24.3}
\def\Monesourcelower{\,22.2}
\def\Mtwosource{\ensuremath{2.59_{-0.09}^{+0.08}}}
\def\Mtwosourcemedian{\,2.59}
\def\Mtwosourceupper{\,2.67}
\def\Mtwosourcelower{\,2.50}
\def\Massratio{\ensuremath{0.112_{-0.009}^{+0.008}}}
\def\Massratiomedian{\,0.112}
\def\Massratioupper{\,0.12}
\def\Massratiolower{\,0.103}
\def\ThetaJN{\,0.8_{-0.2}^{+0.3}}
\def\ThetaJNmedian{\,0.8}
\def\ThetaJNupper{\,1.1}
\def\ThetaJNlower{\,0.6}
\def\FinalMass{\,25.6_{-0.9}^{+1.1}}
\def\FinalMassmedian{\,25.6}
\def\FinalMassupper{\,26.7}
\def\FinalMasslower{\,24.7}
\def\FinalSpin{\,0.28_{-0.02}^{+0.02}}
\def\FinalSpinmedian{0.28}
\def\FinalSpinupper{0.3}
\def\FinalSpinlower{0.26}
\def\Chieff{\,-0.002_{-0.061}^{+0.060}}
\def\Chieffmedian{\,-0.002}
\def\Chieffupper{\,0.058}
\def\Chiefflower{\,-0.063}
\def\Chip{\,0.04_{-0.03}^{+0.04}}
\def\Chipmedian{\,0.04}
\def\Chipupper{\,\,0.08}
\def\Chiplower{\,0.01}
\def\Primaryspin{\,0.03_{-0.03}^{+0.05}}
\def\Primaryspinmedian{\,0.03}
\def\Primaryspinupper{\ensuremath{0.08}}
\def\Primaryspinlower{\,0.0}
\def\Primaryspinninety{\,0.07}
\def\Secondaryspin{\,0.5_{-0.5}^{+0.4}}
\def\Secondaryspinmedian{\,0.5}
\def\Secondaryspinupper{\,0.9}
\def\Secondaryspinlower{\,0.0}
\def\Dlum{\ensuremath{241_{-45}^{+41}}}
\def\Dlummedian{\,241.3}
\def\Dlumupper{\,282.4}
\def\Dlumlower{\,196.3}
\def\Redshift{\,0.053^{+0.009}_{-0.010}}
\def\Redshiftmedian{\,0.053}
\def\Redshiftupper{\,0.062}
\def\Redshiftlower{\,0.043}
\def\rhoL{\,22.2_{-0.2}^{+0.1}}
\def\rhoLmedian{\,22.2}
\def\rhoLupper{\,22.3}
\def\rhoLlower{\,22.0}
\def\rhoH{\,10.7_{-0.2}^{+0.1}}
\def\rhoHmedian{\,10.7}
\def\rhoHupper{\,10.8}
\def\rhoHlower{\,10.5}
\def\rhoV{\,4.2_{-0.6}^{+0.2}}
\def\rhoVmedian{\,4.2}
\def\rhoVupper{\,4.4}
\def\rhoVlower{\,3.6}
\def\rhonetwork{\,25.0_{-0.2}^{+0.1}}
\def\rhonetworkmedian{\,25.0}
\def\rhonetworkupper{\,25.1}
\def\rhonetworklower{\,24.8}
\def\SkyArea{\,18.5}
\def\NinetyPercentVolume{\,39000}
\def\CosThetaJN{\,fix\,me!\,}

\def\DateTime{\,2019 August 14, 21:10:39 UTC}
\def\FirstDetDateTime{\,2019 August 14, 21:11:00 UTC}
\def\LLOInitialSNR{\,21.4}
\def\VirgoInitialSNR{\,4.3}
\def\GCNNotice{\,20}
\def\LHOSNR{\,10.6}
\def\LLOSNR{\,21.6}
\def\VirgoSNR{\,4.5}
\def\UploadTime{\,90}
\def\InitialDist{\,220--330}
\def\InitialSkyArea{\,38}
\def\SecondCircularTime{\,13.5}
\def\SearchDist{\,215--320}
\def\SearchSkyArea{\,23}
\def\GstlalFAR{\,$1$ in $1.3\times 10^3$}
\def\PyCBCFAR{\,$1$ in $8.1$}
\def\GstlalFARExclusive{\, $<1$ in $10^{5}$}
\def\PyCBCFARExclusive{\, $<1$ in $4.2\times 10^4$}
\def\cWBFAR{$<1$ in $10^3$}
\def\PyCBCCombinedSNR{\,19.4}
\def\GstlalCombinedSNR{\,22.2}

\def\HoGWNineteenAugFourteen{\ensuremath{75^{+59}_{-13}\mathrm{~km \,s}^{-1}\mathrm{ Mpc}^{-1}}}
\def\HoGWNineteenAugFourteenMed{$H_0=83^{+55}_{-53}$~km s$^{-1}$ Mpc$^{-1}$}
\def\Hopriorlow{$20$}
\def\Hopriorupp{$140$}
\def\Hopriorfrac{$60\%$}
\def\Hocombined{$69^{+13}_{-7}$~km s$^{-1}$ Mpc$^{-1}$}
\def\HocombinedMed{$74^{+27}_{-20}$~km s$^{-1}$ Mpc$^{-1}$}
\def\HoOtwo{$68^{+16}_{-8}$~km s$^{-1}$ Mpc$^{-1}$}
\def\HoOtwoMed{$74^{+30}_{-22}$~km s$^{-1}$ Mpc$^{-1}$}
\def\HoGWSeventeenAugFourteen{$H_0=75^{+40}_{-32}$~km s$^{-1}$ Mpc$^{-1}$}
\def\GLADEcomplete{$40\%$}
\def\GLADEngal{$472$} 
\def\GLADEpostreg{$90\%$}
\def\HoGWSeventeenAugSeventeen{$H_0=69^{+22}_{-8}$~km s$^{-1}$ Mpc$^{-1}$}
\def\HoGWSeventeenAugSeventeenMed{$H_0=78^{+35}_{-25}$~km s$^{-1}$ Mpc$^{-1}$}
\def\HocombinedGWSeventeenAugSeventeen{$H_0=70^{+17}_{-8}$~km s$^{-1}$ Mpc$^{-1}$}
\def\HocombinedGWSeventeenAugSeventeenMed{$H_0=77^{+33}_{-23}$~km s$^{-1}$ Mpc$^{-1}$}

\def\Lcanon{$616^{+273}_{-158}$}
\def\Rcanon{$12.9^{+0.8}_{-0.7}$}

\def\qGWNineteenAugFourteen{\ensuremath{0.28^{+0.12}_{-0.06}}}
\def\Msolar{\ensuremath{M_{\odot}}}

\def\mergerrate{1--23 Gpc\ensuremath{^{-3}}\,yr\ensuremath{^{-1}}}
\def\mergerratemed{\ensuremath{7^{+16}_{-6}}~Gpc\ensuremath{^{-3}}\,yr\ensuremath{^{-1}}}



\title{GW190814: Gravitational Waves from the Coalescence of a 23\,M$_\odot$ Black Hole \\ with a 2.6\,M$_\odot$ Compact Object} 



\author{R.~Abbott}
\affiliation{LIGO, California Institute of Technology, Pasadena, CA 91125, USA}
\author{T.~D.~Abbott}
\affiliation{Louisiana State University, Baton Rouge, LA 70803, USA}
\author{S.~Abraham}
\affiliation{Inter-University Centre for Astronomy and Astrophysics, Pune 411007, India}
\author{F.~Acernese}
\affiliation{Dipartimento di Farmacia, Universit\`a di Salerno, I-84084 Fisciano, Salerno, Italy}
\affiliation{INFN, Sezione di Napoli, Complesso Universitario di Monte S.Angelo, I-80126 Napoli, Italy}
\author{K.~Ackley}
\affiliation{OzGrav, School of Physics \& Astronomy, Monash University, Clayton 3800, Victoria, Australia}
\author{C.~Adams}
\affiliation{LIGO Livingston Observatory, Livingston, LA 70754, USA}
\author{R.~X.~Adhikari}
\affiliation{LIGO, California Institute of Technology, Pasadena, CA 91125, USA}
\author{V.~B.~Adya}
\affiliation{OzGrav, Australian National University, Canberra, Australian Capital Territory 0200, Australia}
\author{C.~Affeldt}
\affiliation{Max Planck Institute for Gravitational Physics (Albert Einstein Institute), D-30167 Hannover, Germany}
\affiliation{Leibniz Universit\"at Hannover, D-30167 Hannover, Germany}
\author{M.~Agathos}
\affiliation{Theoretisch-Physikalisches Institut, Friedrich-Schiller-Universit\"at Jena, D-07743 Jena, Germany}
\affiliation{University of Cambridge, Cambridge CB2 1TN, UK}
% ...





\collaboration{0}{LIGO Scientific Collaboration and Virgo Collaboration}

\begin{abstract}

We report the observation of a compact binary coalescence involving a \Monesourcelower~--\Monesourceupper 
\,$M_\odot$ black hole and a compact object with a mass of  \Mtwosourcelower~--\Mtwosourceupper\,$M_{\odot}$ (all measurements quoted at the 90\% credible level). The gravitational-wave signal, \eventname{}, was observed during LIGO's and Virgo's third observing run on August 14, 2019 at 21:10:39 UTC and has a signal-to-noise ratio of 25 in the three-detector network. The source was localized to 18.5 deg$^2$ at a distance of \Dlum\,Mpc; no electromagnetic counterpart has been confirmed to date. The source has the most unequal mass ratio yet measured with gravitational waves, \Massratio, and its secondary component is either the lightest black hole or the heaviest neutron star ever discovered in a double compact-object system. The dimensionless spin of the primary black hole is tightly constrained to $\leq \Primaryspinninety$. Tests of general relativity reveal no measurable deviations from the theory, and its prediction of higher-multipole emission is confirmed at high confidence. We estimate a merger rate density of \mergerrate{} for the new class of binary coalescence sources that \eventname{} represents.
Astrophysical models predict that binaries with mass ratios similar to this event can form through several channels, but are unlikely to have formed in globular clusters. 
{However,} the combination of mass ratio, component masses, and the inferred merger rate for this event challenges {all} current models for the formation and mass distribution of compact-object binaries.
\end{abstract}


\date{\today}


\newpage
\section{INTRODUCTION}

The first two observing runs (O1 and O2) with Advanced LIGO~\citep{TheLIGOScientific:2014jea} and Advanced Virgo~\citep{Virgo:2014hva} opened up the field of gravitational-wave astrophysics with the detection of the first binary black hole (BBH) coalescence signal, GW150914~\citep{GW150914}. 
Another nine such events \citep{O1BBH,GWLIGOScientific:2018mvr} were discovered by the LIGO Scientific and Virgo Collaborations (LVC) during this period, and additional events were reported by independent groups~\citep{Venumadhav:2019lyq,Zackay:2019btq,Zackay:2019tzo,Nitz:2019hdf}. 
The first binary neutron star (BNS) coalescence signal, GW170817, was discovered during the second of these observing campaigns \citep{Abbott:2018exr,2019PhRvX...9a1001A}. 
It proved to be a multi-messenger source with emission across the electromagnetic spectrum \citep{LVCmulti2017}, with implications for the origin of short gamma-ray bursts \citep{2017ApJ...848L..13A}, the formation of heavy elements \citep{LVC170817ejecta,ChornockBerger2017,TanvirLevan2017,RosswogSollerman2018,KasliwalKasen2019,WatsonHansen2019}, cosmology \citep{2017Natur.551...85A,LVCO2Cosmo} and fundamental physics \citep{2017ApJ...848L..13A,Abbott:2018lct}.

The first six months of the third observing run (O3) were completed between April 1 and September 30, 2019. The LVC recently reported on the discovery of GW190425, the coalescence signal of what is most likely a BNS
with unusually large chirp mass and total mass compared to the Galactic BNSs known from radio pulsar observations \citep{Abbott:2020uma}. 
Another discovery from O3 is that of GW190412, the first BBH coalescence with an unequivocally unequal mass ratio $q=m_2/m_1$ of $\qGWNineteenAugFourteen$ (all measurements are reported as symmetric 90\% credible intervals around the median of the marginalized posterior distribution, unless otherwise specified). It is also the first event for which higher-multipole gravitational radiation was detected with high significance~\citep{GW190412-discovery}.

Here we report on another O3 detection, \eventname{}, the signal of a compact binary coalescence with the most unequal mass ratio yet measured with gravitational waves: $q = \Massratio$. The signal was first identified in data from two detectors, \llo{} and \virgo{}, on \FirstDetDateTime{}.
Subsequent analysis of data from the full three-detector network revealed a merger signal with signal-to-noise ratio (SNR) of $\simeq 25$.

The primary component of \eventname{} is conclusively a black hole (BH) with mass $m_1 =\Monesource\,\Msolar$. Its dimensionless spin magnitude is constrained to $\chi_1 \leq \Primaryspinninety$. The nature of the \Mtwosource\,\Msolar{} secondary component is unclear. The lack of measurable tidal deformations and the absence of an electromagnetic counterpart are consistent with either a neutron star (NS) or a BH given the event's asymmetric masses and distance of \Dlum\,Mpc. 
However, we show here that comparisons with the maximum NS mass predicted by studies of GW170817's remnant, by current knowledge of the NS equation of state, and by electromagnetic observations of NSs in binary systems indicate that the secondary is likely too heavy to be a NS. Either way, this is an unprecedented source because the secondary's well-constrained mass of \Mtwosourcelower--\Mtwosourceupper\,$\Msolar$ makes it either the lightest BH or the heaviest NS ever observed in a double compact-object system.

As in the case of GW190412, we are able to measure the presence of higher multipoles in the gravitational radiation, and a set of tests of general relativity with the signal reveal no deviations from the theory. Treating this event as a new class of compact binary coalescences, we estimate a merger rate density of \mergerrate{} for \eventname-like events. Forming coalescing compact binaries with this unusual combination of masses at such a rate challenges our current understanding of astrophysical models.  

We report on the status of the detector network and the specifics of the detection in Sections 2 and 3. In Section 4, we estimate physical source properties with a set of waveform models, and we assess statistical and systematic uncertainties. Tests of general relativity are described in Section 5. In Section 6, we calculate the merger rate density and discuss implications for the nature of the secondary component, compact binary formation and cosmology. Section 7 summarizes our findings.

\section{Detector Network}

\begin{figure}
    \centering
    \includegraphics[width=0.5\textwidth]{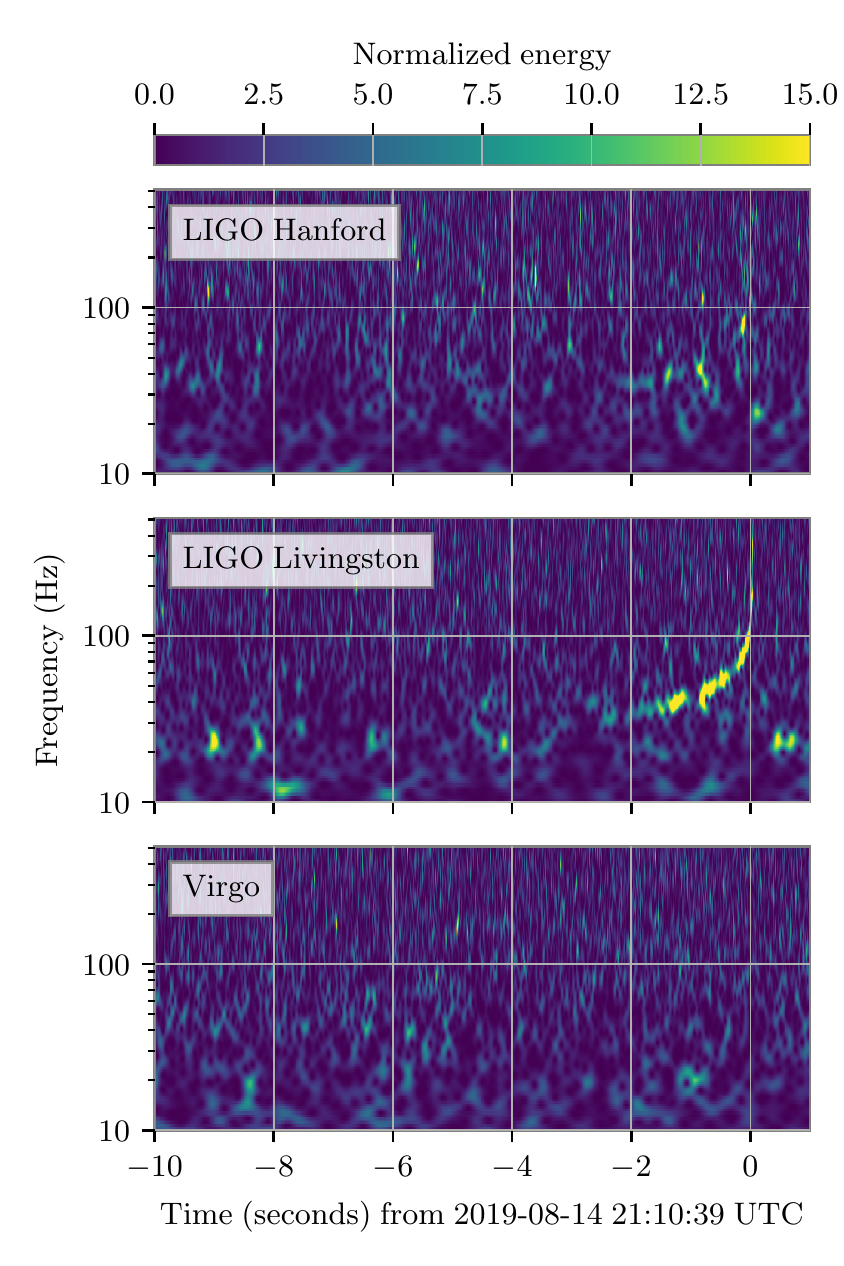}
    \caption{Time--frequency representations~\citep{Chatterji:2004qg} of data containing \eventname{}, observed by \lho{} (top), \llo{} (middle), and \virgo{} (bottom). Times are shown relative to \DateTime{}. Each detector's data are whitened by their respective noise amplitude spectral density and a Q-transform is calculated. The colorbar displays the normalized energy reported by the Q-transform at each frequency. These plots are not used in our detection procedure and are for visualization purposes only.}
    \label{fig:omegascan}
\end{figure}

At the time of \eventname{}, \lho{}, \llo{}~and \virgo were operating with typical O3 sensitivities~\citep{Abbott:2020uma}. Although \lho{} was in a stable operating configuration at the time of \eventname{}, the detector was not in observing mode due to a routine procedure to minimize angular noise coupling to the strain measurement~\citep{Kasprzack:2016}. This same procedure took place at \lho{} around the time of GW170608; we refer the reader to~\cite{Abbott:2017gyy} for details of this procedure. Within a 5 min window around \eventname{}, this procedure was not taking place, therefore \lho{} data for \eventname{} are usable in the nominal range of analyzed frequencies. A time--frequency representation~\citep{Chatterji:2004qg} of the data
from all three detectors around the time of the signal is shown in Figure~\ref{fig:omegascan}. 

We used validation procedures similar to those used to vet previous gravitational-wave events~\citep{Abbott:2016ctn, GWLIGOScientific:2018mvr}. Overall we found no evidence that instrumental or environmental disturbances~\citep{Effler:2014zpa} could account for \eventname{}. However, we did identify low-frequency transient noise due to scattered light at \llo{}, a common source of noise in all three interferometers~\citep{2018RSPTA.37670286N}. Scattered light features in the strain data are produced when a small fraction of the main laser beam reflects off a moving surface and is phase modulated before recombining with the main beam. This recombination can result in excess noise with the morphology of arches in the time--frequency plane; the frequency of this noise is determined by the velocity of the moving surface~\citep{Accadia:2010zzb}. Thunderstorms near \llo{} around the time of \eventname{} resulted in acoustic noise coupling to the detector and caused features in the strain data associated with scattered light~\citep{GWLIGOScientific:2018mvr}. In this instance, this form of noise affects frequencies up to 30~Hz from roughly 22~s to 8~s before and 0.2~s to 1.5~s after the detected time of \eventname{}, as seen in the middle panel of Figure~\ref{fig:omegascan}. Since this noise could bias the estimation of \eventname{}'s source parameters, we used a starting frequency of 30~Hz to analyse \llo{} data. Virgo was operating nominally and there are no quality issues in the Virgo data.

The LIGO and Virgo detectors are calibrated by photon pressure from modulated auxiliary lasers inducing
test-mass motion~\citep{Karki:2016pht, Acernese:2018bfl, Viets:2017yvy}. Over the frequency range 20--2048~Hz, the maximum 1$\sigma$ calibration
uncertainties for strain data used in the analysis of
\eventname{} were 6\% in amplitude and 4 deg in phase for LIGO data, and 5\% in amplitude and 7 deg
in phase for Virgo data. These calibration uncertainties are propagated into the parameter estimation reported in Section~\ref{sec.pe} via marginalization.

\section{Detection}
\subsection{Low-latency Identification of a Candidate Event}
\eventname{} was first identified on \FirstDetDateTime{} as a loud
two-detector event in \llo{} and \virgo{} data 
(SNR \LLOInitialSNR\ and \VirgoInitialSNR)
by the low-latency \gstlal{} matched-filtering search
pipeline for coalescing binaries~\citep{Cannon:2011vi,
Privitera:2013xza,Messick:2016aqy,Sachdev:2019vvd,Hanna:2019ezx}. Matched-filtering searches use 
banks~\citep{Sathyaprakash:1991,Blanchet:1995ez,owen:1995tm,
Owen:1998dk,damour2001dimensional,blanchet2005dimensional,
Cokelaer:2007kx,Harry:2009ea,Brown:2012nn,Ajith:2012mn,
harry:2013tca,Capano:2016dsf,PhysRevD.95.104045,
PhysRevD.99.024048,Indik:2017vqq} of modeled gravitational
waveforms~\citep{Buonanno:1998gg,Arun:2008kb,Blanchet:2013haa,Bohe:2016gbl,
Purrer:2015tud} as filter templates.
A Notice was issued through NASA’s Gamma-ray Coordinates Network
(GCN) \GCNNotice\ min later~\citep{GCNnoticeS190814bv}
with a two-detector source localization computed using the rapid
Bayesian algorithm \bayestar{}~\citep{BAYESTAR} that is shown in
Figure~\ref{fig:skymap}. The event was initially classified as ``MassGap''~\citep{Kapadia:2019uut,LIGOEMFOLLOWUSERGUIDE},
implying that at least one of the binary merger components was
found to have a mass between $3$--$5\,\Msolar$ in the low-latency
analyses.

Other low-latency searches, including the matched-filtering based
\mbta{}~\citep{2016CQGra..33q5012A} and \pycbc{}~\citep{Usman:2015kfa,
Nitz:2017svb,Nitz:2018rgo,pycbc-software} pipelines, could not detect
the event at the time as its SNR in Virgo data was below their
single-detector detection thresholds. Test versions of \mbta{}
and the additional matched-filtering pipeline \spiir{}~\citep{Hooper2012,Liu2012,Guo2018} operating with a lower SNR threshold also identified the event
with consistent attributes.

Shortly thereafter, reanalyses including LIGO Hanford data were
performed using \gstlal{} and \pycbc{}.
A coincident gravitational-wave signal was identified in all
three detectors by both searches, {with SNR \LLOSNR{} in \llo{},
\LHOSNR{} in \lho{}, and \VirgoSNR{} in \virgo{} data (as
measured by \gstlal{}, consistent with SNRs reported by
\pycbc{})}.
Results of these 3-detector analyses were reported in a GCN Circular
within 2.3 hours of the time of the event~\citep{S190814bv,GCN25324}, providing
a 3-detector localization~\citep{BAYESTAR}
constraining the distance to \InitialDist\ Mpc and the
sky area to \InitialSkyArea\ deg${}^2$ at the $90\%$ credible level.
Another GCN Circular~\citep{GCN25333} sent \SecondCircularTime\
hours after the event updated the source localization to
a distance of \SearchDist\ Mpc, the sky area to \SearchSkyArea{}
deg${}^2$, and the source classification to
``NSBH''~\citep{Kapadia:2019uut,LIGOEMFOLLOWUSERGUIDE},
indicating that the secondary had a mass below 3\,\Msolar.
These updated sky localizations are also shown in
Figure~\ref{fig:skymap}. The two disjoint sky localizations arise because the low SNR in the \virgo{} detector (\hspace{-0.5mm}\VirgoSNR{}) means that the data are consistent with two different signal arrival times in that detector.

\subsection{Multi-messenger Follow-up}
Several external groups performed multi-messenger follow-up of the source with observations
across the electromagnetic spectrum~\citep[e.g.,][]{LipunovGorbovskoy2019,GomezHosseinzadeh2019,AntierAgayeva2020,
AndreoniGoldstein2019,DobieStewart2019,WatsonButler2020,
AckleyAmati2020,VieiraRuan2020} and with neutrino observations~\citep[e.g.,][]{GCN25330,GCN25557}. No counterpart
candidates were reported. The non-detection is consistent with the
source's highly unequal mass ratio and low primary
spin~\citep{GCN25324,GCN25333,fernandez2020landscape,morgan2020constraints}. Tentative constraints placed by
multi-messenger studies on the properties of the system, such as
the ejecta mass and maximum primary spin~\citep{AndreoniGoldstein2019,
AckleyAmati2020,KawaguchiShibata2020,CoughlinDietrich2020} or the
circum-merger density \citep{DobieStewart2019} assuming a
neutron-star--black-hole (NSBH) source, may need to be revisited
in light of the updated source parameters we present in
Sec.\,\ref{sec:properties}.

\subsection{Significance}\label{s2:significance}

The significance of \eventname{} was estimated by follow-up
searches using improved calibration and refined data-quality
information that are not available in low latency.
They also used longer stretches of data for better
precision~\citep{O1BBH,Abbott:2016ctn}.
With \lho{} data being usable but not in nominal observing
mode at the time of \eventname{}, we used only data from the \llo{} and
\virgo{} detectors for significance estimation.
\eventname{} was identified as a confident detection in
analyses of detector data collected over the period
from August $7$ to August $15, 2019$ by the two independent
matched-filtering searches \gstlal{}
and \pycbc{}, with SNR values consistent with the
low-latency analyses.
The production version of \pycbc{} for O3
estimates significance only for events that are coincident in 
the \lho{} and \llo{} detectors, and therefore an extended
version~\citep{Davies:2020tsx} was used for \eventname{}
in order to enable the use of Virgo data in significance
estimation.

\gstlal{} and \pycbc{} use different techniques for
estimating the noise background 
and methods of ranking gravitational-wave
candidates. Both use results from searches over non
time-coincident data to improve their noise background
estimation~\citep{Privitera:2013xza,Messick:2016aqy,
Usman:2015kfa}.
Using data from the first six months of O3 and including
all events during this period in the estimation of noise
background, \gstlal{} estimated a false-alarm rate
(FAR) of \GstlalFAR{}~yr for \eventname{}.
Using data from the $8$-day period surrounding
\eventname{} and including this and all quieter events
during this period in noise background estimation,
the extended \pycbc{}
pipeline~\citep{Davies:2020tsx} estimated a FAR for
the event of \PyCBCFAR{}~yr.
The higher FAR estimate from \pycbc{} can 
be attributed to the event being identified by the
pipeline as being quieter than multiple noise events
in \virgo{} data.
As \pycbc{} estimates background statistics using
non-coincident data from both detectors, these louder
noise events in \virgo{} data can form chance
coincidences with the signal in \llo{} data and elevate
the noise background estimate for coincident
events, especially when considering shorter data periods.
All estimated background events that were ranked higher than
\eventname{} by \pycbc{} were indeed confirmed to be
coincidences of the candidate event itself in \llo{} with
random noise events in \virgo{}.
The stated background estimates are therefore
conservative~\citep{Capano:2016uif}.
We also estimate the background excluding the candidate
from the calculation, a procedure that yields a
mean-unbiased estimation of the distribution of noise
events~\citep{Capano:2016uif,TheLIGOScientific:2016qqj}.
In this case, with \gstlal{} we found a FAR of
\GstlalFARExclusive{}~yr while with \pycbc{} we found a FAR
of \PyCBCFARExclusive{}~yr. With both pipelines identifying
\eventname{} as more significant than any event in the
background, the FARs assigned are upper bounds.

\begin{figure}
    \centering
    \includegraphics[width=0.475\textwidth]{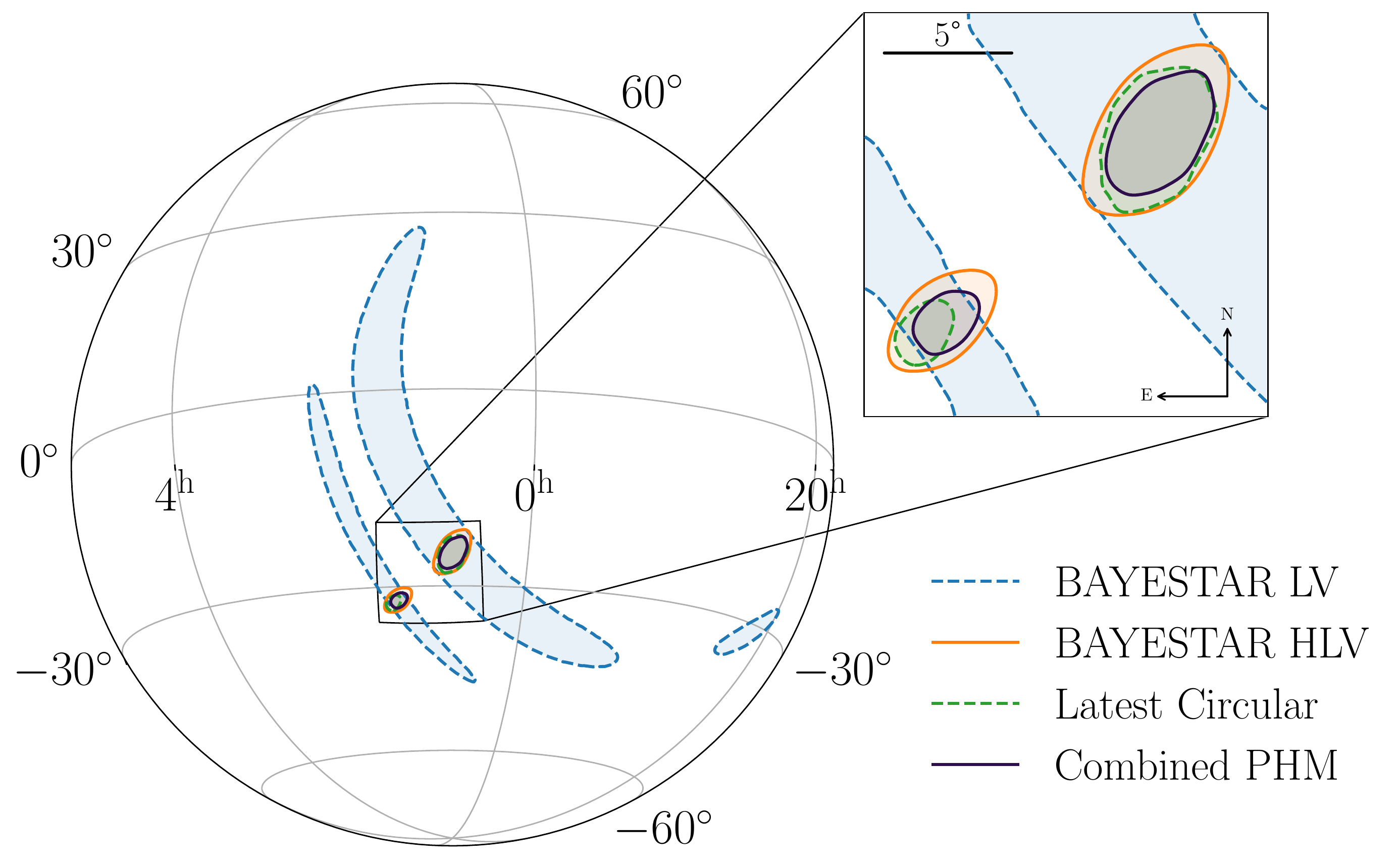}
    \caption{Posterior distributions for the sky location of \eventname{}.
    The contours show the 90\% credible interval for a \llo{}--\virgo{}
    (blue) and \lho{}--\llo{}--\virgo{} (orange) detector network
    based on the rapid localization algorithm \bayestar{}~\citep{BAYESTAR}.
    The sky localization circulated $13.5$ hours after the event, based on a
    \lho{}--\llo{}--\virgo{} analysis with the \lalinference{}
    stochastic sampling software~\citep{PhysRevD.91.042003}, is shown in
    green. The purple contour indicates the final sky localization as
    presented in this paper, which constrains the source to within
    \SkyArea~deg$^2$ at 90\% probability.
    }
    \label{fig:skymap}
\end{figure}

When data from \lho{} were included, \eventname{} was also identified
by the unmodelled coherent Wave Burst (\cwb{})
search that targets generic gravitational-wave transients with
increasing frequency over time without relying on waveform
models~\citep{Klimenko:2008fu,PhysRevD.93.042004,GW150914:obs}.
We found a FAR of \cWBFAR{}~yr of observing time against the
noise background from \lho{} and \llo{} data, consistent with the other searches.

\section{Properties of \eventname}\label{sec.pe}
\begin{table*}[t]
\begin{ruledtabular}
\begin{tabular}{l c c c c c}
& EOBNR PHM & Phenom PHM & Combined \\
\hline
Primary mass $m_1/M_{\odot}$ & $23.2^{+1.0}_{-0.9}$ & $23.2^{+1.3}_{-1.1}$ & $23.2^{+1.1}_{-1.0}$ \\
Secondary mass ${m_2/M_{\odot}}$ & ${2.59^{+0.08}_{-0.08}}$ & $2.58^{+0.09}_{-0.10}$ & $2.59^{+0.08}_{-0.09}$ \\
Mass ratio $q$ & $0.112^{+0.008}_{-0.008}$ & $0.111^{+0.009}_{-0.010}$ & $0.112^{+0.008}_{-0.009}$ \\
Chirp mass $\mathcal{M}/M_{\odot}$ & $6.10^{+0.06}_{-0.05}$ & $6.08^{+0.06}_{-0.05}$ & $6.09^{+0.06}_{-0.06}$ \\
Total mass $M/M_{\odot}$ & $25.8^{+0.9}_{-0.8}$ & $25.8^{+1.2}_{-1.0}$ & $25.8^{+1.0}_{-0.9}$ \\
Final mass ${M_{\mathrm{f}}/M_{\odot}}$ & $25.6^{+1.0}_{-0.8}$ & $25.5^{+1.2}_{-1.0}$ & $25.6^{+1.1}_{-0.9}$ \\
\\
Upper bound on primary spin magnitude $\chi_{1}$ & 0.06 & {0.08} & 0.07 \\
Effective inspiral spin parameter $\chi_{\mathrm{eff}}$ & $0.001^{+0.059}_{-0.056}$ & $-0.005^{+0.061}_{-0.065}$ & ${-0.002^{+0.060}_{-0.061}}$ \\
{Upper bound on effective precession parameter} $\chi_{\mathrm{p}}$ & 0.07 & 0.07 & 0.07 \\
Final spin $\chi_{\mathrm{f}}$ & $0.28^{+0.02}_{-0.02}$ & $0.28^{+0.02}_{-0.03}$ & $0.28^{+0.02}_{-0.02}$ \\
\\
Luminosity distance $D_{\mathrm{L}}/\mathrm{Mpc}$ & $235^{+40}_{-45}$ & $249^{+39}_{-43}$ & $241^{+41}_{-45}$ \\
Source redshift $z$ & $0.051^{+0.008}_{-0.009}$ & $0.054^{+0.008}_{-0.009}$ & ${0.053^{+0.009}_{-0.010}}$ \\
Inclination angle $\Theta/\mathrm{rad}$ & $0.9^{+0.3}_{-0.2}$ & $0.8^{+0.2}_{-0.2}$ & $0.8^{+0.3}_{-0.2}$ \\
\\
Signal to noise ratio in \lho $\rho_{\mathrm{H}}$ & $10.6^{+0.1}_{-0.1}$ & $10.7^{+0.1}_{-0.2}$ & $10.7^{+0.1}_{-0.2}$ \\
Signal to noise ratio in \llo $\rho_{\mathrm{L}}$ & ${22.21^{+0.09}_{-0.15}}$ & $22.16^{+0.09}_{-0.17}$ & $22.18^{+0.10}_{-0.17}$ \\
Signal to noise ratio in \virgo $\rho_{\mathrm{V}}$ & $4.3^{+0.2}_{-0.5}$ & ${4.1^{+0.2}_{-0.6}}$ & $4.2^{+0.2}_{-0.6}$ \\
Network Signal to noise ratio $\rho_{\mathrm{HLV}}$ & $25.0^{+0.1}_{-0.2}$ & $24.9^{+0.1}_{-0.2}$ & $25.0^{+0.1}_{-0.2}$ \\
\end{tabular}
\end{ruledtabular}
\caption{
  \label{table:pe_parameters}
  Source properties of \eventname: We report the median values along with the symmetric 90\% credible intervals for the {\sc{SEOBNRv4PHM}} ({\sc{EOBNR PHM}}) and {\sc{IMRPhenomPv3HM}} ({\sc{Phenom PHM}}) waveform models. The primary spin magnitude and the effective precession is given as the 90\% upper limit. The inclination angle is folded to $[0, \pi/2]$. The  last column is the result of combining the posteriors of each model with equal weight. The sky location of \eventname{} is shown in Figure~\ref{fig:skymap}.
  }
\end{table*}

We infer the physical properties of \eventname~using a coherent Bayesian analysis of the data from \LLO, \LHO and \virgo following the methodology described in Appendix B of~\citet{GWLIGOScientific:2018mvr}. Results presented here are obtained using $16\,$s of data around the time of detection. We use a low-frequency cutoff of $20\,\mathrm{Hz}$ for \LHO and \virgo and $30\,\mathrm{Hz}$ for \LLO for the likelihood evaluations, and we choose uninformative and wide priors, as defined in Appendix B.1 of~\citet{GWLIGOScientific:2018mvr}. The \lalinference~stochastic sampling software~\citep{PhysRevD.91.042003} is the primary tool used to sample the posterior distribution. A parallelized version of the parameter estimation software \bilby~\citep[\pbilby;][]{Smith:2019ucc, 2019ApJS..241...27A} is used for computationally expensive signal models. The power spectral density used in the likelihood calculations is a fair draw estimate calculated with \bayeswave~\citep{Cornish:2014kda, PhysRevD.91.084034}.
\par
This signal is analyzed under two different assumptions: that it represents a BBH, or that it represents a NSBH. For the BBH analyses, two different waveform families are used, one based on the effective-one-body approach~\citep[EOBNR;][]{Bohe:2016gbl, Babak:2016tgq, Cotesta:2018fcv, Ossokine:2020} and the other on a phenomenological approach~\citep[Phenom;][]{2016PhRvD..93d4007K, Husa:2015iqa, London:2017bcn, Khan:2018fmp, Khan:2019kotf}.
\par
For the NSBH analyses, we use BBH waveform models augmented with tidal effects~\citep{SEOBNRNSBH, PhenomNSBH}. Systematic uncertainties due to waveform modeling are expected to be subdominant compared to statistical errors~\citep{huang2020statistical}. When sampling the parameter space with the {\sc{SEOBNRv4\_ROM\_NRTidalv2\_NSBH}}~\citep{SEOBNRNSBH} and {\sc{IMRPhenomNSBH}}~\citep{PhenomNSBH} waveform models, we obtained posterior distributions for the secondary component's tidal deformability $\Lambda_2$ that are uninformative relative to a uniform prior in $\Lambda_2 \in [0,3000]$. The absence of a measurable tidal signature is consistent with the highly unequal mass ratio~\citep{Foucart:2013psa,Kumar:2016zlj} and with the relatively large secondary mass~\citep{Flanagan:2007ix}. The large asymmetry in the masses implies that the binary will merge before the neutron star is tidally disrupted for any expected NS equation of state~\citep{Foucart:2013psa}. Given that the signal carries no discernible information about matter effects, here we present quantitative results only from BBH waveform models.
\par
Our primary analyses include the effect of subdominant multipole moments in precessing waveform template models (PHM): {\sc{IMRPhenomPv3HM}}~\citep[Phenom PHM;][]{Khan:2018fmp, Khan:2019kotf} from the phenomenological family and {\sc{SEOBNRv4PHM}}~\citep[EOBNR PHM;][]{Babak:2016tgq, Ossokine:2020} from the EOBNR family.\footnote{In the co-precessing frame the EOBNR model includes the $(l,m)=(2,\pm 2), (2,\pm 1), (3,\pm 3), (4,\pm 4)$ and $(5,\pm 5)$ multipoles, and the Phenom model includes the $(2,\pm 2), (2,\pm 1), (3,\pm 3), (3,\pm 2), (4,\pm 4)$ and $(4,\pm 3)$ multipoles.} Analyses that assume the spins are aligned with the orbital angular momentum were also performed, either including (Phenom/EOBNR HM) or excluding (Phenom/EOBNR) the effect of subdominant multipole moments. 

\subsection{Properties}\label{sec:properties}
\begin{figure}
    \centering
    \includegraphics[width=0.5\textwidth]{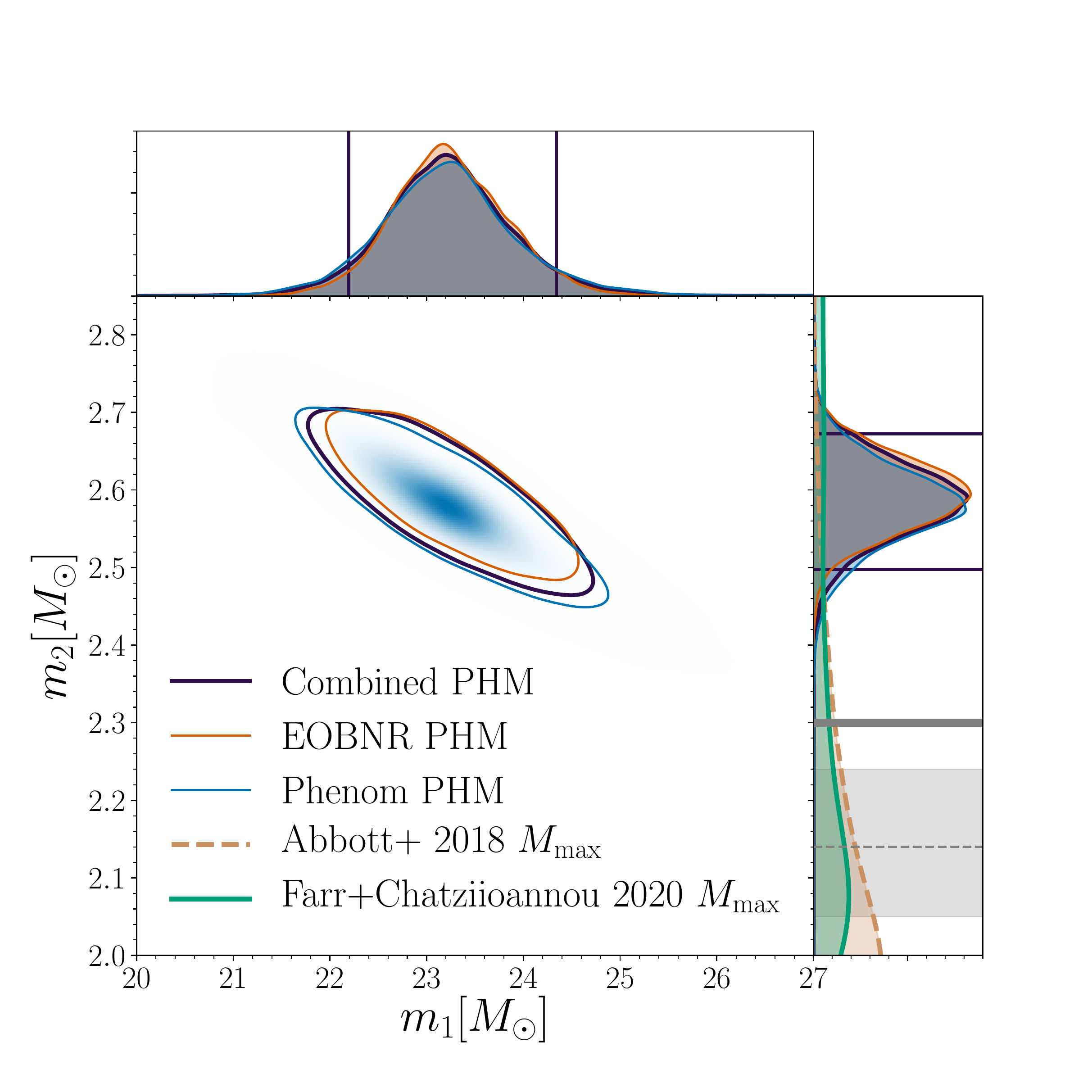}
    \caption{The posterior distribution of the primary and secondary source masses for two waveform models that include precession and subdominant multipole moments. The posterior distribution resulting from combining their samples is also shown. Each contour, as well as the colored horizontal and vertical lines, shows the 90\% credible intervals. The right panel compares $m_2$ to predictions for the maximum NS mass, $M_{\rm max}$ (see Section \ref{sec:astro_implication}). The posterior distribution for $M_{\rm max}$ from the spectral equation of state analysis of GW170817~\citep{LVCeos2018} is shown in orange, and the empirical $M_{\rm max}$ distribution from the population model of~\citet{FarrChatziioannou2020} is shown in green. The grey dashed line and shading represent the measured mass of the heaviest pulsar in the Galaxy~\citep[median and 68\% confidence interval;][]{Cromartie:2019kug}. The solid grey band at $2.3\,M_{\odot}$ is the upper bound on $M_{\rm max}$ from studies of GW170817's merger remnant.
    } \label{fig:different_physics_on_masses}
\end{figure}

From the $\sim$\hspace{1mm}$300$ observed cycles above $20$\,Hz, we are able to tightly constrain the source properties of \eventname. Our analysis shows that \eventname's source is a binary with an unequal mass ratio $q = \Massratio$, with individual source masses $m_1 = \Monesource\,\Msolar$ and $m_2 = \Mtwosource\,\Msolar$, as shown in Figure~\ref{fig:different_physics_on_masses}. A summary of the inferred source properties is given in Table~\ref{table:pe_parameters}. We assume a standard flat $\Lambda$CDM cosmology with Hubble constant $H_{0} = 67.9 \mathrm{\,km\,s^{-1}\,Mpc^{-1}}$~\citep{Ade:2015xua}.
\par
We report detailed results obtained from the two precessing BBH signal models including subdominant multipole moments: Phenom PHM and EOBNR PHM. In order to compare the template models, we compute their Bayes factor ($\log_{10}\mathcal{B}$). We find no significant evidence that one waveform family is preferred over the other as the Bayes factor between Phenom PHM and EOBNR PHM is $\log_{10}\mathcal{B}\simeq 1.0$. As a result, we combine the posterior samples with equal weight, in effect marginalizing over a
discrete set of signal models with a uniform probability. This is shown in the last column of Table~\ref{table:pe_parameters}, and we refer to these values throughout the paper unless stated otherwise.
\par
We find that the secondary mass lies in the range \Mtwosourcelower{}--\Mtwosourceupper{}$\,\Msolar$. This inferred secondary mass exceeds the bounds of the primary component in GW190425~\citep[1.61--2.52\,\Msolar;][]{Abbott:2020uma} and the most massive known pulsar in the Galaxy: $2.14^{+0.10}_{-0.09}\,M_\odot$ at 68.3\% credible interval \citep{Cromartie:2019kug}. Furthermore, the secondary is more massive than bounds on the maximum NS mass from studies of the remnant of GW170817, and from theoretical~\citep{LVCeos2018} and observational estimates~\citep{FarrChatziioannou2020}. The inferred secondary mass is comparable to the putative BH remnant mass of GW170817~\citep{2019PhRvX...9a1001A}.
\par
The primary object is identified as a BH based on its measured mass of $\Monesource\,M_\odot$. Due to accurately observing the frequency evolution over a long inspiral, the chirp mass is well constrained to $\Mchirpsource\,M_{\odot}$. The inferred mass ratio $q=\Massratio$~makes \eventname~only the second gravitational-wave observation with a significantly unequal mass ratio~\citep{GWLIGOScientific:2018mvr, GW190412-discovery}.

\par
\begin{figure}
    \centering
    \includegraphics[width=0.5\textwidth]{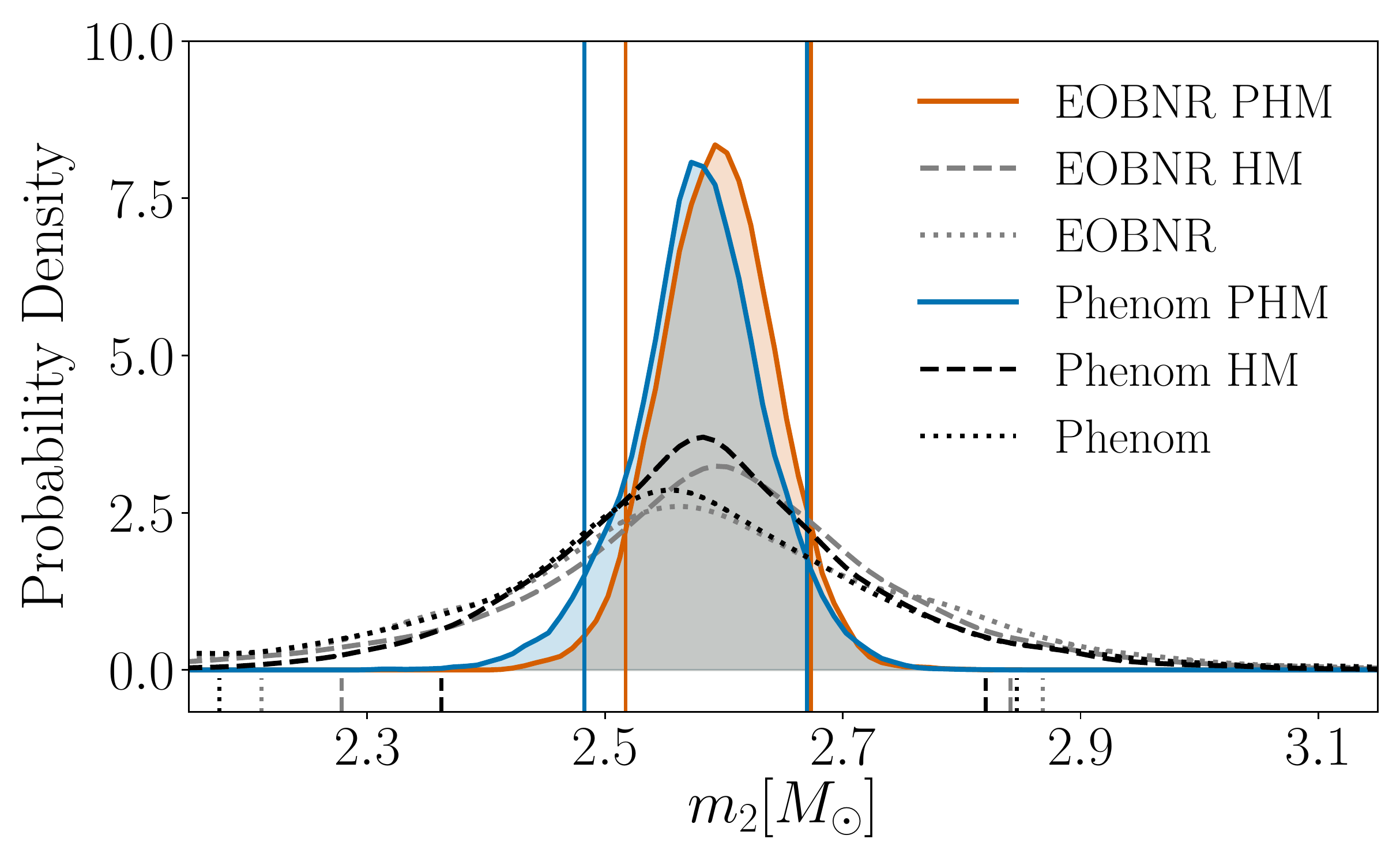}
    \caption{The marginalized posterior distribution for the secondary mass obtained using a suite of waveform models. The vertical lines indicate the 90\% credible bounds for each waveform model. The labels Phenom/EOBNR PHM (generic spin directions + higher multipoles), Phenom/EOBNR HM (aligned-spin + higher multipoles) and Phenom/EOBNR (aligned-spin, quadrupole only) indicate the different physical content in each of the waveform models.}
    \label{fig:waveform_systematics_on_mass_2}
\end{figure}

Given that this system is in a region of the parameter space that has not been explored via gravitational-wave emission previously, we test possible waveform systematics by comparing the Phenom and EOB waveform families. Differences in the inferred secondary mass are shown in Figure~\ref{fig:waveform_systematics_on_mass_2}. The results indicate that the inferred secondary mass is robust to possible waveform systematics, with good agreement between the Phenom PHM and EOBNR PHM signal models. Signal models that exclude higher multipoles or precession do not constrain the secondary mass as well.

\par

\par

The time delay of a signal across a network of gravitational wave detectors, together with the relative amplitude and phase at each detector, allows us to measure the location of the GW source on the sky~\citep{ObservingScenariosArxiv}. We localize \eventname{}'s source to within $\SkyArea~\,\mathrm{deg}^2$ at 90\% probability, as shown in Figure~\ref{fig:skymap}. This is comparable to the localization of GW170817~\citep{Abbott:2018exr, GWLIGOScientific:2018mvr}.
\par
Spins are a fundamental property of BHs. Their magnitude and orientation carry information regarding the evolution history of the binary. The effective inspiral spin parameter $\chi_{\mathrm{eff}}$~\citep{Damour:2001tu, Racine:2008qv, ajith2011inspiral, santamaria2010matching} contains information about the spin components that are perpendicular to the orbital plane. We infer that $\chi_{\mathrm{eff}} = \Chieff$. The tight constraints are consistent with being able to measure the phase evolution from the long inspiral.
\par
Orbital precession occurs when there is a significant spin component in the orbital plane of the binary~\citep{Apostolatos:1994}. We parameterize precession by the effective precession spin parameter $0\leq\chi_{\mathrm{p}}\leq 1$~\citep{2015PhRvD..91b4043S}. This effect is difficult to measure for face-on and face-off systems~\citep{Apostolatos:1994, Buonanno:2002fy, Vitale:2014mka, Vitale:2016avz, Fairhurst:2019srr, Fairhurst:2019_2harm}. \eventname{}~constrains the inclination of the binary to be $\Theta =\ThetaJN$~\,$\mathrm{rad}$. Since the system is neither face-on nor face-off, we are able to put strong constraints on the precession of the system: $\chi_{\mathrm{p}}=\Chip$. This is both the strongest constraint on the amount of precession for any gravitational-wave detection to date, and the first gravitational-wave measurement which conclusively measures near-zero precession~\citep{GWLIGOScientific:2018mvr, Abbott:2020uma,  GW190412-discovery}.
\par
\begin{figure}
    \centering
    \includegraphics[width=0.5\textwidth]{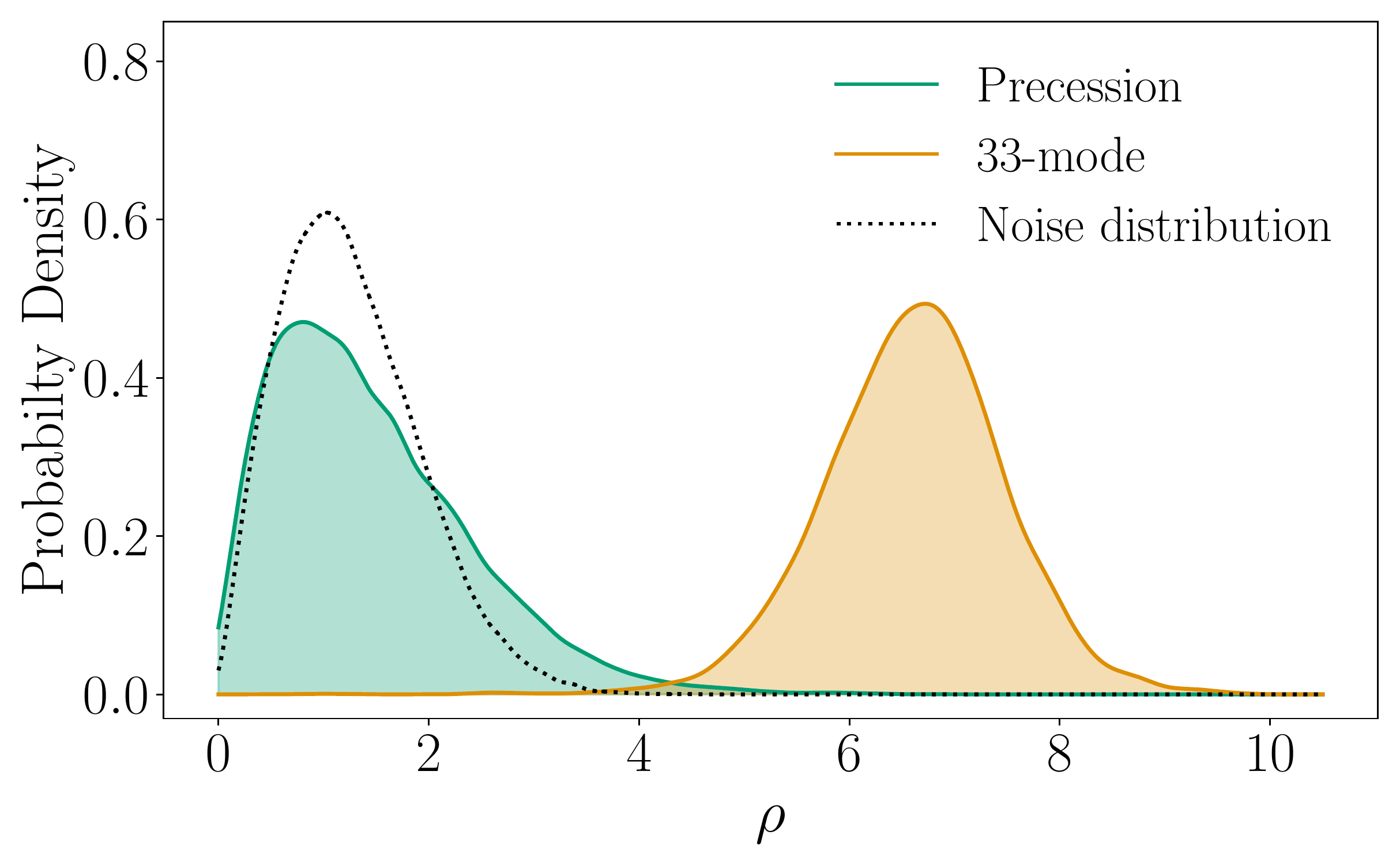}
    \caption{Posterior distributions for the precessing SNR, $\rho_{\mathrm{p}}$ (green) and the optimal SNR in the (3,3) sub-dominant multipole moment, $\rho$ (orange). The grey dotted line shows the expected distribution for Gaussian noise.}
    \label{fig:power_in_different_physics}
\end{figure}

By computing the Bayes factor between a precessing and non-precessing signal model ($\mathrm{log}_{10}\mathcal{B}\sim 0.5$ in favor of precession), we find inconclusive evidence for in-plane spin. This is consistent with the inferred power from precession SNR $\rho_{\mathrm{p}}$~\citep{Fairhurst:2019srr, Fairhurst:2019_2harm}, whose recovered distribution resembles that expected in the absence of any precession in the signal; see Figure~\ref{fig:power_in_different_physics}. The $\rho_{\mathrm{p}}$ calculation assumes a signal dominated by the $\ell=2$ mode; however, we have verified that the contribution of higher harmonics to the measurement of spin precession is subdominant by a factor of 5. The data are therefore consistent with the signal from a non-precessing system.
\par  
{Figure~\ref{fig:waveform_systematics_on_mass_2} shows that signal models including spin-precession effects give tighter constraints on the secondary mass compared to their non-precessing equivalents. 
Signal models that include spin-precession effects can constrain $\chi_{\mathrm{p}}$, whereas non-precessing signal models cannot provide information on in-plane spin components. In all analyses, we assume a prior equivalent to spin orientations being isotropically distributed.
We find that the data are inconsistent with large $\chi_{\mathrm{p}}$ and consistent with any secondary spin. 
Therefore, for precessing signal models the allowed $q$--$\chi_{\mathrm{eff}}$ parameter space is restricted, which helps to break the degeneracy~\citep{Poisson:1995ef, Baird:2012cu, 2016ApJ...825..116F, Baird:2012cu, Ng:2018neg}. 
Consequently, the extra information from constraining $\chi_{\mathrm{p}}$ to small values enables a more precise measurement of the secondary mass.}
\par
\begin{figure}
    \centering
    \includegraphics[width=0.5\textwidth]{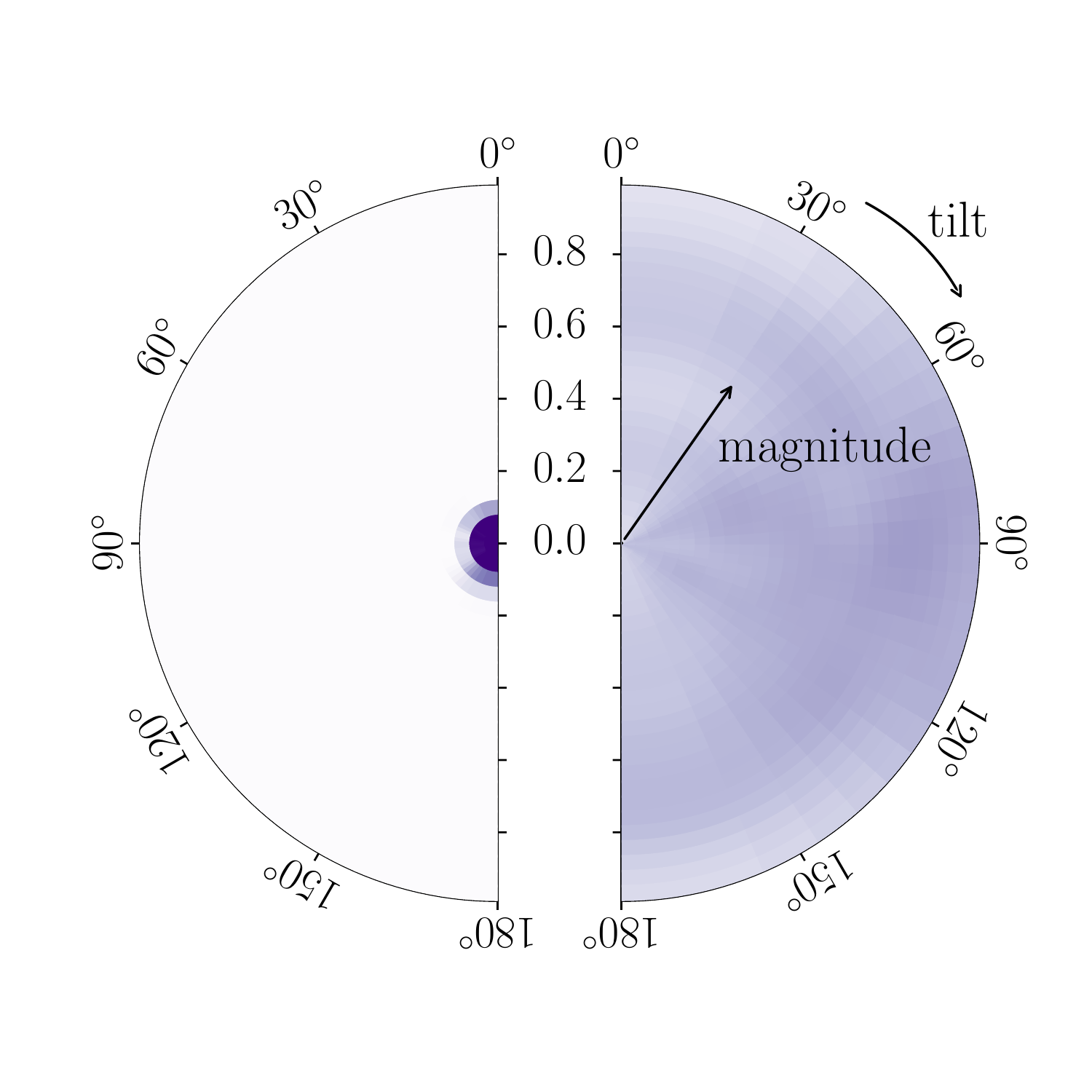}
    \caption{Two-dimensional posterior probability for the tilt-angle and spin-magnitude for the primary object (left) and secondary object (right) based on the Combined samples. The tilt angles are $0^\circ$ for spins aligned and $180^\circ$ for spins anti-aligned with the orbital angular momentum. The tiles are constructed linearly in spin magnitude and the cosine of the tilt angles such that each tile contains identical prior probability. The color indicates the posterior probability per pixel. The probabilities are marginalized over the azimuthal angles.}
    \label{fig:spin_disk}
\end{figure}

The asymmetry in the masses of \eventname{}~means that the spin of the more massive object dominates contributions to $\chi_{\mathrm{eff}}$ and $\chi_{\mathrm{p}}$. As both $\chi_{\mathrm{eff}}$ and $\chi_{\mathrm{p}}$ are tightly constrained, we are able to bound the primary spin of \eventname{}~to be $\chi_{1} \leq\Primaryspinninety$, as shown in Figure~\ref{fig:spin_disk}. This is the strongest constraint on the primary spin for any gravitational-wave event to date~\citep{GWLIGOScientific:2018mvr, Abbott:2020uma, GW190412-discovery}.
\par
The joint posterior probability of the magnitude and orientation of $\chi_{1}$ and $\chi_{2}$ are shown in Figure~\ref{fig:spin_disk}. Deviations from uniform shading indicate a spin property measurement. The primary spin is tightly constrained to small magnitudes, but its orientation is indistinguishable from the prior distribution. The spin of the less massive object, $\chi_{2}$, remains unconstrained; the posterior distribution is broadly consistent with the prior.
\par
The final mass $M_\mathrm{f}$ and final dimensionless spin $\chi_\mathrm{f}$ of the merger remnant are estimated under the assumption that the secondary is a BH. By averaging several fits calibrated to numerical relativity~\citep{Hofmann:2016yih,spinfit-T1600168,Healy:2016lce,Jimenez-Forteza:2016oae}, we infer the final mass and  spin of the remnant BH to be $\FinalMass\,\Msolar$~and $\FinalSpin$, respectively. The final spin is lower than for previous mergers~\citep{GWLIGOScientific:2018mvr, GW190412-discovery}, as expected from the low primary spin and smaller orbital contribution due to the asymmetric masses.

\subsection{Evidence for Higher-order Multipoles} \label{sec:evidence_for_higher_order_modes}

The relative importance of a subdominant multipole moment increases with mass ratio. Each subdominant multipole moment has a different angular dependence on the emission direction. With significant evidence for multipoles other than the dominant $(\ell, m)=(2, 2)$ quadrupole, we gain an independent measurement of the inclination of the source. This allows for the distance-inclination degeneracy to be broken~\citep{Cutler:1994ys, GW150914:PE, Usman:2018imj, Kalaghatgi:2019log}. Measuring higher-order multipoles therefore gives more precise measurements of source parameters~\citep{van2006phenomenology, van2007binary, Kidder:2007rt,Blanchet:2008je, Mishra:2016whh, Kumar:2018hml}.
\par
\aprilevent{} was the first event where there was significant evidence for higher-order multipoles~\citep{Payne:2019wmy,Kumar:2018hml,GW190412-discovery}. \eventname{} exhibits stronger evidence for higher-order multipoles, with $\log_{10}\mathcal{B} \simeq 9.6$ in favor of a higher-multipole vs.~a pure quadrupole model. The $(\ell, m) = (3, 3)$ is the strongest subdominant multipole, with $\log_{10}\mathcal{B} \simeq 9.1$ in favor of a signal model including both the $(\ell, m) = (2, 2)$ and $(3, 3)$ multipole moments. \eventname{}'s stronger evidence for higher multipoles is expected given its more asymmetric masses and the larger network SNR.
\par
{The orthogonal optimal SNR of a subdominant multipole is calculated by decomposing each multipole into components parallel and perpendicular to the dominant harmonic~\citep{Mills:2020, GW190412-discovery}.} We infer that the orthogonal optimal SNR of the $(\ell, m) = (3, 3)$  multipole is~$6.6^{+1.3}_{-1.4}$, as shown in Figure~\ref{fig:power_in_different_physics}. This is the strongest evidence for measuring a subdominant multipole to date~\citep{Payne:2019wmy,Kumar:2018hml,GW190412-discovery}.
\par

\begin{figure}
    \centering
    \includegraphics[width=0.45\textwidth]{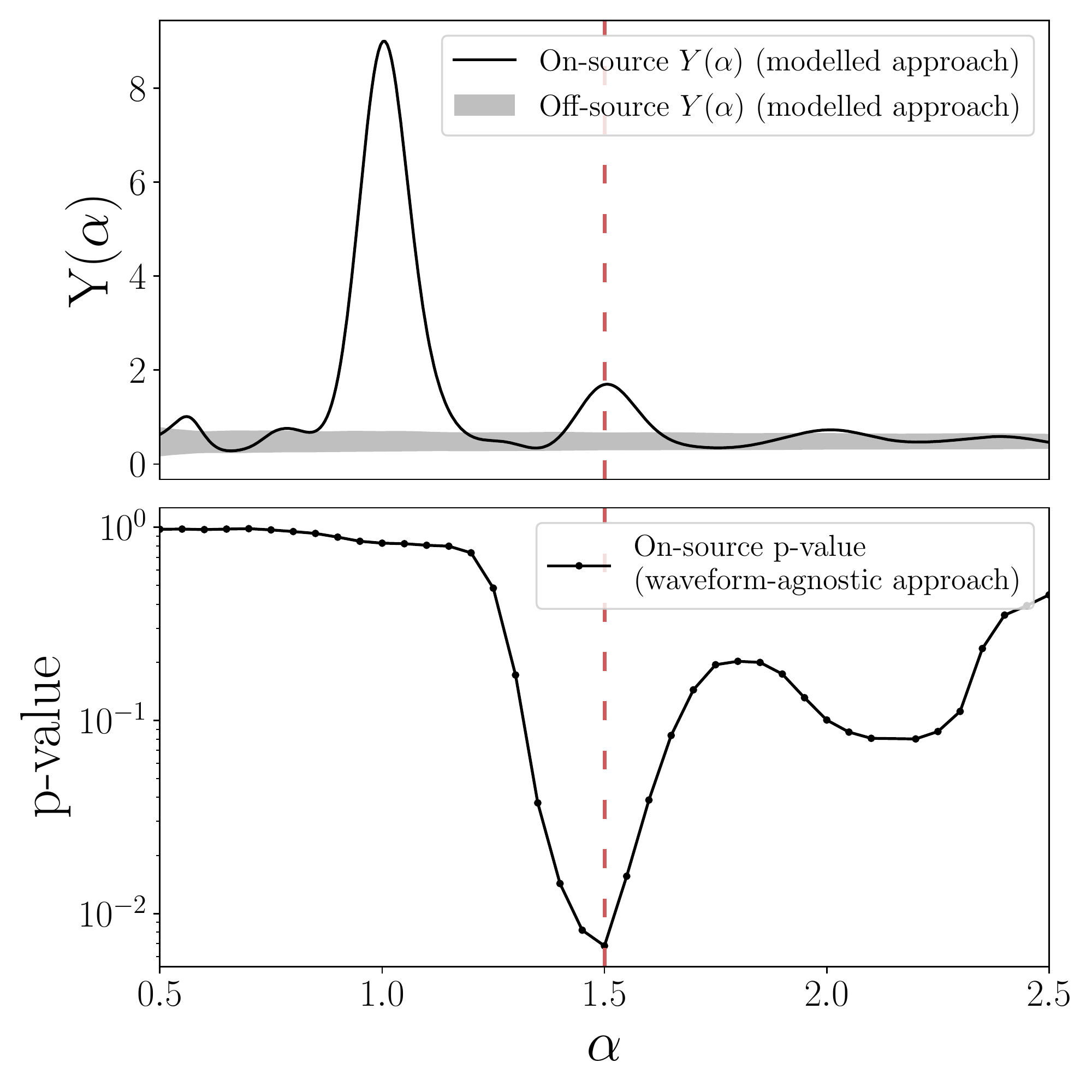}
    \caption{\textit{Top panel}: Variation of $Y(\alpha)$, i.e, the energy in the pixels along the $\alpha$-th track defined by $f_\alpha(t) = \alpha f_{22}(t)$, using the modelled approach. The peaks at $\alpha=1$ and $1.5$ indicate the energies in the $m=2$ and $m=3$ multipoles, respectively. The grey band indicates the 68\% confidence interval on the off-source measurements of $Y(\alpha)$. \textit{Bottom panel}: The variation of p-value of the on-source results, as a function of $\alpha$, using the waveform-agnostic approach. The dip at $\alpha=1.5$ is strong evidence of the presence of the $m=3$ mode in the underlying signal. The red dashed line in both panels corresponds to general relativity's prediction of $\alpha=1.5$ for the $m=3$ mode.}
    \label{fig:detection_higher_modes_inspiral}
\end{figure}

Finally, we perform two complementary analyses involving time--frequency tracks in the data to provide further evidence for the presence of higher multipoles in the signal. In the first approach~\citep[also outlined in][Section 4]{GW190412-discovery} we predict the time--frequency track of the dominant $(2,2)$ multipole in the  \llo{} detector (as seen in Figure~\ref{fig:omegascan}, middle panel) from an EOBNR HM parameter estimation analysis. This analysis collects energies along a time--frequency track which is $\alpha \times f_{22}(t)$, the $(2,2)$ multipole's instantaneous frequency, where $\alpha$ is a dimensionless parameter~\citep{roy2019unveiling,GW190412-discovery}. We find prominent peaks in $Y(\alpha)$, the energy in the pixels along the $\alpha$-th track defined in~\citet{GW190412-discovery}, at $\alpha=1$ and $1.5$, as can be seen from the on-source curve in the top panel of Figure \ref{fig:detection_higher_modes_inspiral}. These peaks correspond to the $m=2$ and $m=3$ multipole predictions in the data containing the signal (on-source data). We also compute a detection statistic $\beta$~\citep{roy2019unveiling} of $10.09$ for the presence of the $m=3$ multipole with a p-value of $< 2.5\times 10^{-4}$, compared to a background distribution estimated over 18 hours of data adjacent to the event (off-source data), where the largest background $\beta$ is $7.59$. The significant difference between on- and off-source values provides much stronger evidence for the presence of higher multipoles than what is reported for GW190412~\citep{GW190412-discovery}.

The second analysis uses waveform-agnostic methods to reconstruct the signal. It then compares the observed coherent signal energy in the \lho{}--\llo{}--\virgo{} network of detectors, as identified by the cWB detection pipeline~\citep{PhysRevD.93.042004}, with the predictions of a waveform model \textit{without} higher multipoles~\citep[EOBNR;][]{Prodi:2020} to investigate if the description of the underlying signal is incomplete if we do not include contributions from the $m=3$ multipole in our waveform model. We compute a test statistic, the squared sum of the coherent residuals estimated over selected time--frequency tracks parameterized in terms of the same $\alpha$ parameter defined in the previous analysis~\citep{roy2019unveiling,GW190412-discovery}. Each time--frequency track centered on $\alpha$ includes frequencies within $[\alpha-0.1, \alpha+0.1] \times f_{22}(t)$, and times within $[t_{\text{merger}} - 0.5 \text{ s}, t_{\text{merger}} - 0.03 \text{ s}]$, where $f_{22}(t)$ and $t_{\text{merger}}$ correspond to the maximum likelihood template from the EOBNR parameter estimation analysis. We further compute a background distribution using simulated signals in off-source data~\citep{Prodi:2020}, and compute p-values for the on-source results as a function of $\alpha$ (Figure~\ref{fig:detection_higher_modes_inspiral}, bottom panel). We find a minimum p--value of $6.8\times 10^{-3}$ at $\alpha=1.5$, providing strong evidence that the disagreement between the actual event and the EOBNR prediction is because of the absence of the $m=3$ multipoles in the waveform model. The local minimum near $\alpha=2$ is not an indication of the $m=4$ multipoles, but rather a statistical fluctuation which is consistent with similar behaviour seen for studies with simulated signals described in detail in~\cite{Prodi:2020}.

Although the two time--frequency analyses are similar in motivation, the latter differs from the former in that it is not restricted to data from just one detector, but rather uses the coherent signal energy across the three-detector network. Both analyses point to strong evidence for the presence of higher multipoles in the signal.

\section{Tests of General Relativity}
\eventname{} is the gravitational-wave event with the most unequal mass ratio to date, and can therefore be used to test general relativity (GR) in a region of parameter space previously unexplored with strong-field tests of GR \citep[][]{GW150914:TGR,LIGOScientific:2019fpa,Abbott:2018lct}. The asymmetric nature of a system excites the higher multipole moments of the gravitational signal, which allows us to test the multipolar structure of gravity~\citep{Kastha_2018,PhysRevD.100.044007,Dhanpal_2019,Islam_2020}. The addition of information from the higher harmonics of a signal also breaks certain degeneracies in the description of the source, and could potentially enable us to place stronger constraints on certain deviations from GR~\citep{van2007binary,van2006phenomenology}. We perform several null tests of GR using \eventname{}. These tests assume \eventname{} is a (quasi-circular) BBH merger as described in GR, and look for inconsistencies between the observed signal and predictions of the theory. An inconsistency might arise from an incomplete understanding of the underlying signal (or noise), and could indicate a non-BBH nature of the signal or a potential departure from GR.

First, as a consistency test of the signal reconstruction, we subtract from the data the maximum likelihood compact binary coalescence waveforms, Phenom~\citep{2016PhRvD..93d4007K}, Phenom HM~\citep{Kalaghatgi:2019log}, Phenom PHM~\citep{Khan:2019kotf}, and EOBNR PHM~\citep{Ossokine:2020} and analyze 4$\,$s of the resulting residual data centered around the time of merger with the morphology-independent transient analysis \bayeswave~\citep{Cornish:2014kda, PhysRevD.91.084034}. We measure the 90\% credible upper limit on the coherent SNR, $\mathrm{\rho}_{90}$, and compare it to the SNR, $\mathrm{\rho}^N_{90}$, recovered by analyzing 175 randomly selected data segments in surrounding time (off--source data) with the same configuration settings. If the residual data are consistent with the noise, we expect $\mathrm{\rho}_{90}$ to be consistent with $\mathrm{\rho}^N_{90}$. We compute the p-value by comparing the distribution of $\mathrm{\rho}^N_{90}$ to $\mathrm{\rho}_{90}$ through $p=P(\mathrm{\rho}^N_{90} < \mathrm{\rho}_{90})$. We obtain p-values of 0.59, 0.82, 0.82, and 0.75 for Phenom, Phenom HM, Phenom PHM, and EOBNR PHM, respectively. Hence, we find no evidence for deviations in the behavior of the residual data stream.

We also look for deviations in the spin-induced quadrupole moments of the binary components. According to the no-hair conjecture~\citep{carter1971axisymmetric,hansen1974multipole} the multipole moments of a Kerr BH are completely described by its mass and spin angular momentum. At leading order in spin, the spin-induced quadrupole moment scalar is \citep{1967ApJ...150.1005H,2012PhRvL.108w1104P}, $Q = -\kappa a^2 m^3$, where $(m,a)$ are the mass and dimensionless spin of the compact object, and $\kappa$ is a dimensionless deformation parameter characterizing deviations in the spin-induced quadrupole moment. Kerr BHs have $\kappa=1$~\citep{1980RvMP...52..299T}, while $\kappa\sim$ 2--14 for NSs (depending on the equation of state) and $\kappa \sim$ 10--150 for spinning boson stars with large self-interaction \citep{ryan1997spinning}. The deformation parameter can even be negative for (slowly-rotating, thin-shelled) gravastars \citep{2016PhRvD..94f4015U}. Hence, an accurate measurement of $\kappa$ sheds light onto the nature of the compact object. For compact binaries, the spin-induced quadrupole moment terms appear at second post-Newtonian order~\citep{poisson1998gravitational}. For Kerr BHs in GR, $\kappa_1=\kappa_2=1$, where $\kappa_1,\,\kappa_2$ are the individual deformation parameters of the primary and secondary compact objects in the binary. Since $\kappa_1$ and $\kappa_2$ are strongly degenerate in the gravitational waveform, we instead measure a linear symmetric combination of these quantities, $\kappa_{\text{s}} = (\kappa_1 + \kappa_2)/2$, which is $1$ for a BBH in GR. The posteriors on $\kappa_{\text{s}}$ are relatively uninformative, and nearly span the prior range of $[0,500]$, with increased support at $\kappa_{\text{s}} = 0$ relative to the prior. The upper bound of the prior was chosen to accommodate all the objects listed above. The result shows that \eventname{} is consistent with having a BBH source described by GR. However, the broad posterior means that we cannot exclude the possibility that one or both components of the source is not a BH. We can attempt to understand this result in terms of the spin measurements for the binary. The measurements of $\kappa_{\text{s}}$ and a non-zero $\chi_{\mathrm{eff}}$ are highly correlated \citep{PhysRevD.100.104019}, and for a system with small $\chi_{\mathrm{eff}}$ the bounds on the measured value of $\kappa_{\text{s}}$ are weak.

Finally, we investigate the source dynamics of the binary through a parameterized test of gravitational waveform generation, where we allow for the coefficients describing the post-Newtonian inspiral of a BBH coalescence to deviate away from their predictions in GR~\citep{PhysRevD.74.024006,Arun_2006,PhysRevD.80.122003,PhysRevD.82.064010,PhysRevD.84.062003,PhysRevD.85.082003,PhysRevD.97.044033}. We use an aligned-spin EOB waveform without higher modes (EOBNR), and find no deviations in the post-Newtonian coefficients from their nominal values in GR. In summary, none of our tests of GR indicate any departure from the predictions of the theory, and \eventname{} is consistent with the description of a compact binary merger in GR.

\section{Astrophysical Implications} \label{sec:astro_implication}

The highly unequal mass ratio of \Massratio~and unusual secondary mass of \Mtwosource\,\Msolar~make the source of \eventname{} unlike any other compact binary coalescence observed so far. The average mass ratio for BBH coalescences detected by the LVC during O1 and O2 is $\simeq 0.9$ \citep{RouletZaldarriaga2019}, and an inference of the underlying population {predicted that 99\% of detectable BBHs} have mass ratios $q \geq 0.5$ \citep{FishbachHolz2019}. However, the paucity of events from O1 and O2 means that this picture is limited. Indeed, the discovery of GW190412 has already changed the picture substantially \citep{GW190412-discovery}.

\eventname{}'s secondary mass lies in the hypothesized lower mass gap of $2.5$--$5\,M_{\odot}$~\citep{1998ApJ...499..367B,2010ApJ...725.1918O, Farr:2010tu,Ozel:2012ax} between known NSs and BHs. It is heavier than the most massive pulsar in the Galaxy \citep{Cromartie:2019kug},  
and almost certainly exceeds the mass of the 1.61--2.52\,\Msolar~primary component of GW190425, which is itself an outlier relative to the Galactic population of BNSs \citep{Abbott:2020uma}. On the other hand, it is comparable in mass to two BH candidates: the $\simeq\,2.7\,\Msolar$ merger remnant of GW170817~\citep{2019PhRvX...9a1001A} and the 2.6--6.1\,\Msolar~compact object (95\% confidence interval) discovered by~\citet{Thompson637}.\footnote{{See~\citet{vandenHeuvelTauris2020} and \citet{ThompsonKochanek2020} for discussion about the interpretation of this observation.}}
It is also comparable to the millisecond pulsar PSR J1748$-$2021B \citep{FreireRansom2008}, whose mass is claimed as $2.74^{+0.21}_{-0.21} M_{\odot}$ at 68\% confidence. However, this estimate, obtained via measurement of the periastron advance, could be inaccurate if the system inclination is low or the pulsar's companion is rapidly rotating \citep{FreireRansom2008}. In sum, it is not clear if \eventname{}'s secondary is a BH or a NS.

\eventname{} poses a challenge for our understanding of the population of merging compact binaries.
In what follows, we estimate the merger rate density of the compact binary subpopulation represented by this source, investigate the nature of its secondary component and possible implications for the NS equation of state, discuss how the system may have formed, and study its implications for cosmology.

\subsection{Merger Rate Density}

Given the unprecedented combination of component masses found in \eventname{}, we take the system to represent a new class of compact binary mergers, and use our analysis of its source properties to estimate a merger rate density for \eventname-like events.
Following a method described in \citet{Kim2003}, we calculate a simple, single-event rate density estimate $\mathcal{R}$ according to our sensitivity to a population of systems drawn from the parameter-estimation posteriors. As in \citet{Abbott:2020uma}, we calculate our surveyed space-time volume $\langle VT\rangle$ semi-analytically, imposing  single-detector and network SNR thresholds of 5 and 10, respectively~\citep{tiwari2018estimation}. 
The semi-analytic $\langle VT\rangle$ for \eventname{} is then multiplied by a calibration factor to match results from the search pipelines assuming a once-per-century FAR threshold. The sensitivity of a search pipeline is estimated using a set of simulated signals. For computational efficiency, this was done using pre-existing search pipeline simulations and the mass properties were not highly optimized. However, given that we are estimating a rate based on a single source, the calibration errors are much smaller than the statistical errors associated with the estimate. The simulated sources were uniformly distributed in comoving volume, component masses, and component spins aligned with the orbital angular momentum.  
For O1 and O2, the simulated BH mass range was 5--100\,\Msolar, but for the first part of O3 we are analyzing here, the injected range was 2.5--40\,\Msolar~(following our updated knowledge of the BH mass distribution); the NS mass range was 1--3\,\Msolar, and component spins are $<0.95$.  
As \eventname{} occurred when LIGO Hanford was not in nominal observing mode, it is not included in the production \pycbc{} results, and we use \gstlal{} results to calculate the merger rate.  

We assume a Poisson likelihood over the astrophysical rate with a single count and we apply a Jeffreys $\mathcal{R}^{-1/2}$ prior to obtain rate posteriors. The analysis was done using samples from the Phenom PHM posterior and separately from the EOBNR PHM posterior, producing the same result in both cases. We find the merger rate density of \eventname{}-like systems to be \mergerratemed.  

As a consistency check, we used the \pycbc{} search results to calculate an upper limit. Repeating the rate calculation with a \pycbc-based $\langle VT\rangle$ calibration and zero event count, we obtain an upper limit consistent (to within 10\%) with the upper limit of the merger rate estimated using \gstlal{} search results. We conclude that the uncertainty in our estimate of the rate density for the class of mergers represented by \eventname~is primarily dominated by Poisson statistics.

\subsection{Nature of the Secondary Component}

The primary mass measurement of \Monesource\,\Msolar~securely identifies the heavier component of \eventname{} as a BH, but the secondary mass of \Mtwosource\,\Msolar~may be compatible with either a NS or a BH depending on the maximum mass supported by the unknown NS equation of state (EOS). The source's asymmetric masses, the non-detection of an electromagnetic counterpart and the lack of a clear signature of tides or spin-induced quadrupole effects in the waveform do not allow us to distinguish between a BBH or a NSBH. 
Instead, we rely on comparisons between $m_2$ and different estimates of the maximum NS mass, $M_{\rm max}$, to indicate the source classification preferred by data: if $m_2 > M_{\rm max}$, then the NSBH scenario is untenable.

While some candidate EOSs from nuclear theory can support nonrotating NSs with masses of up to $\sim 3\,\Msolar$ \citep[e.g.,][]{MullerSerot1996}, such large values of $M_{\rm max}$ are disfavored by the relatively small tidal deformabilities measured in GW170817 \citep{Abbott:2018exr,2019PhRvX...9a1001A}, which correlate with smaller internal pressure gradients as a function of density and hence a lower threshold for gravitational collapse. By adopting a phenomenological model for the EOS, conditioning it on GW170817, and extrapolating the constraints to the high densities relevant for the maximum mass, \citet{LimHolt2019} and \citet{EssickLandry2019} place $M_{\rm max} \lesssim 2.3\,\Msolar$. Similarly, the EOS inference reported in \citet{LVCeos2018}, based on an analysis of GW170817 with a spectral parameterization \citep{Lindblom:2010bb,LindblomIndik2012,LindblomIndik2014} for the EOS, implies a 90\% credible upper bound of $M_{\rm max} \leq 2.43\,\Msolar$, with tenuous but non-zero posterior support beyond $2.6\,\Msolar$. We calculate the corresponding $M_{\rm max}$ posterior distribution, shown in the right panel of Figure~\ref{fig:different_physics_on_masses}, from the GW170817-informed spectral EOS samples used in \citet{LVCeos2018} by reconstructing each EOS from its parameters and computing its maximum mass. Comparison with the $m_2$ posterior suggests that the secondary component of \eventname{}~is probably more massive than this prediction for $M_{\rm max}$: the posterior probability of $m_2 \leq M_{\rm max}$, marginalized over the uncertainty in $m_2$ and $M_{\rm max}$, is only 3\%. 
Nevertheless, the maximum mass predictions from these kinds of EOS inferences come with important caveats: their extrapolations are sensitive to the phenomenological model assumed for the EOS; they use hard $M_{\rm max}$ thresholds on the EOS prior to account for the existence of the heaviest Galactic pulsars, which is known to bias the inferred maximum mass distribution towards the threshold \citep{Miller:2019nzo}; and they predate the NICER observatory's recent simultaneous mass and radius measurement for J0030+0451, which may increase the $M_{\rm max}$ estimates by a few percent \citep{LandryEssick2020} because it favors slightly stiffer EOSs than GW170817 \citep{RaaijmakersRiley2019,RileyWatts2019,MillerLamb2019,JiangTang2019}.

NS mass measurements also inform bounds on $M_{\rm max}$ independently of EOS assumptions. Fitting the known population of NSs in binaries to a double-Gaussian mass distribution with a high-mass cutoff, \citet{AlsingSilva2018} obtained an empirical constraint of $M_{\rm max} \leq 2.6\,\Msolar$ (one-sided 90\% confidence interval). {\citet{FarrChatziioannou2020} recently updated this analysis to include PSR J0740+6620 \citep{Cromartie:2019kug}, which had not been discovered at the time of the original study. Based on samples from the \citet{FarrChatziioannou2020} maximum-mass posterior distribution, which is plotted in the right panel of Figure~\ref{fig:different_physics_on_masses}, we find $M_{\rm max} = 2.25^{+0.81}_{-0.26}\,\Msolar$.
In this case, the posterior probability of $m_2 \leq M_{\rm max}$ is 29\%, again favoring the $m_2 > M_{\rm max}$ scenario, albeit less strongly because of the distribution's long tail up to $\sim 3\,\Msolar$.}
However, the empirical $M_{\rm max}$ prediction is sensitive to selection effects that could potentially bias it \citep{AlsingSilva2018}. In particular, masses are only measurable for binary pulsars, and the mass distribution of isolated NSs could be different. 
Additionally, the discovery of GW190425~\citep{Abbott:2020uma}
should also be taken into account in the population when predicting $M_{\rm max}$. 

Finally, the NS maximum mass is constrained by studies of the merger remnant of GW170817. Although no postmerger gravitational waves were observed \citep{Abbott:2017dke,AbbottAbbott2019longpostmerger}, modeling of the associated kilonova \citep{LVCmulti2017,KasenMetzger2017,2017ApJ...851L..21V,CowperthwaiteBerger2017,LVC170817ejecta} suggests that the merger remnant collapsed to a BH after a brief supramassive or hypermassive NS phase during which it was stabilized by uniform or differential rotation. Assuming this ultimate fate for the merger remnant immediately implies that no NS can be stable above $\sim 2.7\,\Msolar$, but it places a more stringent constraint on NSs that are not rotationally supported. The precise mapping from the collapse threshold mass of the remnant to $M_{\rm max}$ depends on the EOS, but by developing approximate prescriptions based on sequences of rapidly rotating stars for a range of candidate EOSs, $M_\mathrm{max}$ has been bounded below approximately 2.2--2.3\,\Msolar~ \citep{MargalitMetzger2017,RezzollaMost2018,RuizShapiro2018,ShibataZhou2019,GW170817eosMtov}. Although the degree of EOS uncertainty in these results is difficult to quantify precisely, if we take the more conservative $2.3\,\Msolar$ bound at face value, then $m_2$ is almost certainly not a NS: the $m_2$ posterior distribution has negligible support below 2.3\,\Msolar.

Overall, these considerations suggest that \eventname{} is probably not the product of a NSBH coalescence, despite its preliminary classification as such. Nonetheless, the possibility that the secondary component is a NS cannot be completely discounted due to the current uncertainty in $M_{\rm max}$.

There are two further caveats to this assessment. First, because the secondary's spin is unconstrained, it could conceivably be rotating rapidly enough for $m_2$ to exceed $M_{\rm max}$ without triggering gravitational collapse: rapid uniform rotation can stabilize a star up to $\sim 20\%$ more massive than the nonrotating maximum mass \citep{CookShapiro1994}, in which case only the absolute upper bound of $\sim 2.7\,\Msolar$ is relevant. However, it is very unlikely that a NSBH system could merge before dissipating such extreme natal NS spin angular momentum.

Second, our discussion has thus far neglected the possibility that the secondary component is an exotic compact object, such as a boson star \citep{Kaup1968} or a gravastar \citep{Mazur:2004fk}, instead of a NS or a BH. Depending on the model, some exotic compact objects can potentially support masses up to and beyond 2.6\,\Msolar~\citep{CardosoPani2019}. Our analysis does not exclude this hypothesis for the secondary.

\begin{figure}
    \centering
    \includegraphics[width=0.5\textwidth]{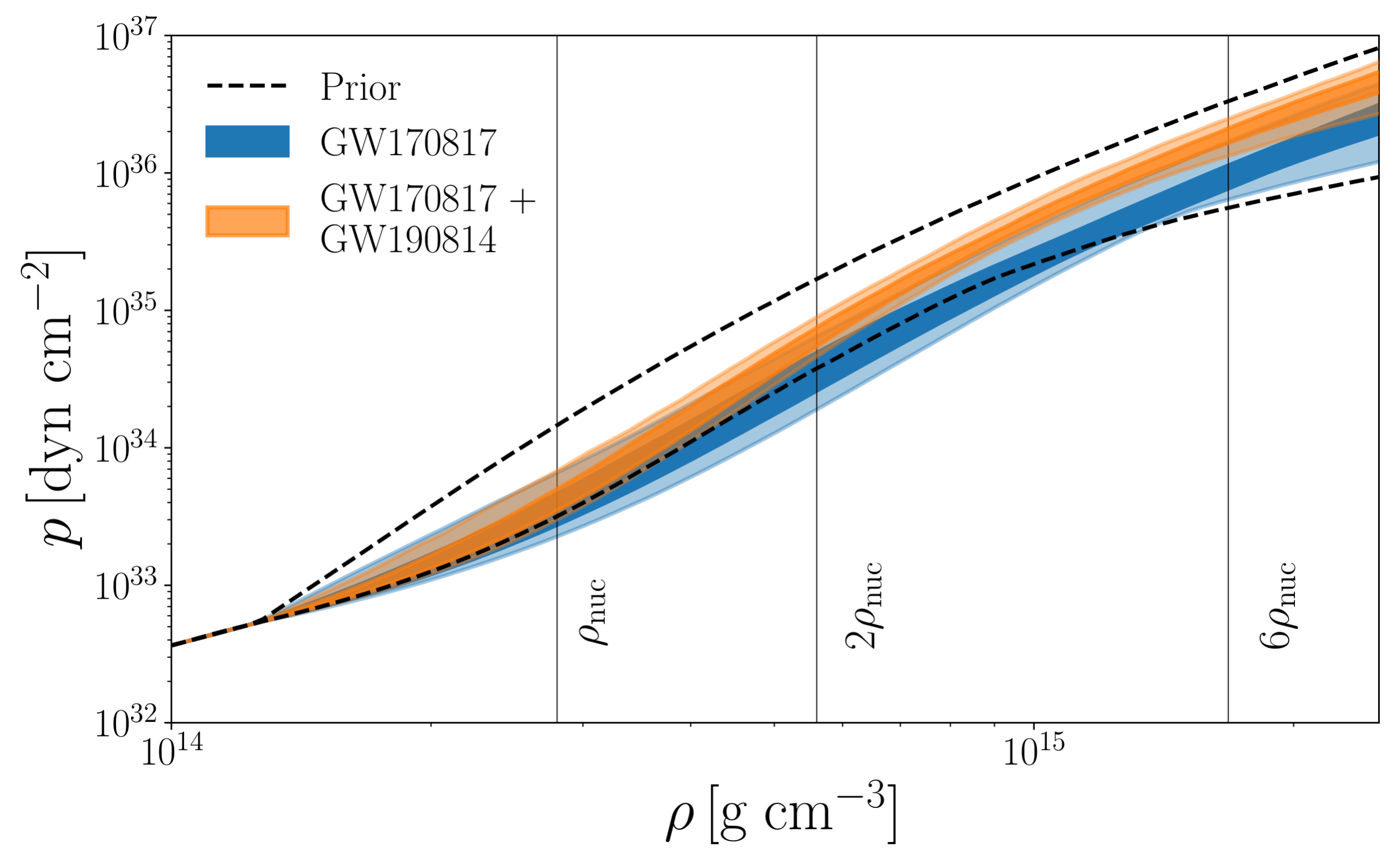}
    \caption{Constraints on the NS EOS assuming \eventname{} was produced by a BBH (blue) or a NSBH (orange) coalescence. The 90\% and 50\% credible contours of the posterior in the pressure-density plane are shown. The constraints are calculated by assuming a spectral decomposition for the EOS, following \citet{LVCeos2018}. The BBH constraints are identical to those from the analysis of GW170817, while for the NSBH case the posterior is reweighted by the probability that each EOS's maximum mass is at least $m_2$. The dashed lines indicate the 90\% credible region of the prior.}
    \label{fig:eos_constraints}
\end{figure}

Since the NSBH scenario cannot be definitively ruled out, we examine \eventname's potential implications for the NS EOS, assuming that the secondary proves to be a NS. 
This would require $M_{\rm max}$ to be no less than $m_2$, a condition that severely constrains the distribution of EOSs compatible with existing astrophysical data. The combined constraints on the EOS from GW170817 and this hypothetical maximum mass information are shown in 
Figure~\ref{fig:eos_constraints}. Specifically, we have taken the spectral EOS distribution conditioned on GW170817 from \citet{LVCeos2018} and  
reweighted each EOS by the probability that its maximum mass is at least as large as $m_2$. 
The updated posterior favors stiffer EOSs, which translates to larger radii for NSs of a given mass.
The corresponding constraints on the radius and tidal deformability of a canonical $1.4\,\Msolar$ NS are $R_{1.4} =\,$\Rcanon\,km and $\Lambda_{1.4} =\,$\Lcanon.

\subsection{Origins of GW190814-like Systems}

The source of \eventname{} represents a previously undetected class of coalescences that has the potential to shed light on the formation of merging compact-object binaries with highly asymmetric masses.  

Electromagnetic observations of Galactic NSs and stellar-mass BHs suggest a dearth of compact objects in the $\sim2.5\,\Msolar$  to $5\,\Msolar$ range~ \citep{1998ApJ...499..367B,2010ApJ...725.1918O,Farr:2010tu,Ozel:2012ax}. Observations of a few candidates with masses in this range seem to disfavor the existence of a gap~ \citep{FreireRansom2008,neustroev2014,giesers2018,Wyrzykowski19massgap, Thompson637}, but whether the mass gap is physical or caused by selection biases is still a matter of debate \cite[e.g.,][]{Kreidberg12massgap}.

From a theoretical point of view, accurately calculating the masses of compact remnants at formation is challenging, because it depends on the complex physics of the supernova  explosion and the details of stellar evolution, especially for the late evolutionary stages of massive stars \citep{janka2012,muller2016,burrows2018,burrows2019}. Whether the models favor the presence of a gap or a smooth transition between NSs and BHs is still unclear, and in fact some models have been developed with the purpose of reproducing this lower mass gap \citep{ugliano2012, fryer2012,kochanek2014,sukhbold2014,ertl2016}. Therefore, our robust discovery of an object with a well-constrained mass in this regime may provide crucial constraints on compact-object formation models. {In fact, \eventname{} demonstrates the need to adjust remnant mass prescriptions previously designed to produce a perceived mass gap. The combination of mass ratio and component masses challenges most results obtained from population synthesis simulations for isolated binaries \citep{dominik2012, dominik2015, marchant2017, giacobbo2018,mapelli2018,Kruckow:2018slo,neijssel2019,mapelli2019,spera2019,OlejakBelczynski2020} }.

Population synthesis models distinguish between NSs and BHs using only a mass threshold, which is generally in the range 2--3\,\Msolar.
Thus, depending on the adopted threshold and on the adopted supernova explosion
model, a \eventname{}-like event may be labeled as either a NSBH merger
or a BBH merger. Most BBH mergers have $q > 0.5$, while the distributions of merging NSBH binaries suggest that systems with $q \lesssim 0.1$ may be up to $\sim 10^3$ times less common than more symmetric ones ($q > 0.1$) and that the mass-ratio distribution peaks at $q \approx 0.2$. Furthermore, models tend to favor mergers of massive ($\gtrsim 1.3\,\Msolar$) NSs with relatively small BHs ($\lesssim 15\,\Msolar$) in environments with sub-solar metallicity ($Z\lesssim 0.5\,Z_{\odot}$). The tendency to disfavor mergers with highly asymmetric masses in isolated binaries may be the consequence of mass transfer \cite[e.g.,][]{postnov2014} and common envelope episodes \cite[e.g.,][]{ivanova2013} that cause systems with initially asymmetric masses to evolve towards more symmetric configurations. {Overall, producing mergers with such unequal masses, with a secondary in the perceived mass gap, and at the rate implied by this discovery is a challenge for current models.}  

{Nevertheless, particular choices of poorly constrained assumptions within rapid population synthesis models may increase}
the number of mergers with $q \lesssim 0.1$ so that the latter may be only a few times less common than {(or even comparable to)} systems with $q \simeq 0.2$ {\citep[e.g.,][]{eldridge2016,eldridge2017,giacobbo2018}.}

Another possibility is that \eventname{} is of dynamical origin. Dynamical exchanges in dense stellar environments tend to pair up massive compact objects with similar masses \cite[e.g.,][]{sigurdsson1993}. This process is effective for globular clusters, where compact-object binaries may undergo tens of exchanges before they get ejected from the cluster~\citep{portegieszwart2000,rodriguez2016,park2017,askar2017,rodriguez2019}. {For such environments, } models predict that most merging
BBHs have $q\simeq 1$~\citep[e.g.,][]{rodriguez2016}, and the formation of NSBH binaries is highly suppressed because BHs dynamically 
dominate the cores over the complete lifetime of the clusters, preventing the 
interactions between BHs and NSs, with the consequence that the merger 
rate of NSBH binaries in globular clusters in the local Universe is $\sim 10^{-2}$--$10^{-1}\, 
\mathrm{Gpc}^{-3}\mathrm{yr}^{-1}$~\citep{clausen2013,ye2020,arcasedda2020}. The rate for \eventname{}-like events, {with a secondary in the perceived mass gap}, is likely even lower. In contrast, the NSBH merger 
rate may be significantly higher in young star clusters {\cite[e.g.,][]{ziosi2014} and the latter can effectively increase the number of progenitors leading to merging compact-object binaries with $q\lesssim 0.15$ \citep{dicarlo2019a,rastello2020}}. Thus, young star clusters may be promising hosts for  \eventname{}-like events, but the parameter space relevant for \eventname{} is mostly unexplored in the context of star clusters.

In dense stellar environments, \eventname{}-like systems may also form from a low-mass merger remnant that acquires a BH companion via dynamical interactions \citep{Gupta:2019nwj}. \cite{Gupta:2019nwj} predicts a population of second-generation BHs in the 2.2--3.8\,\Msolar~range, with a peak in the distribution at $2.6\,\Msolar$, assuming a double-Gaussian mass distribution for the NSs. However, recent dynamical simulations of globular clusters \cite[e.g.,][]{ye2020} find the subsequent merger of such a second-generation BH with a larger stellar-mass BH to be exceedingly rare. A high component spin could be a distinguishing feature of a second-generation compact object, but the uninformative spin posterior for the lighter component of \eventname{} provides no evidence for or against this hypothesis.

A \eventname{}-like merger may also have originated from a hierarchical triple in the field \cite[e.g.,][]{silsbee2017,fragione2019b,antonini2017}, from a wide hierarchical quadruple system \citep{SafarzadehHamers2020}, or from hierarchical triples in galactic centers, where the tertiary body is a supermassive BH~\citep{antonini2012,stephan2019,petrovich2017,hoang2018,fragione2019gc}. Specifically, \citet{SafarzadehHamers2020} explore the possibility that a second-generation remnant with mass $3\,\Msolar$ may merge with a $30\,\Msolar$ BH, catalyzed by a $50\,\Msolar$-BH perturber. The mass-ratio distributions of BBH and NSBH mergers from hierarchical systems are similar to those of field binaries and it is unclear whether hierarchies may enhance the formation of merging compact-object binaries with highly asymmetric masses \citep[e.g.,][]{silsbee2017}.

Disks of gas around supermassive BHs in active galactic nuclei may be promising environments for the formation of \eventname{}-like systems. For such environments, theoretical models show that merging compact-object binaries with asymmetric masses are likely, but cannot necessarily accommodate masses as low as the secondary mass of \eventname{} \cite[e.g.,][]{yang2019}. However, \cite{mckernan2020} show that the median mass ratio of NSBH mergers in active galactic nucleus disks may be as low as {$\sim 0.07$}.

We conclude that {the combination of masses, mass ratio and inferred rate of \eventname{}} is challenging to explain, but {potentially} consistent with multiple formation scenarios. {However, it is not possible to assess the validity of models that produce the right properties but do not make quantitative predictions about formation rates, even at some order-of-magnitude level.} 

Young star clusters and active galactic nucleus disks seem to be more promising hosts for \eventname{}-like mergers, since both these environments {may enhance the formation of either progenitors of or directly merging compact-object binaries with more asymmetric masses to relevant rates.}
In contrast, {globular-cluster models provide more robust predictions, showing that  \eventname{}-like mergers with such asymmetric masses are outliers in the population predictions, even though a revision of the remnant-mass prescription is still needed. Isolated binaries binaries could prove possible progenitors provided similar revisions are implemented. } 
The importance of field multiples remains to be fully explored. Future gravitational-wave observations will provide further insights into the dominance of different channels.

\subsection{Cosmological Implications}
Luminosity distances inferred directly from observed gravitational-wave events can be used with measurements of source redshifts in the electromagnetic spectrum to constrain cosmological parameters~\citep{Schutz:1986gp}. Redshifts can be either obtained directly from counterparts to the gravitational-wave source~\citep{2005ApJ...629...15H}, as was the case for GW170817~\citep{Abbott:2018exr,LVCmulti2017,2017Natur.551...85A}, by cross-correlation of the gravitational-wave localization posterior with catalogs of galaxy redshifts~\citep{DelPozzo:2011yh,Nair:2018ign,Chen:2017rfc,Fishbach:2018gjp,Gray:2019ksv,2019ApJ...876L...7S,LVCO2Cosmo}, {by exploiting information in the neutron star equation of state~\citep{2012PhRvL.108i1101M}, or by using the redshifted masses inferred from the gravitational wave observation and assumptions about the mass distribution of the sources~\citep{1993ApJ...411L...5C,2012PhRvD..85b3535T,2012PhRvD..86b3502T,2019ApJ...883L..42F}}. At current sensitivities, the cosmological parameter to which LIGO--Virgo observations are most sensitive is the Hubble constant, $H_0$. {The gravitational-wave observation of GW170817 provided a posterior on $H_0$ with mode and $68.3\%$ highest posterior density interval of \HoGWSeventeenAugSeventeen~\citep{2017Natur.551...85A,2019PhRvX...9a1001A,LVCO2Cosmo}, assuming a flat prior on $H_0$.}

\eventname{} is the best localized dark siren, i.e., gravitational-wave source without an electromagnetic counterpart, observed to date, 
and so it is a good candidate for the statistical cross-correlation method. {For a fixed reference cosmology~\citep{Ade:2015xua}}, the GLADE galaxy catalog~\citep{2018MNRAS.479.2374D} is approximately {\GLADEcomplete~complete} at the distance of \eventname{} and contains {\GLADEngal~galaxies} within the \GLADEpostreg~posterior credible volume of \eventname{}. To obtain a constrain on $H_0$, we use the methodology described in~\citet{LVCO2Cosmo} and the GLADE catalog. We take a flat prior for $H_0 \in$ [\Hopriorlow, \Hopriorupp]~km s$^{-1}$ Mpc$^{-1}$ and assign a probability to each galaxy that it is the true host of the event that is proportional to its B-band luminosity. Using the posterior distribution on the distance obtained from the combined PHM samples, we obtain $H_0=$~\HoGWNineteenAugFourteen~using \eventname{} alone (mode and $68.3\%$ highest posterior density interval; the median and 90\% symmetric credible interval is \HoGWNineteenAugFourteenMed), which can be compared to \HoGWSeventeenAugFourteen~\citep{2019ApJ...876L...7S} obtained using the dark siren GW170814 alone. The \eventname{} result is the most precise measurement from a single dark siren observation to date, {albeit comparable to the GW170814 result}, which is expected given \eventname{}'s small localization volume ($\sim$\NinetyPercentVolume~Mpc$^{3}$). The result is not very constraining, with the $68.3\%$ highest posterior density interval comprising \Hopriorfrac~of the prior range. 
{Combining the result for \eventname{} with the result obtained from GW170817, we see an improvement over the GW170817-only result, to \HocombinedGWSeventeenAugSeventeen~(the median and 90\% symmetric credible interval is \HocombinedGWSeventeenAugSeventeenMed)}. This result is not yet sufficiently constraining to provide further insight into current tensions in low and high redshift measurements of the Hubble constant~\citep{2019NatAs...3..891V}, but these constraints will continue to improve as further gravitational-wave observations are included~\citep[e.g., projections in][]{Chen:2017rfc,2018PhRvL.121b1303V,Gray:2019ksv,2019PhRvL.122f1105F}.


\section{Conclusions}

During their third observing run, on \DateTime{}, the LIGO and Virgo detectors observed \eventname{}, a novel source unlike any other known compact binary coalescence. Thanks in part to the observation of significant power in subdominant multipoles of the gravitational radiation, and the conclusive measurement of little to no spin precession, we obtain precise measurements of its physical source properties that clearly set it apart from other compact binaries. 

In particular, (a) its mass ratio of $q=\Massratio$ is the most unequal ever observed with gravitational waves, (b) the bound $\chi_1 \leq \Primaryspinninety$ on the spin of the \Monesource\,\Msolar~BH is the strongest constraint on a primary spin for any gravitational-wave source to date, and (c) the secondary mass measurement of \Mtwosource\,\Msolar~makes it the lightest BH or the heaviest NS discovered in a double compact-object system. We find no evidence of measurable tidal effects in the signal, and no electromagnetic counterpart to the gravitational waves has been identified.

Comparisons between the secondary mass and several current estimates of the maximum NS mass suggest that \eventname{} is unlikely to originate in a NSBH coalescence. Nevertheless, the $M_{\rm max}$ estimates are uncertain enough that improved knowledge of the NS EOS or further observations of the astrophysical population of compact objects could alter this assessment. For this reason, we cannot firmly exclude the possibility that the secondary is a NS, nor can we be certain that it is a BH. Regardless, this event sheds new light on the compact-object mass distribution at the interface between known NSs and BHs.

The unique combination of masses and inferred merger rate for this event is difficult to produce in population synthesis models of multi-component systems in galactic fields or dense stellar environments. The discovery of \eventname{} may therefore reshape our understanding of the processes by which the lightest BHs or the most massive NSs form. Based on our rate density estimate, we may reasonably expect to detect more systems of this kind after a year at design sensitivity. This discovery may prove to be the first hint of a larger population that could change our perspective on the formation and mass spectrum of compact objects.

Segments of data containing the signal from all three interferometers, and samples from the posterior distributions for the source parameters, are available from the Gravitational Wave Open Science Center (\href{https://doi.org/10.7935/zzw5-ak90}{ https://doi.org/10.7935/zzw5-ak90}). The software packages used in our analysis are open source. \\

The authors gratefully acknowledge the support of the United States
National Science Foundation (NSF) for the construction and operation of the
LIGO Laboratory and Advanced LIGO as well as the Science and Technology Facilities Council (STFC) of the
United Kingdom, the Max-Planck-Society (MPS), and the State of
Niedersachsen/Germany for support of the construction of Advanced LIGO
and construction and operation of the GEO600 detector.
Additional support for Advanced LIGO was provided by the Australian Research Council.
The authors gratefully acknowledge the Italian Istituto Nazionale di Fisica Nucleare (INFN),
the French Centre National de la Recherche Scientifique (CNRS) and
the Netherlands Organization for Scientific Research,
for the construction and operation of the Virgo detector
and the creation and support  of the EGO consortium.

The authors also gratefully acknowledge research support from these agencies as well as by
the Council of Scientific and Industrial Research of India,
the Department of Science and Technology, India,
the Science \& Engineering Research Board (SERB), India,
the Ministry of Human Resource Development, India,
the Spanish Agencia Estatal de Investigaci\'on,
the Vicepresid\`encia i Conselleria d'Innovaci\'o, Recerca i Turisme and the Conselleria d'Educaci\'o i Universitat del Govern de les Illes Balears,
the Conselleria d'Innovaci\'o, Universitats, Ci\`encia i Societat Digital de la Generalitat Valenciana and
the CERCA Programme Generalitat de Catalunya, Spain,
the National Science Centre of Poland,
the Swiss National Science Foundation (SNSF),
the Russian Foundation for Basic Research,
the Russian Science Foundation,
the European Commission,
the European Regional Development Funds (ERDF),
the Royal Society,
the Scottish Funding Council,
the Scottish Universities Physics Alliance,
the Hungarian Scientific Research Fund (OTKA),
the French Lyon Institute of Origins (LIO),
the Belgian Fonds de la Recherche Scientifique (FRS-FNRS),
Actions de Recherche Concertées (ARC) and
Fonds Wetenschappelijk Onderzoek – Vlaanderen (FWO), Belgium,
the Paris \^{I}le-de-France Region,
the National Research, Development and Innovation Office Hungary (NKFIH),
the National Research Foundation of Korea,
Industry Canada and the Province of Ontario through the Ministry of Economic Development and Innovation,
the Natural Science and Engineering Research Council Canada,
the Canadian Institute for Advanced Research,
the Brazilian Ministry of Science, Technology, Innovations, and Communications,
the International Center for Theoretical Physics South American Institute for Fundamental Research (ICTP-SAIFR),
the Research Grants Council of Hong Kong,
the National Natural Science Foundation of China (NSFC),
the Leverhulme Trust,
the Research Corporation,
the Ministry of Science and Technology (MOST), Taiwan
and
the Kavli Foundation.
The authors gratefully acknowledge the support of the NSF, STFC, INFN and CNRS for provision of computational resources.
Some of the parameter estimation analyses presented in this paper were performed using the supercomputer cluster at the Swinburne University of Technology (OzSTAR and SSTAR).

We would like to thank all of the essential workers who put their health at risk during the COVID-19 pandemic, without whom we would not have been able to complete this work.

\software{The detection of the signal and subsequent significance evaluation were performed
  with the \textsc{GstLAL}-based inspiral software pipeline \citep{Cannon:2011vi,
  Privitera:2013xza, Messick:2016aqy, Sachdev:2019vvd, Hanna:2019ezx}, built on the
  \textsc{LALSuite} software library \citep{LALSuite}, and with the \textsc{PyCBC}
  \citep{Nitz:2018rgo, pycbc-software, Usman:2015kfa} and
  \textsc{MBTAOnline}~\citep{2016CQGra..33q5012A} packages. Parameter estimation was performed with
  the \textsc{LALInference} \citep{PhysRevD.91.042003} and \textsc{LALSimulation} libraries within
  \textsc{LALSuite} \citep{LALSuite}, as well as the \textsc{Bilby} and \textsc{pBilby} Libraries \citep{2019ApJS..241...27A, Smith:2019ucc} and the \textsc{dynesty} nested sampling package \citep{2019arXiv190402180S}. Interpretation and curation of the posterior samples was handled by the \textsc{PESummary} library~\citep{PEsummary}. 
  Estimates of the noise spectra were obtained using \textsc{BayesWave} \citep{Cornish:2014kda, PhysRevD.91.084034}. Plots were prepared with Matplotlib \citep{2007CSE.....9...90H}. The sky map
  plot also used Astropy (\href{http://www.astropy.org}{http://www.astropy.org}) a community-developed core Python
  package for Astronomy \citep{astropy:2013, astropy:2018} and
  \texttt{ligo.skymap} (\href{https://lscsoft.docs.ligo.org/ligo.skymap}{https://lscsoft.docs.ligo.org/ligo.skymap}).}

\bibliographystyle{aasjournal}
\bibliography{references}

\begin{thebibliography}{}
\expandafter\ifx\csname natexlab\endcsname\relax\def\natexlab#1{#1}\fi
\providecommand{\url}[1]{\href{#1}{#1}}
\providecommand{\dodoi}[1]{doi:~\href{http://doi.org/#1}{\nolinkurl{#1}}}
\providecommand{\doeprint}[1]{\href{http://ascl.net/#1}{\nolinkurl{http://ascl.net/#1}}}
\providecommand{\doarXiv}[1]{\href{https://arxiv.org/abs/#1}{\nolinkurl{https://arxiv.org/abs/#1}}}

\bibitem[{{Aasi} {et~al.}(2015){Aasi}, {Abbott}, {Abbott},
  {et~al.}}]{TheLIGOScientific:2014jea}
{Aasi}, J., {Abbott}, B.~P., {Abbott}, R., {et~al.} 2015, \cqg, 32, 074001,
  \dodoi{10.1088/0264-9381/32/7/074001}

\bibitem[{{Abbott} {et~al.}(2016{\natexlab{a}}){Abbott}, {Abbott}, {Abbott},
  {et~al.}}]{GW150914}
{Abbott}, B.~P., {Abbott}, R., {Abbott}, T.~D., {et~al.} 2016{\natexlab{a}},
  \prl, 116, 061102, \dodoi{10.1103/PhysRevLett.116.061102}

\bibitem[{{Abbott} {et~al.}(2016{\natexlab{b}}){Abbott}, {Abbott}, {Abbott},
  {et~al.}}]{O1BBH}
---. 2016{\natexlab{b}}, \prx, 6, 041015, \dodoi{10.1103/PhysRevX.6.041015}

\bibitem[{{Abbott} {et~al.}(2016{\natexlab{c}}){Abbott}, {Abbott}, {Abbott},
  {et~al.}}]{Abbott:2016ctn}
---. 2016{\natexlab{c}}, \cqg, 33, 134001,
  \dodoi{10.1088/0264-9381/33/13/134001}

\bibitem[{{Abbott} {et~al.}(2016{\natexlab{d}}){Abbott}, {Abbott}, {Abbott},
  {et~al.}}]{TheLIGOScientific:2016qqj}
---. 2016{\natexlab{d}}, \prd, 93, 122003, \dodoi{10.1103/PhysRevD.93.122003}

\bibitem[{{Abbott} {et~al.}(2016{\natexlab{e}}){Abbott}, {Abbott}, {Abbott},
  {et~al.}}]{GW150914:obs}
---. 2016{\natexlab{e}}, \prd, 93, 122004, \dodoi{10.1103/PhysRevD.93.122004}

\bibitem[{{Abbott} {et~al.}(2016{\natexlab{f}}){Abbott}, {Abbott}, {Abbott},
  {et~al.}}]{GW150914:PE}
---. 2016{\natexlab{f}}, \prl, 116, 241102,
  \dodoi{10.1103/PhysRevLett.116.241102}

\bibitem[{{Abbott} {et~al.}(2016{\natexlab{g}}){Abbott}, {Abbott}, {Abbott},
  {et~al.}}]{GW150914:TGR}
---. 2016{\natexlab{g}}, \prl, 116, 221101,
  \dodoi{10.1103/PhysRevLett.116.221101}

\bibitem[{{Abbott} {et~al.}(2017{\natexlab{a}}){Abbott}, {Abbott}, {Abbott},
  {et~al.}}]{Abbott:2018exr}
---. 2017{\natexlab{a}}, \prl, 119, 161101,
  \dodoi{10.1103/PhysRevLett.119.161101}

\bibitem[{{Abbott} {et~al.}(2017{\natexlab{b}}){Abbott}, {Abbott}, {Abbott},
  {et~al.}}]{LVCmulti2017}
---. 2017{\natexlab{b}}, \apjl, 848, L12, \dodoi{10.3847/2041-8213/aa91c9}

\bibitem[{{Abbott} {et~al.}(2017{\natexlab{c}}){Abbott}, {Abbott}, {Abbott},
  {et~al.}}]{2017ApJ...848L..13A}
---. 2017{\natexlab{c}}, \apjl, 848, L13, \dodoi{10.3847/2041-8213/aa920c}

\bibitem[{{Abbott} {et~al.}(2017{\natexlab{d}}){Abbott}, {Abbott}, {Abbott},
  {et~al.}}]{LVC170817ejecta}
---. 2017{\natexlab{d}}, \apjl, 850, L39, \dodoi{10.3847/2041-8213/aa9478}

\bibitem[{{Abbott} {et~al.}(2017{\natexlab{e}}){Abbott}, {Abbott}, {Abbott},
  {et~al.}}]{2017Natur.551...85A}
---. 2017{\natexlab{e}}, Nat, 551, 85, \dodoi{10.1038/nature24471}

\bibitem[{{Abbott} {et~al.}(2017{\natexlab{f}}){Abbott}, {Abbott}, {Abbott},
  {et~al.}}]{Abbott:2017gyy}
---. 2017{\natexlab{f}}, \apjl, 851, L35, \dodoi{10.3847/2041-8213/aa9f0c}

\bibitem[{{Abbott} {et~al.}(2017{\natexlab{g}}){Abbott}, {Abbott}, {Abbott},
  {et~al.}}]{Abbott:2017dke}
---. 2017{\natexlab{g}}, \apjl, 851, L16, \dodoi{10.3847/2041-8213/aa9a35}

\bibitem[{{Abbott} {et~al.}(2018){Abbott}, {Abbott}, {Abbott},
  {et~al.}}]{LVCeos2018}
---. 2018, \prl, 121, 161101, \dodoi{10.1103/PhysRevLett.121.161101}

\bibitem[{{Abbott} {et~al.}(2019{\natexlab{a}}){Abbott}, {Abbott}, {Abbott},
  {et~al.}}]{GWLIGOScientific:2018mvr}
---. 2019{\natexlab{a}}, PhRvX, 9, 031040, \dodoi{10.1103/PhysRevX.9.031040}

\bibitem[{{Abbott} {et~al.}(2019{\natexlab{b}}){Abbott}, {Abbott}, {Abbott},
  {et~al.}}]{2019PhRvX...9a1001A}
---. 2019{\natexlab{b}}, \prx, 9, 011001, \dodoi{10.1103/PhysRevX.9.011001}

\bibitem[{{Abbott} {et~al.}(2019{\natexlab{c}}){Abbott}, {Abbott}, {Abbott},
  {et~al.}}]{LVCO2Cosmo}
---. 2019{\natexlab{c}},
  arXiv:\href{https://arxiv.org/abs/1908.06060}{1908.06060}

\bibitem[{{Abbott} {et~al.}(2019{\natexlab{d}}){Abbott}, {Abbott}, {Abbott},
  {et~al.}}]{Abbott:2018lct}
---. 2019{\natexlab{d}}, \prl, 123, 011102,
  \dodoi{10.1103/PhysRevLett.123.011102}

\bibitem[{{Abbott} {et~al.}(2019{\natexlab{e}}){Abbott}, {Abbott}, {Abbott},
  {et~al.}}]{LIGOScientific:2019fpa}
---. 2019{\natexlab{e}}, \prd, 100, 104036, \dodoi{10.1103/PhysRevD.100.104036}

\bibitem[{{Abbott} {et~al.}(2019{\natexlab{f}}){Abbott}, {Abbott}, {Abbott},
  {et~al.}}]{AbbottAbbott2019longpostmerger}
---. 2019{\natexlab{f}}, \apj, 875, 160, \dodoi{10.3847/1538-4357/ab0f3d}

\bibitem[{{Abbott} {et~al.}(2020{\natexlab{a}}){Abbott}, {Abbott}, {Abbott},
  {et~al.}}]{Abbott:2020uma}
---. 2020{\natexlab{a}}, ApJL, 892, L3, \dodoi{10.3847/2041-8213/ab75f5}

\bibitem[{{Abbott} {et~al.}(2020{\natexlab{b}}){Abbott}, {Abbott}, {Abbott},
  {et~al.}}]{ObservingScenariosArxiv}
---. 2020{\natexlab{b}},
  arXiv:\href{https://arxiv.org/abs/1304.0670v10}{1304.0670v10}

\bibitem[{{Abbott} {et~al.}(2020{\natexlab{c}}){Abbott}, {Abbott}, {Abbott},
  {et~al.}}]{GW170817eosMtov}
---. 2020{\natexlab{c}}, CQGra, 37, 045006, \dodoi{10.1088/1361-6382/ab5f7c}

\bibitem[{{Abbott} {et~al.}(2020{\natexlab{d}}){Abbott}, {Abbott}, {Abraham},
  {Acernese}, {Ackley}, {Adams}, {Adhikari}, {Adya}, \&
  et~al.}]{GW190412-discovery}
{Abbott}, R., {Abbott}, T.~D., {Abraham}, S., {et~al.} 2020{\natexlab{d}},
  arXiv:\href{https://arxiv.org/abs/2004.08342}{2004.08342}

\bibitem[{{Accadia} {et~al.}(2010){Accadia}, {Acernese}, {Antonucci}, {Astone},
  {Ballardin}, {et~al.}}]{Accadia:2010zzb}
{Accadia}, T., {Acernese}, F., {Antonucci}, F., {et~al.} 2010, CQGra, 27,
  194011, \dodoi{10.1088/0264-9381/27/19/194011}

\bibitem[{{Acernese} {et~al.}(2018){Acernese}, {Adams}, {Agatsuma},
  {et~al.}}]{Acernese:2018bfl}
{Acernese}, F., {Adams}, T., {Agatsuma}, K., {et~al.} 2018, \cqg, 35, 205004,
  \dodoi{10.1088/1361-6382/aadf1a}

\bibitem[{{Acernese} {et~al.}(2015){Acernese}, {Agathos}, {Agatsuma},
  {et~al.}}]{Virgo:2014hva}
{Acernese}, F., {Agathos}, M., {Agatsuma}, K., {et~al.} 2015, \cqg, 32, 024001,
  \dodoi{10.1088/0264-9381/32/2/024001}

\bibitem[{{Ackley} {et~al.}(2020){Ackley}, {Amati}, {Barbieri}, {Bauer},
  {Benetti}, {Bernardini}, {Bhirombhakdi}, {Botticella}, {Branchesi},
  {Brocato}, \& et~al.}]{AckleyAmati2020}
{Ackley}, K., {Amati}, L., {Barbieri}, C., {et~al.} 2020,
  arXiv:\href{https://arxiv.org/abs/2002.01950}{2002.01950}

\bibitem[{{Adams} {et~al.}(2016){Adams}, {Buskulic}, {Germain}, {Guidi},
  {Marion}, {Montani}, {Mours}, {Piergiovanni}, \&
  {Wang}}]{2016CQGra..33q5012A}
{Adams}, T., {Buskulic}, D., {Germain}, V., {et~al.} 2016, CQGra, 33, 175012,
  \dodoi{10.1088/0264-9381/33/17/175012}

\bibitem[{{Ade} {et~al.}(2016){Ade}, {Aghanim}, {Arnaud},
  {et~al.}}]{Ade:2015xua}
{Ade}, P.~A.~R., {Aghanim}, N., {Arnaud}, M., {et~al.} 2016, \aap, 594, A13,
  \dodoi{10.1051/0004-6361/201525830}

\bibitem[{Ageron {et~al.}(2019)Ageron, Baret, Coleiro, Colomer, Dornic,
  Kouchner, Pradier, \& the ANTARES~Collaboration}]{GCN25330}
Ageron, M., Baret, B., Coleiro, A., {et~al.} 2019, \gcn, 25330.
\newblock \url{https://gcn.gsfc.nasa.gov/gcn3/25330.gcn3}

\bibitem[{Ajith {et~al.}(2014)Ajith, Fotopoulos, Privitera, Neunzert, \&
  Weinstein}]{Ajith:2012mn}
Ajith, P., Fotopoulos, N., Privitera, S., Neunzert, A., \& Weinstein, A.~J.
  2014, PhRvD, 89, 084041, \dodoi{10.1103/PhysRevD.89.084041}

\bibitem[{{Ajith} {et~al.}(2011){Ajith}, {Hannam}, {Husa}, {Chen},
  {Br{\"u}gmann}, {Dorband}, {M{\"u}ller}, {Ohme}, {Pollney}, {Reisswig}, \&
  et~al.}]{ajith2011inspiral}
{Ajith}, P., {Hannam}, M., {Husa}, S., {et~al.} 2011, \prl, 106, 241101,
  \dodoi{10.1103/PhysRevLett.106.241101}

\bibitem[{{Alsing} {et~al.}(2018){Alsing}, {Silva}, \&
  {Berti}}]{AlsingSilva2018}
{Alsing}, J., {Silva}, H.~O., \& {Berti}, E. 2018, \mnras, 478, 1377,
  \dodoi{10.1093/mnras/sty1065}

\bibitem[{{Andreoni} {et~al.}(2020){Andreoni}, {Goldstein}, {Kasliwal},
  {Nugent}, {Zhou}, {Newman}, {Bulla}, {Foucart}, {Hotokezaka}, {Nakar}, \&
  et~al.}]{AndreoniGoldstein2019}
{Andreoni}, I., {Goldstein}, D.~A., {Kasliwal}, M.~M., {et~al.} 2020, \apj,
  890, 131, \dodoi{10.3847/1538-4357/ab6a1b}

\bibitem[{{Antier} {et~al.}(2020){Antier}, {Agayeva}, {Aivazyan}, {Alishov},
  {Arbouch}, {Baransky}, {Barynova}, {Bai}, {Basa}, {Beradze}, \&
  et~al.}]{AntierAgayeva2020}
{Antier}, S., {Agayeva}, S., {Aivazyan}, V., {et~al.} 2020, \mnras, 492, 3904,
  \dodoi{10.1093/mnras/stz3142}

\bibitem[{{Antonini} \& {Perets}(2012)}]{antonini2012}
{Antonini}, F., \& {Perets}, H.~B. 2012, \apj, 757, 27,
  \dodoi{10.1088/0004-637X/757/1/27}

\bibitem[{{Antonini} {et~al.}(2017){Antonini}, {Toonen}, \&
  {Hamers}}]{antonini2017}
{Antonini}, F., {Toonen}, S., \& {Hamers}, A.~S. 2017, \apj, 841, 77,
  \dodoi{10.3847/1538-4357/aa6f5e}

\bibitem[{Apostolatos {et~al.}(1994)Apostolatos, Cutler, Sussman, \&
  Thorne}]{Apostolatos:1994}
Apostolatos, T.~A., Cutler, C., Sussman, G.~J., \& Thorne, K.~S. 1994, \prd,
  49, 6274, \dodoi{10.1103/PhysRevD.49.6274}

\bibitem[{{Arca Sedda}(2020)}]{arcasedda2020}
{Arca Sedda}, M. 2020, CmPhy, 3, 43, \dodoi{10.1038/s42005-020-0310-x}

\bibitem[{Arun {et~al.}(2009)Arun, Buonanno, Faye, \& Ochsner}]{Arun:2008kb}
Arun, K.~G., Buonanno, A., Faye, G., \& Ochsner, E. 2009, \prd, 79, 104023,
  \dodoi{10.1103/PhysRevD.79.104023}

\bibitem[{Arun {et~al.}(2006{\natexlab{a}})Arun, Iyer, Qusailah, \&
  Sathyaprakash}]{PhysRevD.74.024006}
Arun, K.~G., Iyer, B.~R., Qusailah, M. S.~S., \& Sathyaprakash, B.~S.
  2006{\natexlab{a}}, PhRvD, 74, 024006, \dodoi{10.1103/PhysRevD.74.024006}

\bibitem[{Arun {et~al.}(2006{\natexlab{b}})Arun, Iyer, Qusailah, \&
  Sathyaprakash}]{Arun_2006}
---. 2006{\natexlab{b}}, CQGra, 23, L37–L43,
  \dodoi{10.1088/0264-9381/23/9/l01}

\bibitem[{{Ashton} {et~al.}(2019){Ashton}, {H{\"u}bner}, {Lasky}, {Talbot},
  {Ackley}, {Biscoveanu}, {Chu}, {Divakarla}, {Easter}, {Goncharov}, {Hernandez
  Vivanco}, {Harms}, {Lower}, {Meadors}, {Melchor}, {Payne}, {Pitkin},
  {Powell}, {Sarin}, {Smith}, \& {Thrane}}]{2019ApJS..241...27A}
{Ashton}, G., {H{\"u}bner}, M., {Lasky}, P.~D., {et~al.} 2019, \apjs, 241, 27,
  \dodoi{10.3847/1538-4365/ab06fc}

\bibitem[{{Askar} {et~al.}(2017){Askar}, {Szkudlarek}, {Gondek-Rosi{\'n}ska},
  {Giersz}, \& {Bulik}}]{askar2017}
{Askar}, A., {Szkudlarek}, M., {Gondek-Rosi{\'n}ska}, D., {Giersz}, M., \&
  {Bulik}, T. 2017, \mnras, 464, L36, \dodoi{10.1093/mnrasl/slw177}

\bibitem[{{Astropy Collaboration} {et~al.}(2013){Astropy Collaboration},
  {Robitaille}, {Tollerud}, {Greenfield}, {Droettboom}, {Bray}, {Aldcroft},
  {Davis}, {Ginsburg}, {Price-Whelan}, {Kerzendorf}, {Conley}, {Crighton},
  {Barbary}, {Muna}, {Ferguson}, {Grollier}, {Parikh}, {Nair}, {Unther},
  {Deil}, {Woillez}, {Conseil}, {Kramer}, {Turner}, {Singer}, {Fox}, {Weaver},
  {Zabalza}, {Edwards}, {Azalee Bostroem}, {Burke}, {Casey}, {Crawford},
  {Dencheva}, {Ely}, {Jenness}, {Labrie}, {Lim}, {Pierfederici}, {Pontzen},
  {Ptak}, {Refsdal}, {Servillat}, \& {Streicher}}]{astropy:2013}
{Astropy Collaboration}, {Robitaille}, T.~P., {Tollerud}, E.~J., {et~al.} 2013,
  \aap, 558, A33, \dodoi{10.1051/0004-6361/201322068}

\bibitem[{Babak {et~al.}(2017)Babak, Taracchini, \& Buonanno}]{Babak:2016tgq}
Babak, S., Taracchini, A., \& Buonanno, A. 2017, PhRvD, 95, 024010,
  \dodoi{10.1103/PhysRevD.95.024010}

\bibitem[{{Bailyn} {et~al.}(1998){Bailyn}, {Jain}, {Coppi}, \&
  {Orosz}}]{1998ApJ...499..367B}
{Bailyn}, C.~D., {Jain}, R.~K., {Coppi}, P., \& {Orosz}, J.~A. 1998, \apj, 499,
  367, \dodoi{10.1086/305614}

\bibitem[{Baird {et~al.}(2013)Baird, Fairhurst, Hannam, \&
  Murphy}]{Baird:2012cu}
Baird, E., Fairhurst, S., Hannam, M., \& Murphy, P. 2013, PhRvD, 87, 024035,
  \dodoi{10.1103/PhysRevD.87.024035}

\bibitem[{Blanchet(2014)}]{Blanchet:2013haa}
Blanchet, L. 2014, LRR, 17, 2, \dodoi{10.12942/lrr-2014-2}

\bibitem[{{Blanchet} {et~al.}(2005){Blanchet}, {Damour}, {Esposito-Far{\`e}se},
  \& {Iyer}}]{blanchet2005dimensional}
{Blanchet}, L., {Damour}, T., {Esposito-Far{\`e}se}, G., \& {Iyer}, B.~R. 2005,
  \prd, 71, 124004, \dodoi{10.1103/PhysRevD.71.124004}

\bibitem[{{Blanchet} {et~al.}(1995){Blanchet}, {Damour}, {Iyer}, {Will}, \&
  {Wiseman}}]{Blanchet:1995ez}
{Blanchet}, L., {Damour}, T., {Iyer}, B.~R., {Will}, C.~M., \& {Wiseman}, A.~G.
  1995, \prl, 74, 3515, \dodoi{10.1103/PhysRevLett.74.3515}

\bibitem[{Blanchet {et~al.}(2008)Blanchet, Faye, Iyer, \&
  Sinha}]{Blanchet:2008je}
Blanchet, L., Faye, G., Iyer, B.~R., \& Sinha, S. 2008, CQGra, 25, 165003,
  \dodoi{10.1088/0264-9381/25/16/165003, 10.1088/0264-9381/29/23/239501}

\bibitem[{{Boh{\'e}} {et~al.}(2017){Boh{\'e}}, {Shao}, {Taracchini},
  {Buonanno}, {Babak}, {Harry}, {Hinder}, {Ossokine}, {P{\"u}rrer}, {Raymond},
  {Chu}, {Fong}, {Kumar}, {Pfeiffer}, {Boyle}, {Hemberger}, {Kidder},
  {Lovelace}, {Scheel}, \& {Szil{\'a}gyi}}]{Bohe:2016gbl}
{Boh{\'e}}, A., {Shao}, L., {Taracchini}, A., {et~al.} 2017, \prd, 95, 044028,
  \dodoi{10.1103/PhysRevD.95.044028}

\bibitem[{Brown {et~al.}(2013)Brown, Kumar, \& Nitz}]{Brown:2012nn}
Brown, D.~A., Kumar, P., \& Nitz, A.~H. 2013, PhRvD, 87, 082004,
  \dodoi{10.1103/PhysRevD.87.082004}

\bibitem[{Buonanno \& Damour(1999)}]{Buonanno:1998gg}
Buonanno, A., \& Damour, T. 1999, \prd, 59, 084006,
  \dodoi{10.1103/PhysRevD.59.084006}

\bibitem[{Buonanno {et~al.}(2003)}]{Buonanno:2002fy}
Buonanno, A., {et~al.} 2003, \prd, 67, 104025,
  \dodoi{10.1103/PhysRevD.67.104025, 10.1103/PhysRevD.74.029904}

\bibitem[{Burrows {et~al.}(2019)Burrows, Radice, \& Vartanyan}]{burrows2019}
Burrows, A., Radice, D., \& Vartanyan, D. 2019, MNRAS, 485, 3153,
  \dodoi{10.1093/mnras/stz543}

\bibitem[{{Burrows} {et~al.}(2018){Burrows}, {Vartanyan}, {Dolence}, {Skinner},
  \& {Radice}}]{burrows2018}
{Burrows}, A., {Vartanyan}, D., {Dolence}, J.~C., {Skinner}, M.~A., \&
  {Radice}, D. 2018, \ssr, 214, 33, \dodoi{10.1007/s11214-017-0450-9}

\bibitem[{{Cannon} {et~al.}(2012){Cannon}, {Cariou}, {Chapman},
  {et~al.}}]{Cannon:2011vi}
{Cannon}, K., {Cariou}, R., {Chapman}, A., {et~al.} 2012, \apj, 748, 136,
  \dodoi{10.1088/0004-637X/748/2/136}

\bibitem[{Capano {et~al.}(2016{\natexlab{a}})Capano, Dent, Hanna, Hendry, Hu,
  Messenger, \& Veitch}]{Capano:2016uif}
Capano, C., Dent, T., Hanna, C., {et~al.} 2016{\natexlab{a}}, PhRvD, 96,
  082002, \dodoi{10.1103/PhysRevD.96.082002}

\bibitem[{Capano {et~al.}(2016{\natexlab{b}})Capano, Harry, Privitera, \&
  Buonanno}]{Capano:2016dsf}
Capano, C., Harry, I., Privitera, S., \& Buonanno, A. 2016{\natexlab{b}},
  PhRvD, 93, 124007, \dodoi{10.1103/PhysRevD.93.124007}

\bibitem[{{Cardoso} \& {Pani}(2019)}]{CardosoPani2019}
{Cardoso}, V., \& {Pani}, P. 2019, LRR, 22, 4,
  \dodoi{10.1007/s41114-019-0020-4}

\bibitem[{{Carter}(1971)}]{carter1971axisymmetric}
{Carter}, B. 1971, \prl, 26, 331, \dodoi{10.1103/PhysRevLett.26.331}

\bibitem[{Chatterji {et~al.}(2004)}]{Chatterji:2004qg}
Chatterji, S., {et~al.} 2004, \cqg, 21, S1809,
  \dodoi{10.1088/0264-9381/21/20/024}

\bibitem[{Chen {et~al.}(2018)Chen, Fishbach, \& Holz}]{Chen:2017rfc}
Chen, H.-Y., Fishbach, M., \& Holz, D.~E. 2018, Nat, 562, 545,
  \dodoi{10.1038/s41586-018-0606-0}

\bibitem[{{Chernoff} \& {Finn}(1993)}]{1993ApJ...411L...5C}
{Chernoff}, D.~F., \& {Finn}, L.~S. 1993, \apjl, 411, L5,
  \dodoi{10.1086/186898}

\bibitem[{{Chornock} {et~al.}(2017){Chornock}, {Berger}, {Kasen},
  {Cowperthwaite}, {Nicholl}, {Villar}, {Alexand er}, {Blanchard}, {Eftekhari},
  {Fong}, \& et~al.}]{ChornockBerger2017}
{Chornock}, R., {Berger}, E., {Kasen}, D., {et~al.} 2017, \apjl, 848, L19,
  \dodoi{10.3847/2041-8213/aa905c}

\bibitem[{{Clausen} {et~al.}(2013){Clausen}, {Sigurdsson}, \&
  {Chernoff}}]{clausen2013}
{Clausen}, D., {Sigurdsson}, S., \& {Chernoff}, D.~F. 2013, \mnras, 428, 3618,
  \dodoi{10.1093/mnras/sts295}

\bibitem[{Cokelaer(2007)}]{Cokelaer:2007kx}
Cokelaer, T. 2007, \prd, 76, 102004, \dodoi{10.1103/PhysRevD.76.102004}

\bibitem[{{Cook} {et~al.}(1994){Cook}, {Shapiro}, \&
  {Teukolsky}}]{CookShapiro1994}
{Cook}, G.~B., {Shapiro}, S.~L., \& {Teukolsky}, S.~A. 1994, \apj, 424, 823,
  \dodoi{10.1086/173934}

\bibitem[{Cornish {et~al.}(2011)Cornish, Sampson, Yunes, \&
  Pretorius}]{PhysRevD.84.062003}
Cornish, N., Sampson, L., Yunes, N., \& Pretorius, F. 2011, PhRvD, 84, 062003,
  \dodoi{10.1103/PhysRevD.84.062003}

\bibitem[{Cornish \& Littenberg(2015)}]{Cornish:2014kda}
Cornish, N.~J., \& Littenberg, T.~B. 2015, \cqg, 32, 135012,
  \dodoi{10.1088/0264-9381/32/13/135012}

\bibitem[{Cotesta {et~al.}(2018)Cotesta, Buonanno, Bohé, Taracchini, Hinder,
  \& Ossokine}]{Cotesta:2018fcv}
Cotesta, R., Buonanno, A., Bohé, A., {et~al.} 2018, PhRvD, 98, 084028,
  \dodoi{10.1103/PhysRevD.98.084028}

\bibitem[{{Coughlin} {et~al.}(2020){Coughlin}, {Dietrich}, {Antier}, {Bulla},
  {Foucart}, {Hotokezaka}, {Raaijmakers}, {Hinderer}, \&
  {Nissanke}}]{CoughlinDietrich2020}
{Coughlin}, M.~W., {Dietrich}, T., {Antier}, S., {et~al.} 2020, \mnras, 492,
  863, \dodoi{10.1093/mnras/stz3457}

\bibitem[{{Cowperthwaite} {et~al.}(2017){Cowperthwaite}, {Berger}, {Villar},
  {Metzger}, {Nicholl}, {Chornock}, {Blanchard}, {Fong}, {Margutti},
  {Soares-Santos}, \& et~al.}]{CowperthwaiteBerger2017}
{Cowperthwaite}, P.~S., {Berger}, E., {Villar}, V.~A., {et~al.} 2017, \apjl,
  848, L17, \dodoi{10.3847/2041-8213/aa8fc7}

\bibitem[{{Cromartie} {et~al.}(2019){Cromartie}, {Fonseca}, {Ransom},
  {Demorest}, {Arzoumanian}, {Blumer}, {Brook}, {DeCesar}, {Dolch}, {Ellis},
  {Ferdman}, {Ferrara}, {Garver-Daniels}, {Gentile}, {Jones}, {Lam}, {Lorimer},
  {Lynch}, {McLaughlin}, {Ng}, {Nice}, {Pennucci}, {Spiewak}, {Stairs},
  {Stovall}, {Swiggum}, \& {Zhu}}]{Cromartie:2019kug}
{Cromartie}, H.~T., {Fonseca}, E., {Ransom}, S.~M., {et~al.} 2019, NatAs, 439,
  \dodoi{10.1038/s41550-019-0880-2}

\bibitem[{Cutler \& Flanagan(1994)}]{Cutler:1994ys}
Cutler, C., \& Flanagan, E.~E. 1994, \prd, 49, 2658,
  \dodoi{10.1103/PhysRevD.49.2658}

\bibitem[{{D{\'a}lya} {et~al.}(2018){D{\'a}lya}, {Galg{\'o}czi}, {Dobos},
  {Frei}, {Heng}, {Macas}, {Messenger}, {Raffai}, \& {de
  Souza}}]{2018MNRAS.479.2374D}
{D{\'a}lya}, G., {Galg{\'o}czi}, G., {Dobos}, L., {et~al.} 2018, \mnras, 479,
  2374, \dodoi{10.1093/mnras/sty1703}

\bibitem[{Damour(2001)}]{Damour:2001tu}
Damour, T. 2001, Phys. Rev., D64, 124013, \dodoi{10.1103/PhysRevD.64.124013}

\bibitem[{{Damour} {et~al.}(2001){Damour}, {Jaranowski}, \&
  {Sch{\"a}fer}}]{damour2001dimensional}
{Damour}, T., {Jaranowski}, P., \& {Sch{\"a}fer}, G. 2001, PhLB, 513, 147,
  \dodoi{10.1016/S0370-2693(01)00642-6}

\bibitem[{Davies {et~al.}(2020)Davies, Dent, Tápai, Harry, McIsaac, \&
  Nitz}]{Davies:2020tsx}
Davies, G.~S., Dent, T., Tápai, M., {et~al.} 2020,
  arXiv:\href{https://arxiv.org/abs/2002.08291}{2002.08291}

\bibitem[{Del~Pozzo(2012)}]{DelPozzo:2011yh}
Del~Pozzo, W. 2012, \prd, 86, 043011, \dodoi{10.1103/PhysRevD.86.043011}

\bibitem[{Dhanpal {et~al.}(2019)Dhanpal, Ghosh, Mehta, Ajith, \&
  Sathyaprakash}]{Dhanpal_2019}
Dhanpal, S., Ghosh, A., Mehta, A.~K., Ajith, P., \& Sathyaprakash, B. 2019,
  PhRvD, 99, \dodoi{10.1103/physrevd.99.104056}

\bibitem[{{Di Carlo} {et~al.}(2019){Di Carlo}, {Giacobbo}, {Mapelli},
  {Pasquato}, {Spera}, {Wang}, \& {Haardt}}]{dicarlo2019a}
{Di Carlo}, U.~N., {Giacobbo}, N., {Mapelli}, M., {et~al.} 2019, \mnras, 487,
  2947, \dodoi{10.1093/mnras/stz1453}

\bibitem[{{Dobie} {et~al.}(2019){Dobie}, {Stewart}, {Murphy}, {Lenc}, {Wang},
  {Kaplan}, {Andreoni}, {Banfield}, {Brown}, {Corsi}, {De}, {Goldstein},
  {Hallinan}, {Hotan}, {Hotokezaka}, {Jaodand}, {Karambelkar}, {Kasliwal},
  {McConnell}, {Mooley}, {Moss}, {Newman}, {Perley}, {Prakash}, {Pritchard},
  {Sadler}, {Sharma}, {Ward}, {Whiting}, \& {Zhou}}]{DobieStewart2019}
{Dobie}, D., {Stewart}, A., {Murphy}, T., {et~al.} 2019, \apjl, 887, L13,
  \dodoi{10.3847/2041-8213/ab59db}

\bibitem[{{Dominik} {et~al.}(2012){Dominik}, {Belczynski}, {Fryer}, {Holz},
  {Berti}, {Bulik}, {Mand el}, \& {O'Shaughnessy}}]{dominik2012}
{Dominik}, M., {Belczynski}, K., {Fryer}, C., {et~al.} 2012, \apj, 759, 52,
  \dodoi{10.1088/0004-637X/759/1/52}

\bibitem[{{Dominik} {et~al.}(2015){Dominik}, {Berti}, {O'Shaughnessy},
  {Mandel}, {Belczynski}, {Fryer}, {Holz}, {Bulik}, \&
  {Pannarale}}]{dominik2015}
{Dominik}, M., {Berti}, E., {O'Shaughnessy}, R., {et~al.} 2015, \apj, 806, 263,
  \dodoi{10.1088/0004-637X/806/2/263}

\bibitem[{Effler {et~al.}(2015)Effler, Schofield, Frolov, Gonzalez, Kawabe,
  Smith, Birch, \& McCarthy}]{Effler:2014zpa}
Effler, A., Schofield, R. M.~S., Frolov, V.~V., {et~al.} 2015, CQGra, 32,
  035017, \dodoi{10.1088/0264-9381/32/3/035017}

\bibitem[{{Eldridge} \& {Stanway}(2016)}]{eldridge2016}
{Eldridge}, J.~J., \& {Stanway}, E.~R. 2016, \mnras, 462, 3302,
  \dodoi{10.1093/mnras/stw1772}

\bibitem[{{Eldridge} {et~al.}(2017){Eldridge}, {Stanway}, {Xiao}, {McClelland
  }, {Taylor}, {Ng}, {Greis}, \& {Bray}}]{eldridge2017}
{Eldridge}, J.~J., {Stanway}, E.~R., {Xiao}, L., {et~al.} 2017, \pasa, 34,
  e058, \dodoi{10.1017/pasa.2017.51}

\bibitem[{{Ertl} {et~al.}(2016){Ertl}, {Janka}, {Woosley}, {Sukhbold}, \&
  {Ugliano}}]{ertl2016}
{Ertl}, T., {Janka}, H.~T., {Woosley}, S.~E., {Sukhbold}, T., \& {Ugliano}, M.
  2016, \apj, 818, 124, \dodoi{10.3847/0004-637X/818/2/124}

\bibitem[{{Essick} {et~al.}(2020){Essick}, {Landry}, \&
  {Holz}}]{EssickLandry2019}
{Essick}, R., {Landry}, P., \& {Holz}, D.~E. 2020, \prd, 101, 063007,
  \dodoi{10.1103/PhysRevD.101.063007}

\bibitem[{{Fairhurst} {et~al.}(2019{\natexlab{a}})}]{Fairhurst:2019srr}
{Fairhurst}, S., {et~al.} 2019{\natexlab{a}},
  arXiv:\href{https://arxiv.org/abs/1908.00555}{1908.00555}

\bibitem[{{Fairhurst} {et~al.}(2019{\natexlab{b}})}]{Fairhurst:2019_2harm}
---. 2019{\natexlab{b}},
  arXiv:\href{https://arxiv.org/abs/1908.05707}{1908.05707}

\bibitem[{{Farr} {et~al.}(2016){Farr}, {Berry}, {Farr}, {Haster}, {Middleton},
  {Cannon}, {Graff}, {Hanna}, {Mandel}, {Pankow}, {Price}, {Sidery}, {Singer},
  {Urban}, {Vecchio}, {Veitch}, \& {Vitale}}]{2016ApJ...825..116F}
{Farr}, B., {Berry}, C. P.~L., {Farr}, W.~M., {et~al.} 2016, \apj, 825, 116,
  \dodoi{10.3847/0004-637X/825/2/116}

\bibitem[{{Farr} \& {Chatziioannou}(2020)}]{FarrChatziioannou2020}
{Farr}, W.~M., \& {Chatziioannou}, K. 2020, Research Notes of the American
  Astronomical Society, 4, 65, \dodoi{10.3847/2515-5172/ab9088}

\bibitem[{{Farr} {et~al.}(2019){Farr}, {Fishbach}, {Ye}, \&
  {Holz}}]{2019ApJ...883L..42F}
{Farr}, W.~M., {Fishbach}, M., {Ye}, J., \& {Holz}, D.~E. 2019, \apjl, 883,
  L42, \dodoi{10.3847/2041-8213/ab4284}

\bibitem[{Farr {et~al.}(2011)Farr, Sravan, Cantrell, Kreidberg, Bailyn, Mandel,
  \& Kalogera}]{Farr:2010tu}
Farr, W.~M., Sravan, N., Cantrell, A., {et~al.} 2011, ApJ, 741, 103,
  \dodoi{10.1088/0004-637X/741/2/103}

\bibitem[{{Feeney} {et~al.}(2019){Feeney}, {Peiris}, {Williamson}, {Nissanke},
  {Mortlock}, {Alsing}, \& {Scolnic}}]{2019PhRvL.122f1105F}
{Feeney}, S.~M., {Peiris}, H.~V., {Williamson}, A.~R., {et~al.} 2019, \prl,
  122, 061105, \dodoi{10.1103/PhysRevLett.122.061105}

\bibitem[{Fern\'{a}ndez {et~al.}(2020)Fern\'{a}ndez, Foucart, \&
  Lippuner}]{fernandez2020landscape}
Fern\'{a}ndez, R., Foucart, F., \& Lippuner, J. 2020, The landscape of disk
  outflows from black hole - neutron star mergers.
\newblock \doarXiv{2005.14208}

\bibitem[{Fishbach {et~al.}(2019)Fishbach, Gray, Magaña~Hernandez, Qi, \&
  Sur}]{Fishbach:2018gjp}
Fishbach, M., Gray, R., Magaña~Hernandez, I., Qi, H., \& Sur, A. 2019, ApJ,
  871, L13, \dodoi{10.3847/2041-8213/aaf96e}

\bibitem[{{Fishbach} \& {Holz}(2020)}]{FishbachHolz2019}
{Fishbach}, M., \& {Holz}, D.~E. 2020, \apjl, 891, L27,
  \dodoi{10.3847/2041-8213/ab7247}

\bibitem[{Flanagan \& Hinderer(2008)}]{Flanagan:2007ix}
Flanagan, E.~E., \& Hinderer, T. 2008, \prd, 77, 021502,
  \dodoi{10.1103/PhysRevD.77.021502}

\bibitem[{Foucart {et~al.}(2013)Foucart, Buchman, Duez, Grudich, Kidder,
  MacDonald, Mroue, Pfeiffer, Scheel, \& Szilagyi}]{Foucart:2013psa}
Foucart, F., Buchman, L., Duez, M.~D., {et~al.} 2013, PhRvD, 88, 064017,
  \dodoi{10.1103/PhysRevD.88.064017}

\bibitem[{{Fragione} {et~al.}(2019){Fragione}, {Grishin}, {Leigh}, {Perets}, \&
  {Perna}}]{fragione2019gc}
{Fragione}, G., {Grishin}, E., {Leigh}, N. W.~C., {Perets}, H.~B., \& {Perna},
  R. 2019, \mnras, 488, 47, \dodoi{10.1093/mnras/stz1651}

\bibitem[{{Fragione} \& {Loeb}(2019)}]{fragione2019b}
{Fragione}, G., \& {Loeb}, A. 2019, \mnras, 486, 4443,
  \dodoi{10.1093/mnras/stz1131}

\bibitem[{{Freire} {et~al.}(2008){Freire}, {Ransom}, {B{\'e}gin}, {Stairs},
  {Hessels}, {Frey}, \& {Camilo}}]{FreireRansom2008}
{Freire}, P. C.~C., {Ransom}, S.~M., {B{\'e}gin}, S., {et~al.} 2008, \apj, 675,
  670, \dodoi{10.1086/526338}

\bibitem[{{Fryer} {et~al.}(2012){Fryer}, {Belczynski}, {Wiktorowicz},
  {Dominik}, {Kalogera}, \& {Holz}}]{fryer2012}
{Fryer}, C.~L., {Belczynski}, K., {Wiktorowicz}, G., {et~al.} 2012, \apj, 749,
  91, \dodoi{10.1088/0004-637X/749/1/91}

\bibitem[{{Giacobbo} \& {Mapelli}(2018)}]{giacobbo2018}
{Giacobbo}, N., \& {Mapelli}, M. 2018, \mnras, 480, 2011,
  \dodoi{10.1093/mnras/sty1999}

\bibitem[{{Giesers} {et~al.}(2018){Giesers}, {Dreizler}, {Husser}, {Kamann},
  {Anglada Escud{\'e}}, {Brinchmann}, {Carollo}, {Roth}, {Weilbacher}, \&
  {Wisotzki}}]{giesers2018}
{Giesers}, B., {Dreizler}, S., {Husser}, T.-O., {et~al.} 2018, \mnras, 475,
  L15, \dodoi{10.1093/mnrasl/slx203}

\bibitem[{{Gomez} {et~al.}(2019){Gomez}, {Hosseinzadeh}, {Cowperthwaite},
  {Villar}, {Berger}, {Gardner}, {Alexand er}, {Blanchard}, {Chornock},
  {Drout}, {Eftekhari}, {Fong}, {Gill}, {Margutti}, {Nicholl}, {Paterson}, \&
  {Williams}}]{GomezHosseinzadeh2019}
{Gomez}, S., {Hosseinzadeh}, G., {Cowperthwaite}, P.~S., {et~al.} 2019, \apjl,
  884, L55, \dodoi{10.3847/2041-8213/ab4ad5}

\bibitem[{{Gray} {et~al.}(2020){Gray}, {Hernandez}, {Qi}, {Sur}, {Brady},
  {Chen}, {Farr}, {Fishbach}, {Gair}, {Ghosh}, {Holz}, {Mastrogiovanni},
  {Messenger}, {Steer}, \& {Veitch}}]{Gray:2019ksv}
{Gray}, R., {Hernandez}, I.~M., {Qi}, H., {et~al.} 2020, \prd, 101, 122001,
  \dodoi{10.1103/PhysRevD.101.122001}

\bibitem[{{Guo} {et~al.}(2018){Guo}, {Chu}, {Chung}, {Du}, {Wen}, \&
  {Gu}}]{Guo2018}
{Guo}, X., {Chu}, Q., {Chung}, S.~K., {et~al.} 2018, Computer Physics
  Communications, 231, 62, \dodoi{10.1016/j.cpc.2018.05.002}

\bibitem[{{Gupta} {et~al.}(2020){Gupta}, {Gerosa}, {Arun}, {Berti}, {Farr}, \&
  {Sathyaprakash}}]{Gupta:2019nwj}
{Gupta}, A., {Gerosa}, D., {Arun}, K.~G., {et~al.} 2020, \prd, 101, 103036,
  \dodoi{10.1103/PhysRevD.101.103036}

\bibitem[{{Hanna} {et~al.}(2020){Hanna}, {Caudill}, {Messick}, {Reza},
  {Sachdev}, {Tsukada}, {Cannon}, {Blackburn}, {Creighton}, {Fong}, \&
  et~al.}]{Hanna:2019ezx}
{Hanna}, C., {Caudill}, S., {Messick}, C., {et~al.} 2020, \prd, 101, 022003,
  \dodoi{10.1103/PhysRevD.101.022003}

\bibitem[{{Hansen}(1974)}]{hansen1974multipole}
{Hansen}, R.~O. 1974, JMP, 15, 46, \dodoi{10.1063/1.1666501}

\bibitem[{Harry {et~al.}(2014)}]{harry:2013tca}
Harry, I., {et~al.} 2014, \prd, 89, 024010, \dodoi{10.1103/PhysRevD.89.024010}

\bibitem[{Harry {et~al.}(2009)Harry, Allen, \& Sathyaprakash}]{Harry:2009ea}
Harry, I.~W., Allen, B., \& Sathyaprakash, B.~S. 2009, \prd, 80, 104014,
  \dodoi{10.1103/PhysRevD.80.104014}

\bibitem[{{Hartle}(1967)}]{1967ApJ...150.1005H}
{Hartle}, J.~B. 1967, \apj, 150, 1005, \dodoi{10.1086/149400}

\bibitem[{Healy \& Lousto(2017)}]{Healy:2016lce}
Healy, J., \& Lousto, C.~O. 2017, PhRvD, 95, 024037,
  \dodoi{10.1103/PhysRevD.95.024037}

\bibitem[{{Hoang} {et~al.}(2018){Hoang}, {Naoz}, {Kocsis}, {Rasio}, \&
  {Dosopoulou}}]{hoang2018}
{Hoang}, B.-M., {Naoz}, S., {Kocsis}, B., {Rasio}, F.~A., \& {Dosopoulou}, F.
  2018, \apj, 856, 140, \dodoi{10.3847/1538-4357/aaafce}

\bibitem[{Hofmann {et~al.}(2016)Hofmann, Barausse, \&
  Rezzolla}]{Hofmann:2016yih}
Hofmann, F., Barausse, E., \& Rezzolla, L. 2016, ApJL, 825, L19,
  \dodoi{10.3847/2041-8205/825/2/L19}

\bibitem[{{Holz} \& {Hughes}(2005)}]{2005ApJ...629...15H}
{Holz}, D.~E., \& {Hughes}, S.~A. 2005, \apj, 629, 15, \dodoi{10.1086/431341}

\bibitem[{{Hooper} {et~al.}(2012){Hooper}, {Chung}, {Luan}, {Blair}, {Chen}, \&
  {Wen}}]{Hooper2012}
{Hooper}, S., {Chung}, S.~K., {Luan}, J., {et~al.} 2012, \prd, 86, 024012,
  \dodoi{10.1103/PhysRevD.86.024012}

\bibitem[{Hoy \& Raymond(2020)}]{PEsummary}
Hoy, C., \& Raymond, V. 2020.
\newblock \doarXiv{2006.06639}

\bibitem[{Huang {et~al.}(2020)Huang, Haster, Vitale, Varma, Foucart, \&
  Biscoveanu}]{huang2020statistical}
Huang, Y., Haster, C.-J., Vitale, S., {et~al.} 2020, Statistical and systematic
  uncertainties in extracting the source properties of neutron star - black
  hole binaries with gravitational waves.
\newblock \doarXiv{2005.11850}

\bibitem[{{Hunter}(2007)}]{2007CSE.....9...90H}
{Hunter}, J.~D. 2007, CSE, 9, 90, \dodoi{10.1109/MCSE.2007.55}

\bibitem[{{Husa} {et~al.}(2016){Husa}, {Khan}, {Hannam}, {P{\"u}rrer}, {Ohme},
  {Forteza}, \& {Boh{\'e}}}]{Husa:2015iqa}
{Husa}, S., {Khan}, S., {Hannam}, M., {et~al.} 2016, \prd, 93, 044006,
  \dodoi{10.1103/PhysRevD.93.044006}

\bibitem[{Indik {et~al.}(2018)Indik, Fehrmann, Harke, Krishnan, \&
  Nielsen}]{Indik:2017vqq}
Indik, N., Fehrmann, H., Harke, F., Krishnan, B., \& Nielsen, A.~B. 2018, \prd,
  97, 124008, \dodoi{10.1103/PhysRevD.97.124008}

\bibitem[{Islam {et~al.}(2020)Islam, Mehta, Ghosh, Varma, Ajith, \&
  Sathyaprakash}]{Islam_2020}
Islam, T., Mehta, A.~K., Ghosh, A., {et~al.} 2020, PhRvD, 101,
  \dodoi{10.1103/physrevd.101.024032}

\bibitem[{{Ivanova} {et~al.}(2013){Ivanova}, {Justham}, {Chen}, {De Marco},
  {Fryer}, {Gaburov}, {Ge}, {Glebbeek}, {Han}, {Li}, {Lu}, {Marsh},
  {Podsiadlowski}, {Potter}, {Soker}, {Taam}, {Tauris}, {van den Heuvel}, \&
  {Webbink}}]{ivanova2013}
{Ivanova}, N., {Justham}, S., {Chen}, X., {et~al.} 2013, \aapr, 21, 59,
  \dodoi{10.1007/s00159-013-0059-2}

\bibitem[{Janka(2012)}]{janka2012}
Janka, H.-T. 2012, ARNPS, 62, 407, \dodoi{10.1146/annurev-nucl-102711-094901}

\bibitem[{{Jiang} {et~al.}(2020){Jiang}, {Tang}, {Wang}, {Fan}, \&
  {Wei}}]{JiangTang2019}
{Jiang}, J.-L., {Tang}, S.-P., {Wang}, Y.-Z., {Fan}, Y.-Z., \& {Wei}, D.-M.
  2020, \apj, 892, 55, \dodoi{10.3847/1538-4357}

\bibitem[{Jim\'enez-Forteza {et~al.}(2017)Jim\'enez-Forteza, Keitel, Husa,
  Hannam, Khan, \& P\"urrer}]{Jimenez-Forteza:2016oae}
Jim\'enez-Forteza, X., Keitel, D., Husa, S., {et~al.} 2017, PhRvD, 95, 064024,
  \dodoi{10.1103/PhysRevD.95.064024}

\bibitem[{{Johnson-McDaniel} {et~al.}(2016)}]{spinfit-T1600168}
{Johnson-McDaniel}, N.~K., {et~al.} 2016, Determining the final spin of a
  binary black hole system including in-plane spins: Method and checks of
  accuracy, Tech. Rep. {LIGO}-T1600168, {LIGO} Project.
\newblock \url{https://dcc.ligo.org/T1600168/public}

\bibitem[{{Kalaghatgi} {et~al.}(2020){Kalaghatgi}, {Hannam}, \&
  {Raymond}}]{Kalaghatgi:2019log}
{Kalaghatgi}, C., {Hannam}, M., \& {Raymond}, V. 2020, \prd, 101, 103004,
  \dodoi{10.1103/PhysRevD.101.103004}

\bibitem[{Kapadia {et~al.}(2020)}]{Kapadia:2019uut}
Kapadia, S.~J., {et~al.} 2020, CQGra, 37, 045007,
  \dodoi{10.1088/1361-6382/ab5f2d}

\bibitem[{{Karki} {et~al.}(2016){Karki}, {Tuyenbayev}, {Kandhasamy},
  {et~al.}}]{Karki:2016pht}
{Karki}, S., {Tuyenbayev}, D., {Kandhasamy}, S., {et~al.} 2016, RScI, 87,
  114503, \dodoi{10.1063/1.4967303}

\bibitem[{{Kasen} {et~al.}(2017){Kasen}, {Metzger}, {Barnes}, {Quataert}, \&
  {Ramirez-Ruiz}}]{KasenMetzger2017}
{Kasen}, D., {Metzger}, B., {Barnes}, J., {Quataert}, E., \& {Ramirez-Ruiz}, E.
  2017, Nat, 551, 80, \dodoi{10.1038/nature24453}

\bibitem[{{Kasliwal} {et~al.}(2019){Kasliwal}, {Kasen}, {Lau}, {Perley},
  {Rosswog}, {Ofek}, {Hotokezaka}, {Chary}, {Sollerman}, {Goobar}, \&
  et~al.}]{KasliwalKasen2019}
{Kasliwal}, M.~M., {Kasen}, D., {Lau}, R.~M., {et~al.} 2019, \mnras, L14,
  \dodoi{10.1093/mnrasl/slz007}

\bibitem[{{Kasprzack} \& {Yu}(2017)}]{Kasprzack:2016}
{Kasprzack}, M., \& {Yu}, H. 2017, Beam Position from Angle to Length
  minimization, Tech. Rep. {LIGO}-T1600397, {LIGO} Project.
\newblock \url{https://dcc.ligo.org/T1600397/public}

\bibitem[{Kastha {et~al.}(2018)Kastha, Gupta, Arun, Sathyaprakash, \& Van
  Den~Broeck}]{Kastha_2018}
Kastha, S., Gupta, A., Arun, K., Sathyaprakash, B., \& Van Den~Broeck, C. 2018,
  PhRvD, 98, \dodoi{10.1103/physrevd.98.124033}

\bibitem[{Kastha {et~al.}(2019)Kastha, Gupta, Arun, Sathyaprakash, \& Van
  Den~Broeck}]{PhysRevD.100.044007}
Kastha, S., Gupta, A., Arun, K.~G., Sathyaprakash, B.~S., \& Van Den~Broeck, C.
  2019, PhRvD, 100, 044007, \dodoi{10.1103/PhysRevD.100.044007}

\bibitem[{{Kaup}(1968)}]{Kaup1968}
{Kaup}, D.~J. 1968, PhRv, 172, 1331, \dodoi{10.1103/PhysRev.172.1331}

\bibitem[{{Kawaguchi} {et~al.}(2020){Kawaguchi}, {Shibata}, \&
  {Tanaka}}]{KawaguchiShibata2020}
{Kawaguchi}, K., {Shibata}, M., \& {Tanaka}, M. 2020, \apj, 893, 153,
  \dodoi{10.3847/1538-4357/ab8309}

\bibitem[{Khan {et~al.}(2019)Khan, Chatziioannou, Hannam, \&
  Ohme}]{Khan:2018fmp}
Khan, S., Chatziioannou, K., Hannam, M., \& Ohme, F. 2019, PhRvD, 100, 024059,
  \dodoi{10.1103/PhysRevD.100.024059}

\bibitem[{{Khan} {et~al.}(2016){Khan}, {Husa}, {Hannam}, {Ohme}, {P{\"u}rrer},
  {Forteza}, \& {Boh{\'e}}}]{2016PhRvD..93d4007K}
{Khan}, S., {Husa}, S., {Hannam}, M., {et~al.} 2016, \prd, 93, 044007,
  \dodoi{10.1103/PhysRevD.93.044007}

\bibitem[{{Khan} {et~al.}(2020){Khan}, {Ohme}, {Chatziioannou}, \&
  {Hannam}}]{Khan:2019kotf}
{Khan}, S., {Ohme}, F., {Chatziioannou}, K., \& {Hannam}, M. 2020, \prd, 101,
  024056, \dodoi{10.1103/PhysRevD.101.024056}

\bibitem[{Kidder(2008)}]{Kidder:2007rt}
Kidder, L.~E. 2008, PhRvD, 77, 044016, \dodoi{10.1103/PhysRevD.77.044016}

\bibitem[{{Kim} {et~al.}(2003){Kim}, {Kalogera}, \& {Lorimer}}]{Kim2003}
{Kim}, C., {Kalogera}, V., \& {Lorimer}, D.~R. 2003, \apj, 584, 985,
  \dodoi{10.1086/345740}

\bibitem[{Klimenko {et~al.}(2008)}]{Klimenko:2008fu}
Klimenko, S., {et~al.} 2008, \cqg, 25, 114029

\bibitem[{Klimenko {et~al.}(2016)}]{PhysRevD.93.042004}
---. 2016, \prd, 93, 042004, \dodoi{10.1103/PhysRevD.93.042004}

\bibitem[{{Kochanek}(2014)}]{kochanek2014}
{Kochanek}, C.~S. 2014, \apj, 785, 28, \dodoi{10.1088/0004-637X/785/1/28}

\bibitem[{{Kreidberg} {et~al.}(2012){Kreidberg}, {Bailyn}, {Farr}, \&
  {Kalogera}}]{Kreidberg12massgap}
{Kreidberg}, L., {Bailyn}, C.~D., {Farr}, W.~M., \& {Kalogera}, V. 2012, \apj,
  757, 36, \dodoi{10.1088/0004-637X/757/1/36}

\bibitem[{Krishnendu {et~al.}(2019)Krishnendu, Saleem, Samajdar, Arun,
  Del~Pozzo, \& Mishra}]{PhysRevD.100.104019}
Krishnendu, N.~V., Saleem, M., Samajdar, A., {et~al.} 2019, PhRvD, 100, 104019,
  \dodoi{10.1103/PhysRevD.100.104019}

\bibitem[{Kruckow {et~al.}(2018)Kruckow, Tauris, Langer, Kramer, \&
  Izzard}]{Kruckow:2018slo}
Kruckow, M.~U., Tauris, T.~M., Langer, N., Kramer, M., \& Izzard, R.~G. 2018,
  \mnras, 481, 1908, \dodoi{10.1093/mnras/sty2190}

\bibitem[{Kumar {et~al.}(2017)Kumar, Pürrer, \& Pfeiffer}]{Kumar:2016zlj}
Kumar, P., Pürrer, M., \& Pfeiffer, H.~P. 2017, PhRvD, 95, 044039,
  \dodoi{10.1103/PhysRevD.95.044039}

\bibitem[{Kumar {et~al.}(2019)Kumar, Blackman, Field, Scheel, Galley, Boyle,
  Kidder, Pfeiffer, Szilagyi, \& Teukolsky}]{Kumar:2018hml}
Kumar, P., Blackman, J., Field, S.~E., {et~al.} 2019, PhRvD, 99, 124005,
  \dodoi{10.1103/PhysRevD.99.124005}

\bibitem[{{Landry} {et~al.}(2020){Landry}, {Essick}, \&
  {Chatziioannou}}]{LandryEssick2020}
{Landry}, P., {Essick}, R., \& {Chatziioannou}, K. 2020, \prd, 101, 123007,
  \dodoi{10.1103/PhysRevD.101.123007}

\bibitem[{Li {et~al.}(2012)Li, Del~Pozzo, Vitale, Van Den~Broeck, Agathos,
  Veitch, Grover, Sidery, Sturani, \& Vecchio}]{PhysRevD.85.082003}
Li, T. G.~F., Del~Pozzo, W., Vitale, S., {et~al.} 2012, PhRvD, 85, 082003,
  \dodoi{10.1103/PhysRevD.85.082003}

\bibitem[{{LIGO Scientific Collaboration}(2018)}]{LALSuite}
{LIGO Scientific Collaboration}. 2018, {LIGO Algorithm Library},
  \dodoi{10.7935/GT1W-FZ16}

\bibitem[{{LIGO Scientific Collaboration, Virgo
  Collaboration}(2019{\natexlab{a}})}]{GCNnoticeS190814bv}
{LIGO Scientific Collaboration, Virgo Collaboration}. 2019{\natexlab{a}}, \gcn.
\newblock \url{https://gcn.gsfc.nasa.gov/notices_l/S190814bv.lvc}

\bibitem[{{LIGO Scientific Collaboration, Virgo
  Collaboration}(2019{\natexlab{b}})}]{LIGOEMFOLLOWUSERGUIDE}
---. 2019{\natexlab{b}}, Public Alerts User Guide.
\newblock \url{https://emfollow.docs.ligo.org/userguide/content.html}

\bibitem[{{LIGO Scientific Collaboration, Virgo
  Collaboration}(2019{\natexlab{c}})}]{S190814bv}
---. 2019{\natexlab{c}}, GraceDB, S190814bv.
\newblock \url{https://gracedb.ligo.org/superevents/S190814bv/}

\bibitem[{{LIGO Scientific Collaboration, Virgo
  Collaboration}(2019{\natexlab{d}})}]{GCN25324}
---. 2019{\natexlab{d}}, \gcn, 25324.
\newblock \url{https://gcn.gsfc.nasa.gov/other/GW190814bv.gcn3}

\bibitem[{{LIGO Scientific Collaboration, Virgo
  Collaboration}(2019{\natexlab{e}})}]{GCN25333}
---. 2019{\natexlab{e}}, \gcn, 25333.
\newblock \url{https://gcn.gsfc.nasa.gov/gcn3/25333.gcn3}

\bibitem[{{Lim} \& {Holt}(2019)}]{LimHolt2019}
{Lim}, Y., \& {Holt}, J.~W. 2019, EPJA, 55, 209,
  \dodoi{10.1140/epja/i2019-12917-9}

\bibitem[{Lindblom(2010)}]{Lindblom:2010bb}
Lindblom, L. 2010, \prd, 82, 103011, \dodoi{10.1103/PhysRevD.82.103011}

\bibitem[{{Lindblom} \& {Indik}(2012)}]{LindblomIndik2012}
{Lindblom}, L., \& {Indik}, N.~M. 2012, \prd, 86, 084003,
  \dodoi{10.1103/PhysRevD.86.084003}

\bibitem[{{Lindblom} \& {Indik}(2014)}]{LindblomIndik2014}
---. 2014, \prd, 89, 064003, \dodoi{10.1103/PhysRevD.89.064003}

\bibitem[{{Lipunov} {et~al.}(2019){Lipunov}, {Gorbovskoy}, {Kornilov},
  {Tyurina}, {Balanutsa}, {Kuznetsov}, {Balakin}, {Vladimirov}, {Vlasenko},
  {Gorbunov}, \& et~al.}]{LipunovGorbovskoy2019}
{Lipunov}, V., {Gorbovskoy}, E., {Kornilov}, V., {et~al.} 2019, GCN, 25354.
\newblock \url{https://gcn.gsfc.nasa.gov/gcn/gcn3/25354.gcn3}

\bibitem[{Littenberg \& Cornish(2015)}]{PhysRevD.91.084034}
Littenberg, T.~B., \& Cornish, N.~J. 2015, \prd, 91, 084034,
  \dodoi{10.1103/PhysRevD.91.084034}

\bibitem[{{Liu} {et~al.}(2012){Liu}, {Du}, {Chung}, {Hooper}, {Blair}, \&
  {Wen}}]{Liu2012}
{Liu}, Y., {Du}, Z., {Chung}, S.~K., {et~al.} 2012, Classical and Quantum
  Gravity, 29, 235018, \dodoi{10.1088/0264-9381/29/23/235018}

\bibitem[{London {et~al.}(2018)London, Khan, Fauchon-Jones, García, Hannam,
  Husa, Jiménez-Forteza, Kalaghatgi, Ohme, \& Pannarale}]{London:2017bcn}
London, L., Khan, S., Fauchon-Jones, E., {et~al.} 2018, PhRvL., 120, 161102,
  \dodoi{10.1103/PhysRevLett.120.161102}

\bibitem[{{Mapelli} \& {Giacobbo}(2018)}]{mapelli2018}
{Mapelli}, M., \& {Giacobbo}, N. 2018, \mnras, 479, 4391,
  \dodoi{10.1093/mnras/sty1613}

\bibitem[{{Mapelli} {et~al.}(2019){Mapelli}, {Giacobbo}, {Santoliquido}, \&
  {Artale}}]{mapelli2019}
{Mapelli}, M., {Giacobbo}, N., {Santoliquido}, F., \& {Artale}, M.~C. 2019,
  \mnras, 487, 2, \dodoi{10.1093/mnras/stz1150}

\bibitem[{{Marchant} {et~al.}(2017){Marchant}, {Langer}, {Podsiadlowski},
  {Tauris}, {de Mink}, {Mandel}, \& {Moriya}}]{marchant2017}
{Marchant}, P., {Langer}, N., {Podsiadlowski}, P., {et~al.} 2017, \aap, 604,
  A55, \dodoi{10.1051/0004-6361/201630188}

\bibitem[{{Margalit} \& {Metzger}(2017)}]{MargalitMetzger2017}
{Margalit}, B., \& {Metzger}, B.~D. 2017, \apjl, 850, L19,
  \dodoi{10.3847/2041-8213/aa991c}

\bibitem[{Matas {et~al.}(2020)}]{SEOBNRNSBH}
Matas, A., {et~al.} 2020,
  arXiv:\href{https://arxiv.org/abs/2004.10001}{2004.10001}

\bibitem[{Mazur \& Mottola(2004)}]{Mazur:2004fk}
Mazur, P.~O., \& Mottola, E. 2004, PNAS, 101, 9545,
  \dodoi{10.1073/pnas.0402717101}

\bibitem[{{McKernan} {et~al.}(2020){McKernan}, {Ford}, \&
  {O'Shaughnessy}}]{mckernan2020}
{McKernan}, B., {Ford}, K.~E.~S., \& {O'Shaughnessy}, R. 2020,
  arXiv:\href{https://arxiv.org/abs/2002.00046}{2002.00046}

\bibitem[{Meidam {et~al.}(2018)Meidam, Tsang, Goldstein, Agathos, Ghosh,
  Haster, Raymond, Samajdar, Schmidt, Smith, Blackburn, Del~Pozzo, Field, Li,
  P\"urrer, Van Den~Broeck, Veitch, \& Vitale}]{PhysRevD.97.044033}
Meidam, J., Tsang, K.~W., Goldstein, J., {et~al.} 2018, PhRvD, 97, 044033,
  \dodoi{10.1103/PhysRevD.97.044033}

\bibitem[{{Messenger} \& {Read}(2012)}]{2012PhRvL.108i1101M}
{Messenger}, C., \& {Read}, J. 2012, \prl, 108, 091101,
  \dodoi{10.1103/PhysRevLett.108.091101}

\bibitem[{{Messick} {et~al.}(2017){Messick}, {Blackburn}, {Brady},
  {et~al.}}]{Messick:2016aqy}
{Messick}, C., {Blackburn}, K., {Brady}, P., {et~al.} 2017, \prd, 95, 042001,
  \dodoi{10.1103/PhysRevD.95.042001}

\bibitem[{{Miller} {et~al.}(2020){Miller}, {Chirenti}, \&
  {Lamb}}]{Miller:2019nzo}
{Miller}, M.~C., {Chirenti}, C., \& {Lamb}, F.~K. 2020, \apj, 888, 12,
  \dodoi{10.3847/1538-4357/ab4ef9}

\bibitem[{{Miller} {et~al.}(2019){Miller}, {Lamb}, {Dittmann}, {Bogdanov},
  {Arzoumanian}, {Gendreau}, {Guillot}, {Harding}, {Ho}, {Lattimer}, \&
  et~al.}]{MillerLamb2019}
{Miller}, M.~C., {Lamb}, F.~K., {Dittmann}, A.~J., {et~al.} 2019, \apjl, 887,
  L24, \dodoi{10.3847/2041-8213/ab50c5}

\bibitem[{Mills \& Fairhurst(2020)}]{Mills:2020}
Mills, J.~C., \& Fairhurst, S. 2020, Measuring gravitational-wave subdominant
  multipoles, Tech. Rep. LIGO-P2000136.
\newblock \url{https://https://dcc.ligo.org/P2000136/public}

\bibitem[{Mishra {et~al.}(2010)Mishra, Arun, Iyer, \&
  Sathyaprakash}]{PhysRevD.82.064010}
Mishra, C.~K., Arun, K.~G., Iyer, B.~R., \& Sathyaprakash, B.~S. 2010, PhRvD,
  82, 064010, \dodoi{10.1103/PhysRevD.82.064010}

\bibitem[{Mishra {et~al.}(2016)Mishra, Kela, Arun, \& Faye}]{Mishra:2016whh}
Mishra, C.~K., Kela, A., Arun, K.~G., \& Faye, G. 2016, PhRvD, 93, 084054,
  \dodoi{10.1103/PhysRevD.93.084054}

\bibitem[{Morgan {et~al.}(2020)Morgan, Soares-Santos, Annis, Herner, Garcia,
  Palmese, Drlica-Wagner, Kessler, Garcia-Bellido, Sherman, Allam, Bechtol,
  Bom, Brout, Butler, Butner, Cartier, Chen, Conselice, Cook, Davis, Doctor,
  Farr, Figueiredo, Finley, Foley, Galarza, Gill, Gruendl, Holz, Kuropatkin,
  Lidman, Lin, Malik, Mann, Marriner, Marshall, Martinez-Vazquez, Meza,
  Neilsen, Nicolaou, E., Paz-Chinchon, Points, Quirola, Rodriguez, Sako,
  Scolnic, Smith, Sobreira, Tucker, Vivas, Wiesner, Wood, Yanny, Zenteno,
  Abbott, Aguena, Avila, Bertin, Bhargava, Brooks, Burke, Rosell, Kind,
  Carretero, da~Costa, Costanzi, Vicente, Desai, Diehl, Doel, Eifler, Everett,
  Flaugher, Frieman, Gaztanaga, Gerdes, Gruen, Gschwend, Gutierrez, Hartley,
  Hinton, Hollowood, Honscheid, James, Kuehn, Lahav, Lima, Maia, March, Miquel,
  Ogando, Plazas, Roodman, Sanchez, Scarpine, Schubnell, Serrano,
  Sevilla-Noarbe, Suchyta, \& Tarle}]{morgan2020constraints}
Morgan, R., Soares-Santos, M., Annis, J., {et~al.} 2020, Constraints on the
  Physical Properties of S190814bv through Simulations based on DECam Follow-up
  Observations by the Dark Energy Survey.
\newblock \doarXiv{2006.07385}

\bibitem[{{M{\"u}ller}(2016)}]{muller2016}
{M{\"u}ller}, B. 2016, \pasa, 33, e048, \dodoi{10.1017/pasa.2016.40}

\bibitem[{{M{\"u}ller} \& {Serot}(1996)}]{MullerSerot1996}
{M{\"u}ller}, H., \& {Serot}, B.~D. 1996, \nphysa, 606, 508,
  \dodoi{10.1016/0375-9474(96)00187-X}

\bibitem[{Nair {et~al.}(2018)Nair, Bose, \& Saini}]{Nair:2018ign}
Nair, R., Bose, S., \& Saini, T.~D. 2018, PhRvD, 98, 023502,
  \dodoi{10.1103/PhysRevD.98.023502}

\bibitem[{{Neijssel} {et~al.}(2019){Neijssel}, {Vigna-G{\'o}mez}, {Stevenson},
  {Barrett}, {Gaebel}, {Broekgaarden}, {de Mink}, {Sz{\'e}csi}, {Vinciguerra},
  \& {Mandel}}]{neijssel2019}
{Neijssel}, C.~J., {Vigna-G{\'o}mez}, A., {Stevenson}, S., {et~al.} 2019,
  \mnras, 490, 3740, \dodoi{10.1093/mnras/stz2840}

\bibitem[{{Neustroev} {et~al.}(2014){Neustroev}, {Veledina}, {Poutanen},
  {Zharikov}, {Tsygankov}, {Sjoberg}, \& {Kajava}}]{neustroev2014}
{Neustroev}, V.~V., {Veledina}, A., {Poutanen}, J., {et~al.} 2014, \mnras, 445,
  2424, \dodoi{10.1093/mnras/stu1924}

\bibitem[{Ng {et~al.}(2018)Ng, Vitale, Zimmerman, Chatziioannou, Gerosa, \&
  Haster}]{Ng:2018neg}
Ng, K. K.~Y., Vitale, S., Zimmerman, A., {et~al.} 2018, PhRvD, 98, 083007,
  \dodoi{10.1103/PhysRevD.98.083007}

\bibitem[{Nitz {et~al.}(2019)Nitz, Harry, Brown, Biwer, Willis, Canton, Capano,
  Pekowsky, Dent, Williamson, {et~al.}}]{pycbc-software}
Nitz, A., Harry, I., Brown, D., {et~al.} 2019, gwastro/pycbc: PyCBC Release
  v1.15.2, \dodoi{10.5281/zenodo.3596447}

\bibitem[{Nitz {et~al.}(2018)Nitz, Dal~Canton, Davis, \& Reyes}]{Nitz:2018rgo}
Nitz, A.~H., Dal~Canton, T., Davis, D., \& Reyes, S. 2018, \prd, 98, 024050,
  \dodoi{10.1103/PhysRevD.98.024050}

\bibitem[{{Nitz} {et~al.}(2017){Nitz}, {Dent}, {Dal Canton}, {Fairhurst}, \&
  {Brown}}]{Nitz:2017svb}
{Nitz}, A.~H., {Dent}, T., {Dal Canton}, T., {Fairhurst}, S., \& {Brown}, D.~A.
  2017, \apj, 849, 118, \dodoi{10.3847/1538-4357/aa8f50}

\bibitem[{{Nitz} {et~al.}(2020){Nitz}, {Dent}, {Davies}, {Kumar}, {Capano},
  {Harry}, {Mozzon}, {Nuttall}, {Lundgren}, \& {T{\'a}pai}}]{Nitz:2019hdf}
{Nitz}, A.~H., {Dent}, T., {Davies}, G.~S., {et~al.} 2020, \apj, 891, 123,
  \dodoi{10.3847/1538-4357/ab733f}

\bibitem[{{Nuttall}(2018)}]{2018RSPTA.37670286N}
{Nuttall}, L.~K. 2018, RSPTA, 376, 20170286, \dodoi{10.1098/rsta.2017.0286}

\bibitem[{{Olejak} {et~al.}(2020){Olejak}, {Belczynski}, {Holz}, {Lasota},
  {Bulik}, \& {Miller}}]{OlejakBelczynski2020}
{Olejak}, A., {Belczynski}, K., {Holz}, D.~E., {et~al.} 2020,
  arXiv:\href{https://arxiv.org/abs/2004.11866}{2004.11866}

\bibitem[{Ossokine {et~al.}(2020)}]{Ossokine:2020}
Ossokine, S., {et~al.} 2020,
  arXiv:\href{https://arxiv.org/abs/2004.09442}{2004.09442}

\bibitem[{Owen(1996)}]{owen:1995tm}
Owen, B.~J. 1996, \prd, 53, 6749, \dodoi{10.1103/PhysRevD.53.6749}

\bibitem[{Owen \& Sathyaprakash(1999)}]{Owen:1998dk}
Owen, B.~J., \& Sathyaprakash, B.~S. 1999, \prd, 60, 022002,
  \dodoi{10.1103/PhysRevD.60.022002}

\bibitem[{{{\"O}zel} {et~al.}(2010){{\"O}zel}, {Psaltis}, {Narayan}, \&
  {McClintock}}]{2010ApJ...725.1918O}
{{\"O}zel}, F., {Psaltis}, D., {Narayan}, R., \& {McClintock}, J.~E. 2010,
  \apj, 725, 1918, \dodoi{10.1088/0004-637X/725/2/1918}

\bibitem[{{\"{O}zel} {et~al.}(2012){\"{O}zel}, {Psaltis}, {Narayan}, \&
  {Villarreal}}]{Ozel:2012ax}
{\"{O}zel}, F., {Psaltis}, D., {Narayan}, R., \& {Villarreal}, A.~S. 2012,
  \apj, 757, 55, \dodoi{10.1088/0004-637X/757/1/55}

\bibitem[{{Pappas} \& {Apostolatos}(2012)}]{2012PhRvL.108w1104P}
{Pappas}, G., \& {Apostolatos}, T.~A. 2012, \prl, 108, 231104,
  \dodoi{10.1103/PhysRevLett.108.231104}

\bibitem[{{Park} {et~al.}(2017){Park}, {Kim}, {Lee}, {Bae}, \&
  {Belczynski}}]{park2017}
{Park}, D., {Kim}, C., {Lee}, H.~M., {Bae}, Y.-B., \& {Belczynski}, K. 2017,
  \mnras, 469, 4665, \dodoi{10.1093/mnras/stx1015}

\bibitem[{Payne {et~al.}(2019)Payne, Talbot, \& Thrane}]{Payne:2019wmy}
Payne, E., Talbot, C., \& Thrane, E. 2019, PhRvD, 100, 123017,
  \dodoi{10.1103/PhysRevD.100.123017}

\bibitem[{{Petrovich} \& {Antonini}(2017)}]{petrovich2017}
{Petrovich}, C., \& {Antonini}, F. 2017, \apj, 846, 146,
  \dodoi{10.3847/1538-4357/aa8628}

\bibitem[{{Poisson}(1998)}]{poisson1998gravitational}
{Poisson}, E. 1998, \prd, 57, 5287, \dodoi{10.1103/PhysRevD.57.5287}

\bibitem[{Poisson \& Will(1995)}]{Poisson:1995ef}
Poisson, E., \& Will, C.~M. 1995, PhRvD, 52, 848,
  \dodoi{10.1103/PhysRevD.52.848}

\bibitem[{{Portegies Zwart} \& {McMillan}(2000)}]{portegieszwart2000}
{Portegies Zwart}, S.~F., \& {McMillan}, S. L.~W. 2000, \apjl, 528, L17,
  \dodoi{10.1086/312422}

\bibitem[{{Postnov} \& {Yungelson}(2014)}]{postnov2014}
{Postnov}, K.~A., \& {Yungelson}, L.~R. 2014, LRR, 17, 3,
  \dodoi{10.12942/lrr-2014-3}

\bibitem[{{Price-Whelan} {et~al.}(2018){Price-Whelan}, {Sip{\H{o}}cz},
  {G{\"u}nther}, {Lim}, {Crawford}, {Conseil}, {Shupe}, {Craig}, {Dencheva},
  {Ginsburg}, {VanderPlas}, {Bradley}, {P{\'e}rez-Su{\'a}rez}, {de Val-Borro},
  {Paper Contributors}, {Aldcroft}, {Cruz}, {Robitaille}, {Tollerud},
  {Coordination Committee}, {Ardelean}, {Babej}, {Bach}, {Bachetti}, {Bakanov},
  {Bamford}, {Barentsen}, {Barmby}, {Baumbach}, {Berry}, {Biscani}, {Boquien},
  {Bostroem}, {Bouma}, {Brammer}, {Bray}, {Breytenbach}, {Buddelmeijer},
  {Burke}, {Calderone}, {Cano Rodr{\'\i}guez}, {Cara}, {Cardoso}, {Cheedella},
  {Copin}, {Corrales}, {Crichton}, {D{\textquoteright}Avella}, {Deil},
  {Depagne}, {Dietrich}, {Donath}, {Droettboom}, {Earl}, {Erben}, {Fabbro},
  {Ferreira}, {Finethy}, {Fox}, {Garrison}, {Gibbons}, {Goldstein}, {Gommers},
  {Greco}, {Greenfield}, {Groener}, {Grollier}, {Hagen}, {Hirst}, {Homeier},
  {Horton}, {Hosseinzadeh}, {Hu}, {Hunkeler}, {Ivezi{\'c}}, {Jain}, {Jenness},
  {Kanarek}, {Kendrew}, {Kern}, {Kerzendorf}, {Khvalko}, {King}, {Kirkby},
  {Kulkarni}, {Kumar}, {Lee}, {Lenz}, {Littlefair}, {Ma}, {Macleod},
  {Mastropietro}, {McCully}, {Montagnac}, {Morris}, {Mueller}, {Mumford},
  {Muna}, {Murphy}, {Nelson}, {Nguyen}, {Ninan}, {N{\"o}the}, {Ogaz}, {Oh},
  {Parejko}, {Parley}, {Pascual}, {Patil}, {Patil}, {Plunkett}, {Prochaska},
  {Rastogi}, {Reddy Janga}, {Sabater}, {Sakurikar}, {Seifert}, {Sherbert},
  {Sherwood-Taylor}, {Shih}, {Sick}, {Silbiger}, {Singanamalla}, {Singer},
  {Sladen}, {Sooley}, {Sornarajah}, {Streicher}, {Teuben}, {Thomas},
  {Tremblay}, {Turner}, {Terr{\'o}n}, {van Kerkwijk}, {de la Vega}, {Watkins},
  {Weaver}, {Whitmore}, {Woillez}, {Zabalza}, \& {Contributors}}]{astropy:2018}
{Price-Whelan}, A.~M., {Sip{\H{o}}cz}, B.~M., {G{\"u}nther}, H.~M., {et~al.}
  2018, \aj, 156, 123, \dodoi{10.3847/1538-3881/aabc4f}

\bibitem[{{Privitera} {et~al.}(2014){Privitera}, {Mohapatra}, {Ajith},
  {et~al.}}]{Privitera:2013xza}
{Privitera}, S., {Mohapatra}, S.~R.~P., {Ajith}, P., {et~al.} 2014, \prd, 89,
  024003, \dodoi{10.1103/PhysRevD.89.024003}

\bibitem[{{Prodi} {et~al.}(2020){Prodi}, {Vedovato}, {Drago}, {Klimenko},
  {Lazzaro}, {Miani}, {Milotti}, {Salemi}, \& {Tiwari}}]{Prodi:2020}
{Prodi}, G., {Vedovato}, G., {Drago}, M., {et~al.} 2020, {Technical note on the
  measurement of inspiral higher order modes by coherentWaveBurst in GW190814},
  Tech. Rep. LIGO-T2000124, {LIGO} Project.
\newblock \url{https://dcc.ligo.org/T2000124/public}

\bibitem[{Pürrer(2016)}]{Purrer:2015tud}
Pürrer, M. 2016, PhRvD, 93, 064041, \dodoi{10.1103/PhysRevD.93.064041}

\bibitem[{{Raaijmakers} {et~al.}(2019){Raaijmakers}, {Riley}, {Watts}, {Greif},
  {Morsink}, {Hebeler}, {Schwenk}, {Hinderer}, {Nissanke}, {Guillot}, \&
  et~al.}]{RaaijmakersRiley2019}
{Raaijmakers}, G., {Riley}, T.~E., {Watts}, A.~L., {et~al.} 2019, \apjl, 887,
  L22, \dodoi{10.3847/2041-8213/ab451a}

\bibitem[{Racine(2008)}]{Racine:2008qv}
Racine, E. 2008, \prd, 78, 044021, \dodoi{10.1103/PhysRevD.78.044021}

\bibitem[{{Rastello} {et~al.}(2020){Rastello}, {Mapelli}, {Di Carlo},
  {Giacobbo}, {Santoliquido}, {Spera}, \& {Ballone}}]{rastello2020}
{Rastello}, S., {Mapelli}, M., {Di Carlo}, U.~N., {et~al.} 2020,
  arXiv:\href{https://arxiv.org/abs/2003.02277}{2003.02277}

\bibitem[{{Rezzolla} {et~al.}(2018){Rezzolla}, {Most}, \&
  {Weih}}]{RezzollaMost2018}
{Rezzolla}, L., {Most}, E.~R., \& {Weih}, L.~R. 2018, \apjl, 852, L25,
  \dodoi{10.3847/2041-8213/aaa401}

\bibitem[{{Riley} {et~al.}(2019){Riley}, {Watts}, {Bogdanov}, {Ray}, {Ludlam},
  {Guillot}, {Arzoumanian}, {Baker}, {Bilous}, {Chakrabarty}, \&
  et~al.}]{RileyWatts2019}
{Riley}, T.~E., {Watts}, A.~L., {Bogdanov}, S., {et~al.} 2019, \apjl, 887, L21,
  \dodoi{10.3847/2041-8213/ab481c}

\bibitem[{{Rodriguez} {et~al.}(2016){Rodriguez}, {Chatterjee}, \&
  {Rasio}}]{rodriguez2016}
{Rodriguez}, C.~L., {Chatterjee}, S., \& {Rasio}, F.~A. 2016, \prd, 93, 084029,
  \dodoi{10.1103/PhysRevD.93.084029}

\bibitem[{{Rodriguez} {et~al.}(2019){Rodriguez}, {Zevin}, {Amaro-Seoane},
  {Chatterjee}, {Kremer}, {Rasio}, \& {Ye}}]{rodriguez2019}
{Rodriguez}, C.~L., {Zevin}, M., {Amaro-Seoane}, P., {et~al.} 2019, \prd, 100,
  043027, \dodoi{10.1103/PhysRevD.100.043027}

\bibitem[{{Rosswog} {et~al.}(2018){Rosswog}, {Sollerman}, {Feindt}, {Goobar},
  {Korobkin}, {Wollaeger}, {Fremling}, \& {Kasliwal}}]{RosswogSollerman2018}
{Rosswog}, S., {Sollerman}, J., {Feindt}, U., {et~al.} 2018, \aap, 615, A132,
  \dodoi{10.1051/0004-6361/201732117}

\bibitem[{{Roulet} \& {Zaldarriaga}(2019)}]{RouletZaldarriaga2019}
{Roulet}, J., \& {Zaldarriaga}, M. 2019, \mnras, 484, 4216,
  \dodoi{10.1093/mnras/stz226}

\bibitem[{Roy {et~al.}(2019)Roy, Sengupta, \& Ajith}]{PhysRevD.99.024048}
Roy, S., Sengupta, A.~S., \& Ajith, P. 2019, \prd, 99, 024048,
  \dodoi{10.1103/PhysRevD.99.024048}

\bibitem[{{Roy} {et~al.}(2019){Roy}, {Sengupta}, \& {Arun}}]{roy2019unveiling}
{Roy}, S., {Sengupta}, A.~S., \& {Arun}, K.~G. 2019,
  arXiv:\href{https://arxiv.org/abs/1910.04565}{1910.04565}

\bibitem[{Roy {et~al.}(2017)Roy, Sengupta, \& Thakor}]{PhysRevD.95.104045}
Roy, S., Sengupta, A.~S., \& Thakor, N. 2017, \prd, 95, 104045,
  \dodoi{10.1103/PhysRevD.95.104045}

\bibitem[{{Ruiz} {et~al.}(2018){Ruiz}, {Shapiro}, \&
  {Tsokaros}}]{RuizShapiro2018}
{Ruiz}, M., {Shapiro}, S.~L., \& {Tsokaros}, A. 2018, \prd, 97, 021501,
  \dodoi{10.1103/PhysRevD.97.021501}

\bibitem[{{Ryan}(1997)}]{ryan1997spinning}
{Ryan}, F.~D. 1997, \prd, 55, 6081, \dodoi{10.1103/PhysRevD.55.6081}

\bibitem[{{Sachdev} {et~al.}(2019){Sachdev}, {Caudill}, {Fong}, {Lo},
  {Messick}, {et~al.}}]{Sachdev:2019vvd}
{Sachdev}, S., {Caudill}, S., {Fong}, H., {et~al.} 2019,
  arXiv:\href{https://arxiv.org/abs/1901.08580}{1901.08580}

\bibitem[{{Safarzadeh} {et~al.}(2020){Safarzadeh}, {Hamers}, {Loeb}, \&
  {Berger}}]{SafarzadehHamers2020}
{Safarzadeh}, M., {Hamers}, A.~S., {Loeb}, A., \& {Berger}, E. 2020, \apjl,
  888, L3, \dodoi{10.3847/2041-8213/ab5dc8}

\bibitem[{{Santamar{\'\i}a} {et~al.}(2010){Santamar{\'\i}a}, {Ohme}, {Ajith},
  {Br{\"u}gmann}, {Dorband}, {Hannam}, {Husa}, {M{\"o}sta}, {Pollney},
  {Reisswig}, \& et~al.}]{santamaria2010matching}
{Santamar{\'\i}a}, L., {Ohme}, F., {Ajith}, P., {et~al.} 2010, \prd, 82,
  064016, \dodoi{10.1103/PhysRevD.82.064016}

\bibitem[{Sathyaprakash \& Dhurandhar(1991)}]{Sathyaprakash:1991}
Sathyaprakash, B., \& Dhurandhar, S. 1991, \prd, 44, 3819,
  \dodoi{10.1103/PhysRevD.44.3819}

\bibitem[{{Schmidt} {et~al.}(2015){Schmidt}, {Ohme}, \&
  {Hannam}}]{2015PhRvD..91b4043S}
{Schmidt}, P., {Ohme}, F., \& {Hannam}, M. 2015, \prd, 91, 024043,
  \dodoi{10.1103/PhysRevD.91.024043}

\bibitem[{Schutz(1986)}]{Schutz:1986gp}
Schutz, B.~F. 1986, Nat, 323, 310, \dodoi{10.1038/323310a0}

\bibitem[{{Shibata} {et~al.}(2019){Shibata}, {Zhou}, {Kiuchi}, \&
  {Fujibayashi}}]{ShibataZhou2019}
{Shibata}, M., {Zhou}, E., {Kiuchi}, K., \& {Fujibayashi}, S. 2019, \prd, 100,
  023015, \dodoi{10.1103/PhysRevD.100.023015}

\bibitem[{{Sigurdsson} \& {Hernquist}(1993)}]{sigurdsson1993}
{Sigurdsson}, S., \& {Hernquist}, L. 1993, Nat, 364, 423,
  \dodoi{10.1038/364423a0}

\bibitem[{{Silsbee} \& {Tremaine}(2017)}]{silsbee2017}
{Silsbee}, K., \& {Tremaine}, S. 2017, \apj, 836, 39,
  \dodoi{10.3847/1538-4357/aa5729}

\bibitem[{Singer \& Price(2016)}]{BAYESTAR}
Singer, L.~P., \& Price, L. 2016, \prd, 93, 024013,
  \dodoi{10.1103/PhysRevD.93.024013}

\bibitem[{{Smith} \& {Ashton}(2019)}]{Smith:2019ucc}
{Smith}, R., \& {Ashton}, G. 2019,
  arXiv:\href{https://arxiv.org/abs/1909.11873}{1909.11873}

\bibitem[{{Soares-Santos} {et~al.}(2019){Soares-Santos}, {Palmese}, {Hartley},
  {et~al.}}]{2019ApJ...876L...7S}
{Soares-Santos}, M., {Palmese}, A., {Hartley}, W., {et~al.} 2019, \apjl, 876,
  L7, \dodoi{10.3847/2041-8213/ab14f1}

\bibitem[{{Speagle}(2020)}]{2019arXiv190402180S}
{Speagle}, J.~S. 2020, {\mnras}, 493, 3132–3158,
  \dodoi{10.1093/mnras/staa278}

\bibitem[{{Spera} {et~al.}(2019){Spera}, {Mapelli}, {Giacobbo}, {Trani},
  {Bressan}, \& {Costa}}]{spera2019}
{Spera}, M., {Mapelli}, M., {Giacobbo}, N., {et~al.} 2019, \mnras, 485, 889,
  \dodoi{10.1093/mnras/stz359}

\bibitem[{{Stephan} {et~al.}(2019){Stephan}, {Naoz}, {Ghez}, {Morris},
  {Ciurlo}, {Do}, {Breivik}, {Coughlin}, \& {Rodriguez}}]{stephan2019}
{Stephan}, A.~P., {Naoz}, S., {Ghez}, A.~M., {et~al.} 2019, \apj, 878, 58,
  \dodoi{10.3847/1538-4357/ab1e4d}

\bibitem[{{Sukhbold} \& {Woosley}(2014)}]{sukhbold2014}
{Sukhbold}, T., \& {Woosley}, S.~E. 2014, \apj, 783, 10,
  \dodoi{10.1088/0004-637X/783/1/10}

\bibitem[{{Tanvir} {et~al.}(2017){Tanvir}, {Levan},
  {Gonz{\'a}lez-Fern{\'a}ndez}, {Korobkin}, {Mandel}, {Rosswog}, {Hjorth},
  {D'Avanzo}, {Fruchter}, {Fryer}, \& et~al.}]{TanvirLevan2017}
{Tanvir}, N.~R., {Levan}, A.~J., {Gonz{\'a}lez-Fern{\'a}ndez}, C., {et~al.}
  2017, \apjl, 848, L27, \dodoi{10.3847/2041-8213/aa90b6}

\bibitem[{{Taylor} \& {Gair}(2012)}]{2012PhRvD..86b3502T}
{Taylor}, S.~R., \& {Gair}, J.~R. 2012, \prd, 86, 023502,
  \dodoi{10.1103/PhysRevD.86.023502}

\bibitem[{{Taylor} {et~al.}(2012){Taylor}, {Gair}, \&
  {Mandel}}]{2012PhRvD..85b3535T}
{Taylor}, S.~R., {Gair}, J.~R., \& {Mandel}, I. 2012, \prd, 85, 023535,
  \dodoi{10.1103/PhysRevD.85.023535}

\bibitem[{{The IceCube Collaboration}(2019)}]{GCN25557}
{The IceCube Collaboration}. 2019, \gcn, 25557.
\newblock \url{https://gcn.gsfc.nasa.gov/gcn3/25557.gcn3}

\bibitem[{Thompson {et~al.}(2020)Thompson, Fauchon-Jones, Khan, Nitoglia,
  Pannarale, Dietrich, \& Hannam}]{PhenomNSBH}
Thompson, J.~E., Fauchon-Jones, E., Khan, S., {et~al.} 2020,
  arXiv:\href{https://arxiv.org/abs/2002.08383}{2002.08383}

\bibitem[{Thompson {et~al.}(2019)Thompson, Kochanek, Stanek, Badenes, Post,
  Jayasinghe, Latham, Bieryla, Esquerdo, Berlind, Calkins, Tayar, Lindegren,
  Johnson, Holoien, Auchettl, \& Covey}]{Thompson637}
Thompson, T.~A., Kochanek, C.~S., Stanek, K.~Z., {et~al.} 2019, Sci, 366, 637,
  \dodoi{10.1126/science.aau4005}

\bibitem[{{Thompson} {et~al.}(2020){Thompson}, {Kochanek}, {Stanek}, {Badenes},
  {Jayasinghe}, {Tayar}, {Johnson}, {Holoien}, {Auchettl}, \&
  {Covey}}]{ThompsonKochanek2020}
{Thompson}, T.~A., {Kochanek}, C.~S., {Stanek}, K.~Z., {et~al.} 2020, Sci, 368,
  eaba4356, \dodoi{10.1126/science.aba4356}

\bibitem[{{Thorne}(1980)}]{1980RvMP...52..299T}
{Thorne}, K.~S. 1980, RvMP, 52, 299, \dodoi{10.1103/RevModPhys.52.299}

\bibitem[{{Tiwari}(2018)}]{tiwari2018estimation}
{Tiwari}, V. 2018, CQGra, 35, 145009, \dodoi{10.1088/1361-6382/aac89d}

\bibitem[{{Uchikata} {et~al.}(2016){Uchikata}, {Yoshida}, \&
  {Pani}}]{2016PhRvD..94f4015U}
{Uchikata}, N., {Yoshida}, S., \& {Pani}, P. 2016, \prd, 94, 064015,
  \dodoi{10.1103/PhysRevD.94.064015}

\bibitem[{{Ugliano} {et~al.}(2012){Ugliano}, {Janka}, {Marek}, \&
  {Arcones}}]{ugliano2012}
{Ugliano}, M., {Janka}, H.-T., {Marek}, A., \& {Arcones}, A. 2012, \apj, 757,
  69, \dodoi{10.1088/0004-637X/757/1/69}

\bibitem[{Usman {et~al.}(2019)Usman, Mills, \& Fairhurst}]{Usman:2018imj}
Usman, S.~A., Mills, J.~C., \& Fairhurst, S. 2019, ApJ, 877, 82,
  \dodoi{10.3847/1538-4357/ab0b3e}

\bibitem[{{Usman} {et~al.}(2016){Usman}, {Nitz}, {Harry}, {Biwer}, {Brown},
  {et~al.}}]{Usman:2015kfa}
{Usman}, S.~A., {Nitz}, A.~H., {Harry}, I.~W., {et~al.} 2016, \cqg, 33, 215004,
  \dodoi{10.1088/0264-9381/33/21/215004}

\bibitem[{{Van Den Broeck} \&
  {Sengupta}(2007{\natexlab{a}})}]{van2006phenomenology}
{Van Den Broeck}, C., \& {Sengupta}, A.~S. 2007{\natexlab{a}}, CQGra, 24, 155,
  \dodoi{10.1088/0264-9381/24/1/009}

\bibitem[{{Van Den Broeck} \& {Sengupta}(2007{\natexlab{b}})}]{van2007binary}
---. 2007{\natexlab{b}}, CQGra, 24, 1089, \dodoi{10.1088/0264-9381/24/5/005}

\bibitem[{{van den Heuvel} \& {Tauris}(2020)}]{vandenHeuvelTauris2020}
{van den Heuvel}, E. P.~J., \& {Tauris}, T.~M. 2020, Sci, 368, eaba3282,
  \dodoi{10.1126/science.aba3282}

\bibitem[{{Veitch} {et~al.}(2015){Veitch}, {Raymond}, {Farr},
  {et~al.}}]{PhysRevD.91.042003}
{Veitch}, J., {Raymond}, V., {Farr}, B., {et~al.} 2015, \prd, 91, 042003,
  \dodoi{10.1103/PhysRevD.91.042003}

\bibitem[{{Venumadhav} {et~al.}(2020){Venumadhav}, {Zackay}, {Roulet}, {Dai},
  \& {Zaldarriaga}}]{Venumadhav:2019lyq}
{Venumadhav}, T., {Zackay}, B., {Roulet}, J., {Dai}, L., \& {Zaldarriaga}, M.
  2020, \prd, 101, 083030, \dodoi{10.1103/PhysRevD.101.083030}

\bibitem[{{Verde} {et~al.}(2019){Verde}, {Treu}, \&
  {Riess}}]{2019NatAs...3..891V}
{Verde}, L., {Treu}, T., \& {Riess}, A.~G. 2019, NatAs, 3, 891,
  \dodoi{10.1038/s41550-019-0902-0}

\bibitem[{{Vieira} {et~al.}(2020){Vieira}, {Ruan}, {Haggard}, {Drout}, {Nynka},
  {Boyce}, {Spekkens}, {Safi-Harb}, {Carlberg}, {Fern{\'a}ndez}, {Piro},
  {Afsariardchi}, \& {Moon}}]{VieiraRuan2020}
{Vieira}, N., {Ruan}, J.~J., {Haggard}, D., {et~al.} 2020, \apj, 895, 96,
  \dodoi{10.3847/1538-4357/ab917d}

\bibitem[{{Viets} {et~al.}(2018){Viets}, {Urban}, {et~al.}}]{Viets:2017yvy}
{Viets}, A.~D.~{Wade}, M., {Urban}, A.~L., {et~al.} 2018, \cqg, 35, 095015,
  \dodoi{10.1088/1361-6382/aab658}

\bibitem[{{Villar} {et~al.}(2017){Villar}, {Guillochon}, {Berger}, {Metzger},
  {Cowperthwaite}, {Nicholl}, {Alexand er}, {Blanchard}, {Chornock},
  {Eftekhari}, {Fong}, {Margutti}, \& {Williams}}]{2017ApJ...851L..21V}
{Villar}, V.~A., {Guillochon}, J., {Berger}, E., {et~al.} 2017, \apjl, 851,
  L21, \dodoi{10.3847/2041-8213/aa9c84}

\bibitem[{{Vitale} \& {Chen}(2018)}]{2018PhRvL.121b1303V}
{Vitale}, S., \& {Chen}, H.-Y. 2018, \prl, 121, 021303,
  \dodoi{10.1103/PhysRevLett.121.021303}

\bibitem[{Vitale {et~al.}(2017)Vitale, Lynch, Raymond, Sturani, Veitch, \&
  Graff}]{Vitale:2016avz}
Vitale, S., Lynch, R., Raymond, V., {et~al.} 2017, PhRvD, 95, 064053,
  \dodoi{10.1103/PhysRevD.95.064053}

\bibitem[{Vitale {et~al.}(2014)Vitale, Lynch, Veitch, Raymond, \&
  Sturani}]{Vitale:2014mka}
Vitale, S., Lynch, R., Veitch, J., Raymond, V., \& Sturani, R. 2014, PhRvL,
  112, 251101, \dodoi{10.1103/PhysRevLett.112.251101}

\bibitem[{{Watson} {et~al.}(2020){Watson}, {Butler}, {Lee}, {Becerra},
  {Pereyra}, {Angeles}, {Farah}, {Figueroa}, {G{\'o}nzalez-Buitrago},
  {Quir{\'o}s}, {Ru{\'\i}z-D{\'\i}az-Soto}, {Tejada de Vargas}, {Tinoco}, \&
  {Wolfram}}]{WatsonButler2020}
{Watson}, A.~M., {Butler}, N.~R., {Lee}, W.~H., {et~al.} 2020, \mnras, 159,
  \dodoi{10.1093/mnras/staa161}

\bibitem[{{Watson} {et~al.}(2019){Watson}, {Hansen}, {Selsing}, {Koch},
  {Malesani}, {Andersen}, {Fynbo}, {Arcones}, {Bauswein}, {Covino}, \&
  et~al.}]{WatsonHansen2019}
{Watson}, D., {Hansen}, C.~J., {Selsing}, J., {et~al.} 2019, Nat, 574, 497,
  \dodoi{10.1038/s41586-019-1676-3}

\bibitem[{{Wyrzykowski} \& {Mandel}(2020)}]{Wyrzykowski19massgap}
{Wyrzykowski}, {\L}., \& {Mandel}, I. 2020, \aap, 636, A20,
  \dodoi{10.1051/0004-6361/201935842}

\bibitem[{{Yang} {et~al.}(2019){Yang}, {Bartos}, {Haiman}, {Kocsis},
  {M{\'a}rka}, {Stone}, \& {M{\'a}rka}}]{yang2019}
{Yang}, Y., {Bartos}, I., {Haiman}, Z., {et~al.} 2019, \apj, 876, 122,
  \dodoi{10.3847/1538-4357/ab16e3}

\bibitem[{{Ye} {et~al.}(2020){Ye}, {Fong}, {Kremer}, {Rodriguez}, {Chatterjee},
  {Fragione}, \& {Rasio}}]{ye2020}
{Ye}, C.~S., {Fong}, W.-F., {Kremer}, K., {et~al.} 2020, \apjl, 888, L10,
  \dodoi{10.3847/2041-8213/ab5dc5}

\bibitem[{Yunes \& Pretorius(2009)}]{PhysRevD.80.122003}
Yunes, N., \& Pretorius, F. 2009, PhRvD, 80, 122003,
  \dodoi{10.1103/PhysRevD.80.122003}

\bibitem[{{Zackay} {et~al.}(2019{\natexlab{a}}){Zackay}, {Dai}, {Venumadhav},
  {Roulet}, \& {Zaldarriaga}}]{Zackay:2019btq}
{Zackay}, B., {Dai}, L., {Venumadhav}, T., {Roulet}, J., \& {Zaldarriaga}, M.
  2019{\natexlab{a}}, arXiv e-prints, arXiv:1910.09528.
\newblock \doarXiv{1910.09528}

\bibitem[{{Zackay} {et~al.}(2019{\natexlab{b}}){Zackay}, {Venumadhav}, {Dai},
  {Roulet}, \& {Zaldarriaga}}]{Zackay:2019tzo}
{Zackay}, B., {Venumadhav}, T., {Dai}, L., {Roulet}, J., \& {Zaldarriaga}, M.
  2019{\natexlab{b}}, \prd, 100, 023007, \dodoi{10.1103/PhysRevD.100.023007}

\bibitem[{{Ziosi} {et~al.}(2014){Ziosi}, {Mapelli}, {Branchesi}, \&
  {Tormen}}]{ziosi2014}
{Ziosi}, B.~M., {Mapelli}, M., {Branchesi}, M., \& {Tormen}, G. 2014, \mnras,
  441, 3703, \dodoi{10.1093/mnras/stu824}

\end{thebibliography}

\allauthors
\end{document}